\documentclass[12pt,oneside,reqno,titlepage, hyperfootnotes=false]{amsart}
\usepackage{lmodern}
\usepackage[T1]{fontenc}
\usepackage[latin9]{inputenc}
\usepackage{color}
\usepackage{array}
\usepackage{bm}
\usepackage{amstext}
\usepackage{amsthm}
\usepackage{amssymb}
\usepackage{graphicx}
\usepackage[a4paper]{geometry}
\usepackage{setspace}
\usepackage{wasysym}
\usepackage[authoryear]{natbib}
\usepackage[pdfusetitle,
 bookmarks=true,bookmarksnumbered=false,bookmarksopen=false,
 breaklinks=false,pdfborder={0 0 0},pdfborderstyle={},backref=false,colorlinks=true]
 {hyperref}

\makeatletter

\providecommand{\tabularnewline}{\\}

\numberwithin{equation}{section}
\numberwithin{figure}{section}

\usepackage{amsfonts}
\usepackage{amstext}
\usepackage{amsthm}

\usepackage{hyperref}
\hypersetup{
    colorlinks = true,
    citecolor = black
}

\usepackage{threeparttable}
\usepackage{graphicx}
\usepackage{enumerate}
\usepackage{listings}
\usepackage{subcaption}

\DeclareMathOperator*{\argmax}{arg\,max}

\usepackage{float}

\newtheorem{thm}{Theorem}

\newtheorem{lem}{Lemma}
\newtheorem{definition}{Definition}
\newtheorem*{lem*}{Lemma}
\newtheorem*{thm*}{Theorem}
\newtheorem{cor}{Corollary}
\newtheorem*{asm1}{Assumption 1}
\newtheorem*{asm2}{Assumption 2}
\newtheorem*{asm3}{Assumption 3}
\newtheorem*{asm4}{Assumption 4}
\newtheorem*{asm5}{Assumption 5}
\newtheorem*{asm6}{Assumption 6}

\theoremstyle{definition}

\makeatother

\begin{document}
\title{Continuous time asymptotic representations for adaptive experiments}
\author{Karun Adusumilli$^\dagger$}
\begin{abstract}
This article develops a continuous-time asymptotic framework for analyzing
adaptive experiments---settings in which data collection and treatment
assignment evolve dynamically in response to incoming information.
A key challenge in analyzing fully adaptive experiments, where the
assignment policy is updated after each observation, is that the sequence
of policy rules often lack a well-defined asymptotic limit. To address
this, we focus instead on the empirical allocation process, which
captures the (normalized) number of observations assigned to each
treatment over time. We show that, under general conditions, any adaptive
experiment and its associated empirical allocation process can be
approximated by a limit experiment defined by Gaussian diffusions
with unknown drifts and a corresponding continuous-time allocation
process. This limit representation facilitates the analysis of optimal
decision rules by reducing the dimensionality of the state-space and
exploiting the tractability of Gaussian diffusions. We apply the framework
to derive optimal estimators, analyze in-sample regret for adaptive
experiments, and construct e-processes for anytime-valid inference.
Notably, we introduce the first definition of any-time and any-experiment
valid inference for multi-treatment settings.
\end{abstract}

\thanks{\textit{This version}: \today{}\\
\thispagestyle{empty}\\
$^\dagger$Department of Economics, University of Pennsylvania}
\maketitle

\section{Introduction \label{sec:Introduction}}

Adaptive experiments are experiments where data is collected and analyzed
continuously, allowing for adjustments or decisions to be made on
an ongoing basis. Originating from early work by \citet{wald1947sequential},
\citet{arrow1949bayes}, and others, adaptive experimentation has
evolved to encompass a wide range of strategies, including bandit
experiments, A/B testing, costly sampling, and best-arm identification.
These strategies are now widely used across various fields. For instance,
tech companies frequently employ bandit algorithms and A/B testing
for applications such as web interface optimization, dynamic pricing,
and targeted ad placement (see, \citealp{bouneffouf2019survey} for
a survey of applications). In clinical trials, multi-stage or group
sequential designs (\citealp{wassmer2016group}) have become standard,
allowing early termination of experiments when strong evidence of
positive or negative effects emerges. Economics has also seen a growing
adoption of adaptive experimentation. For example, \citet{kasy2019adaptive}
and \citet{finan2022reinforcing} develop new sequential experimentation
strategies for use in development contexts. More recently, \citet{chapman2024dynamically}
introduced an optimal dynamic strategy for eliciting economic preferences.

Despite these growing number of applications, determining optimal
decision rules in adaptive experiments remains challenging due to
the interactive nature of the data generating process. For instance,
there is currently no well-established notion of optimal point estimation
following an adaptive experiment, nor of inference procedures that
remain valid at any point during the experiment. In this article,
we address these challenges by deriving a continuous-time asymptotic
representation for adaptive experiments. Specifically, we model the
sequential data collection process using an empirical allocation process,
which, at any given time $t$ specifies the fraction of time allocated
to observing outcomes from a particular treatment. We then show that
this empirical allocation process weakly converges to an allocation
process in a limiting experiment, where signals consist of multiple
Gaussian processes with unknown drifts, each corresponding to a different
treatment arm. Then, by characterizing optimal decisions in the limit
experiment, we can construct asymptotically optimal decision rules
in the original experiment. 

The limit experiment greatly simplifies the characterization of optimal
decisions due to two key properties. First, its state space is considerably
smaller than that of the original experiment. In the limit experiment,
the sufficient statistics are the past sample paths of the signal
processes. However, in many applications, we show that these statistics
can be further reduced to just the current values of the signal and
allocation processes, resulting in a state-space dimension of $2K$,
where $K$ is the number of treatments. In contrast, the state space
of the actual experiment encompasses all collected observations, making
it substantially larger and more complex. Second, the limit experiment
is far more tractable to analyze, as the properties of Gaussian diffusions
in continuous time are well understood, allowing us to easily characterize
optimal decisions in that setting.

To illustrate the broad applicability of our framework, we use our
representation theorem to derive optimal decision rules for several
fundamental aspects of adaptive experimentation. In particular, we
address: (1) the construction of optimal estimators following adaptive
experiments, (2) the analysis of in-sample regret, and (3) the development
of e-processes for anytime-valid inference. An e-process is a nonnegative
supermartingale (under the null hypothesis) that continuously tracks
statistical evidence against the null over the course of an experiment.
We introduce the first definition of an e-process for multi-treatment
adaptive experiments. This enables the design of algorithm-free anytime-valid
tests, i.e., tests that maintain correct size even when the sampling
strategy used in the adaptive experiment is unknown.

In each of the above applications, our asymptotic framework significantly
simplifies the decision problem, making it much more tractable. For
example, in the estimation problem, we find that all optimal Bayes
estimators share a common form that is independent of how the experiment
was conducted. Furthermore, we derive explicit expressions for these
estimators under Gaussian priors. 

\subsection{Related literature}

In an important prior work, \citet{hirano2023asymptotic} develop
an asymptotic representation theorem for batched adaptive experiments,
where sampling strategies are updated only a finite number of times
over the course of the experiment. In this setting, they show that
the policy rule and any statistic from the finite-sample experiment
can be matched to a corresponding rule and statistic in a limit experiment
involving Gaussian signals from each batch.

Our asymptotic representation theorem is very different in terms of
both the scope and the formulation. It applies to fully adaptive experiments
and is expressed in terms of allocation processes rather than sequences
of policy rules. This shift is essential, for in fully adaptive experiments,
sequences of policy rules generally fail to admit weak limits. At
the same time, our theory is more specialized in certain respects.
It does not directly provide asymptotic representations for arbitrary
statistics; instead, it characterizes the joint evolution of the score
and the empirical allocation processes, which together fully determine
the limit experiment. Obtaining representations for arbitrary statistics
is more difficult in our setting because there is no straightforward
coupling method for continuous-time processes that respects the required
informational constraints (for example, ensuring that an anytime-valid
test depends only on observed data).

However, for many applications such comprehensive representation theorems
are not always necessary. In practice, lower bounds on losses or risk
can be established using our representation theorem and standard change-of-measure
arguments. For instance, in the case of point estimation, we show
that the frequentist risk of any sequence of estimators is asymptotically
bounded below by the risk of an estimator in the limit experiment
that depends only on the terminal values of the signal and allocation
processes. While this result does not establish a one-to-one mapping
from finite-sample estimators to their counterparts in the limit experiment,
such stronger representation is not essential for deriving lower bounds
or constructing asymptotically optimal estimators.

Beyond these technical contributions, the broader value of our approach
is conceptual. This article provides the first general definition
of adaptive experiments in continuous time through the lens of allocation
processes. This formulation not only simplifies the analysis of adaptive
experiments---continuous time being more tractable than discrete
time---but also yields a sharp characterization of sufficient statistics.
As discussed earlier, in many applications, including point estimation
and anytime-valid inference, the sufficient statistics reduce to the
current values of the signal and allocation processes, resulting in
a finite-dimensional state-space. More generally, allocation processes
offer a natural and flexible framework for representing adaptive experiments,
and our formulation of e-processes is indeed most naturally articulated
in terms of these processes.

In terms of the style of asymptotic approximations, this article is
most closely related to Le Cam's \citeyearpar{LeCam1979} work on
stopping time representations. We extend his framework and build on
his proof techniques to handle the additional complexities introduced
by adaptive sampling.

The systematic study of Gaussian process approximations for adaptive
experiments was initiated by \citet{fan2021diffusion} and \citet{wager2021diffusion},
who introduced diffusion asymptotics to analyze the behavior of adaptive
algorithms such as Thompson Sampling. \citet{kalvit2021closer} extended
this approach to cover UCB algorithms, and \citet{adusumilli2025optimal}
further generalized the framework to characterize optimal bandit algorithms
under both Bayesian and minimax regret criteria. \citet{adusumilli2025optimal}
also established that likelihood ratio processes from finite-sample
adaptive experiments converge uniformly to their diffusion counterparts.
While this convergence is algorithm-agnostic and well-suited for analyzing
Bayesian decision criteria, it is insufficient for tasks such as anytime-valid
inference, as it lacks a representation explicitly connecting the
finite-sample algorithms to a suitable counterpart in the diffusion
limit. This article fills that gap by providing precisely such a representation.

Finally, this article also builds on earlier work by this author \citep{adusumilli2023optimal}
on optimal testing following adaptive experiments. The analysis of
applications such as point estimation and anytime-valid inference
draws on the strategies and proof techniques developed in that work.

\section{Adaptive experiments, Policy rules and Allocation processes\label{sec:Diffusion-asymptotics-and}}

\subsection{An illustrative example\label{subsec:An-illustrative-example}}

We begin with a simple illustration involving two-armed bandits to
motivate our theoretical analysis.

Consider a scenario in which the goal is to identify the better-performing
version of a website, denoted by variants $a=0,1$, each with an unknown
average click-through rate $\theta^{(a)}$. The observed outcome from
each variant is a binary draw from $\text{Bernoulli}(\theta^{(a)})$.
To determine the optimal variant, we run a bandit algorithm for $n$
rounds, sequentially assigning users to one of the two alternatives.

Two of the most widely used bandit algorithms in practice are Thompson
Sampling (TS) and the Upper Confidence Bound (UCB) algorithm. Both
use accumulated data to guide user allocation but differ in their
approach.

The TS algorithm begins with a prior distribution over each $\theta^{(a)}$.
Given the Bernoulli outcome model, a Beta prior is standard; for this
illustration, we will take it to be $\textrm{Beta}(1,1)$. At each
round, the algorithm samples a draw of $\{\theta^{(a)}\}_{a}$ from
its posterior distribution given the past observations, and then allocates
the user to the variant with the highest draw.

In contrast, the UCB algorithm computes an index
\[
\widehat{\text{UCB}}_{j}^{(a)}=\hat{\theta}_{j}^{(a)}+\sqrt{\frac{2\ln(j/n)}{N_{j}^{(a)}}},
\]
where $\hat{\theta}_{j}^{(a)}$ is the sample mean of outcomes for
variant $a$, and $N_{j}^{(a)}$ is the number of users allocated
to variant $a$ prior to round $j$. The user is assigned to the variant
with the higher index.

Define time $t:=j/n\in[0,1]$ as the fraction of the experiment completed.
For either algorithm, let
\[
q_{n,1}(t):=\frac{1}{n}\sum_{j=1}^{\lfloor nt\rfloor}\mathbb{I}\{A_{j}=1\}
\]
denote the number of assignments to variant 1 up to time $t$, normalized
by $n$.

\begin{figure}
\includegraphics[totalheight=5cm,height=4cm]{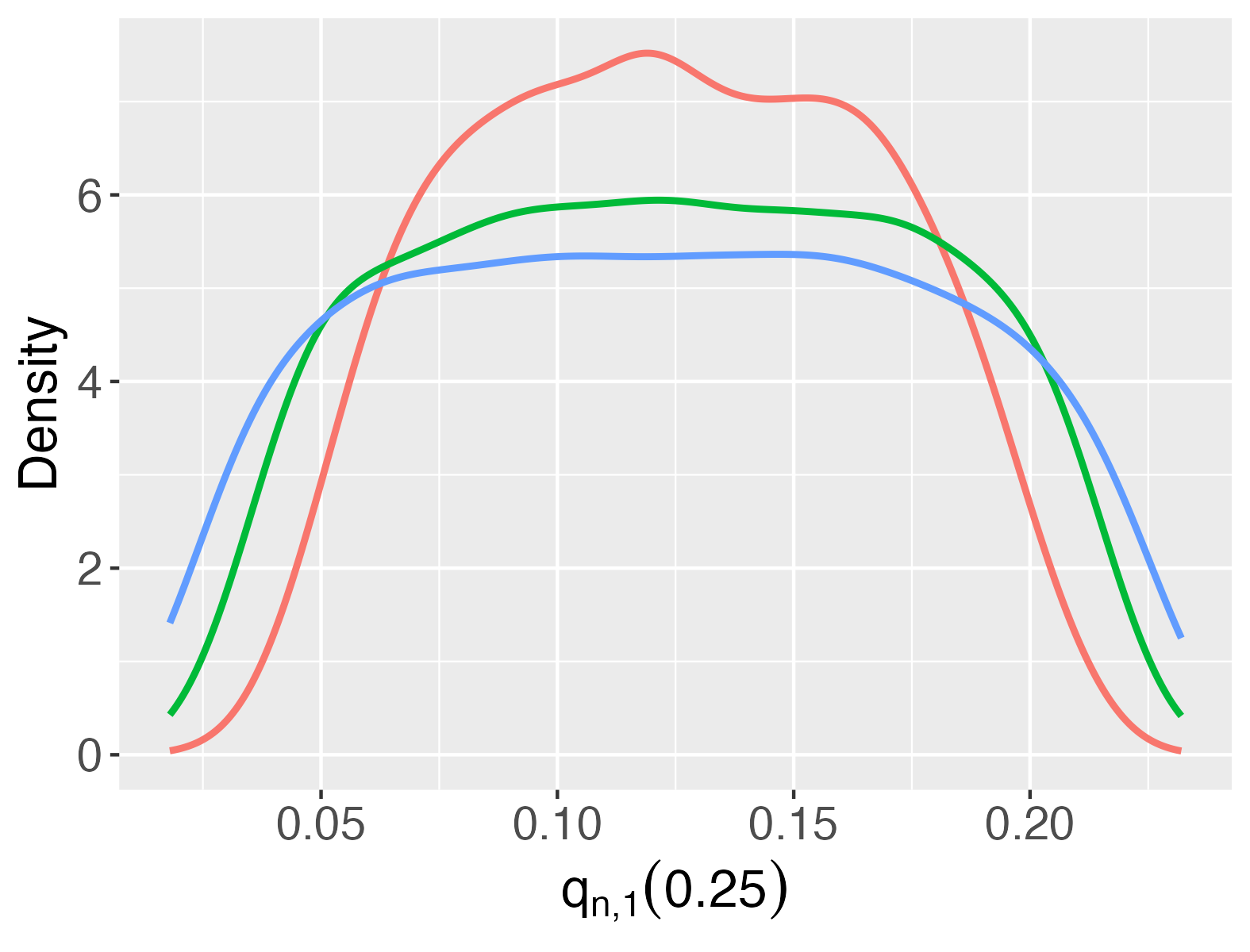}~\includegraphics[totalheight=5cm,height=4cm]{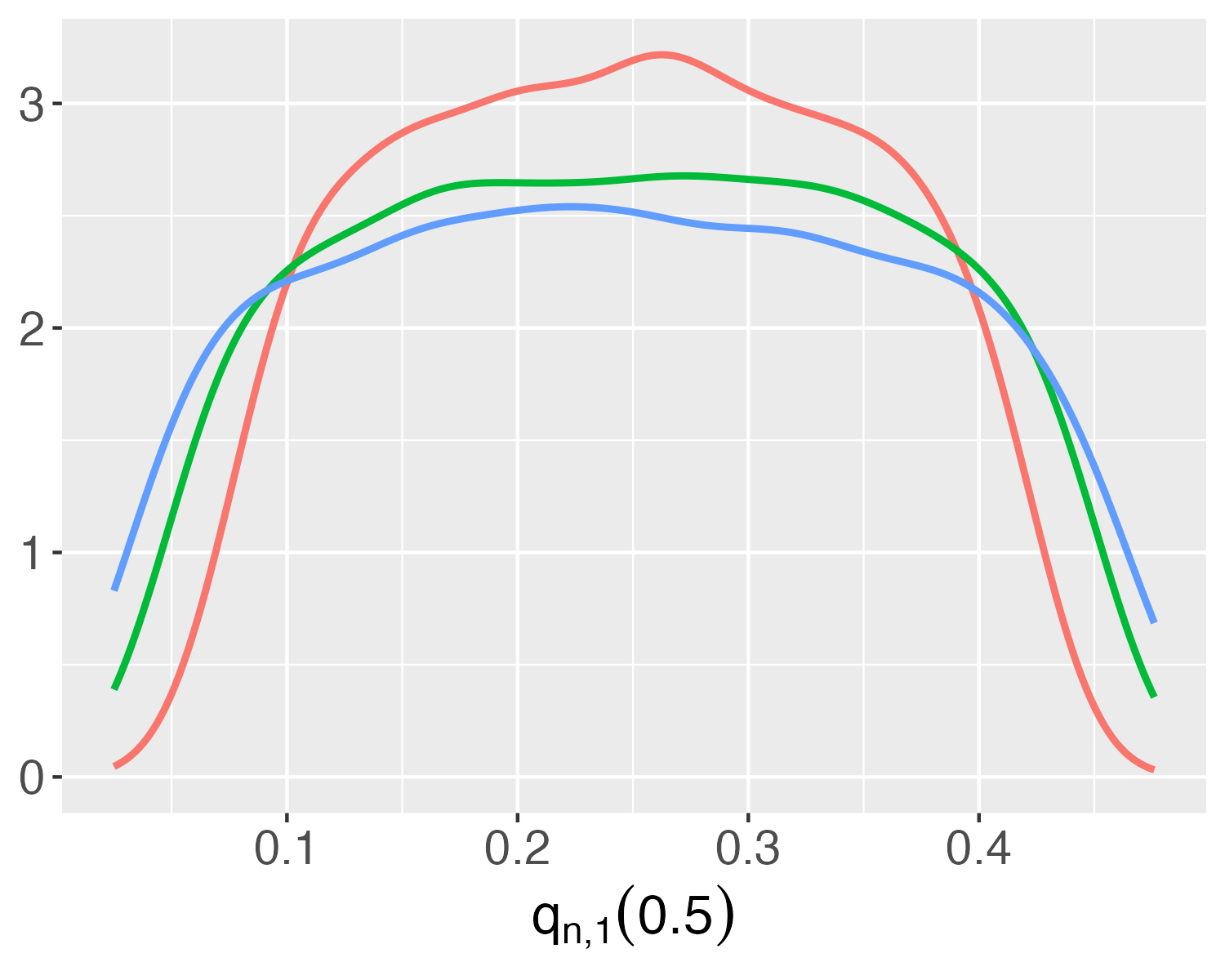}~\includegraphics[totalheight=5cm,height=4cm]{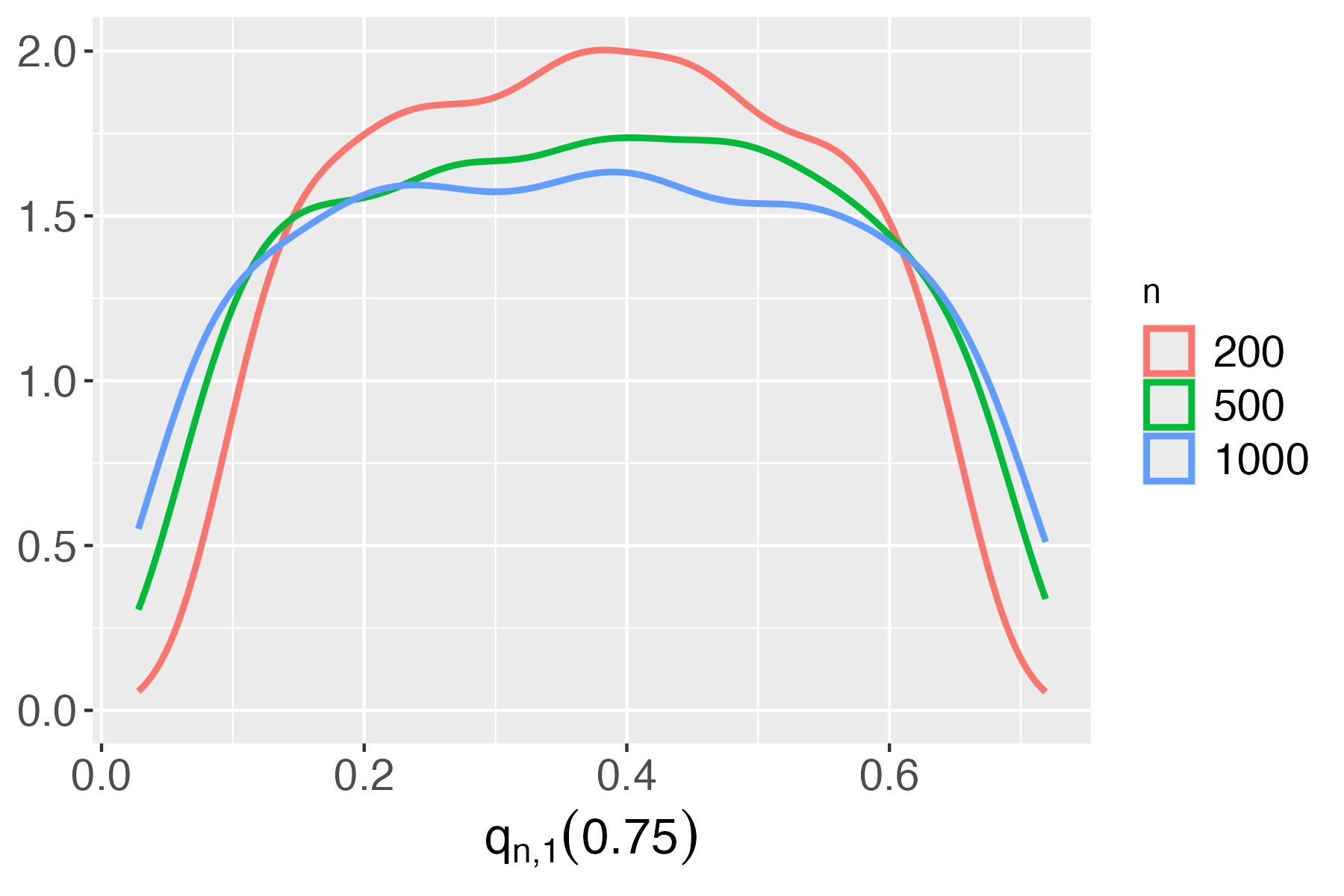}

{\scriptsize Note: Results from 2-armed bandit experiment with $Y^{(a)}\sim\textrm{Bernoulli}(\theta^{(a)})$
and $\theta^{(1)}=\theta^{(0)}=0.1$. }{\scriptsize\par}

\caption{Distribution of $q_{n,1}(t)$ under Thompson Sampling\label{fig:q1_plots_TS}}
\end{figure}

\begin{figure}
\includegraphics[totalheight=5cm,height=4cm]{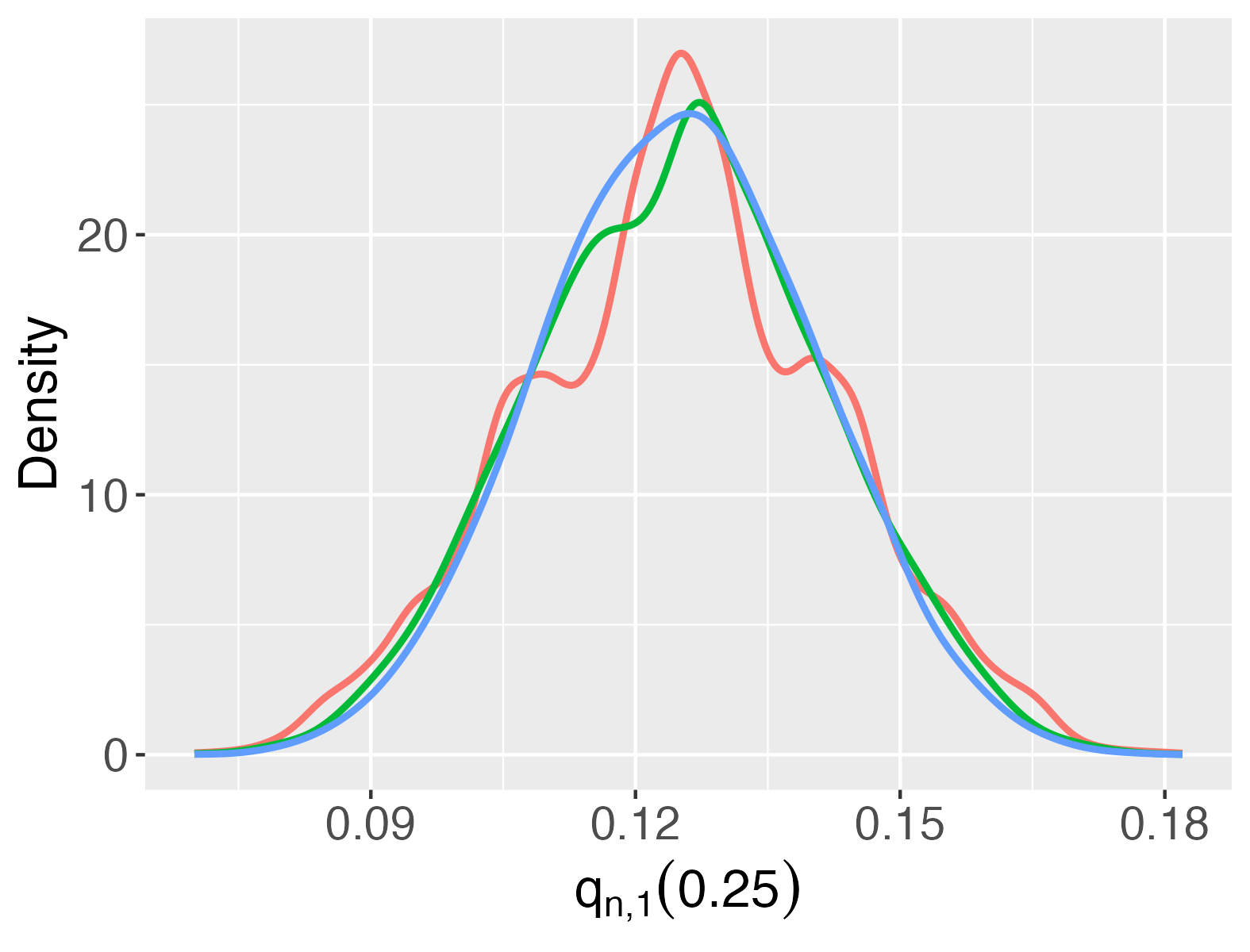}~\includegraphics[totalheight=5cm,height=4cm]{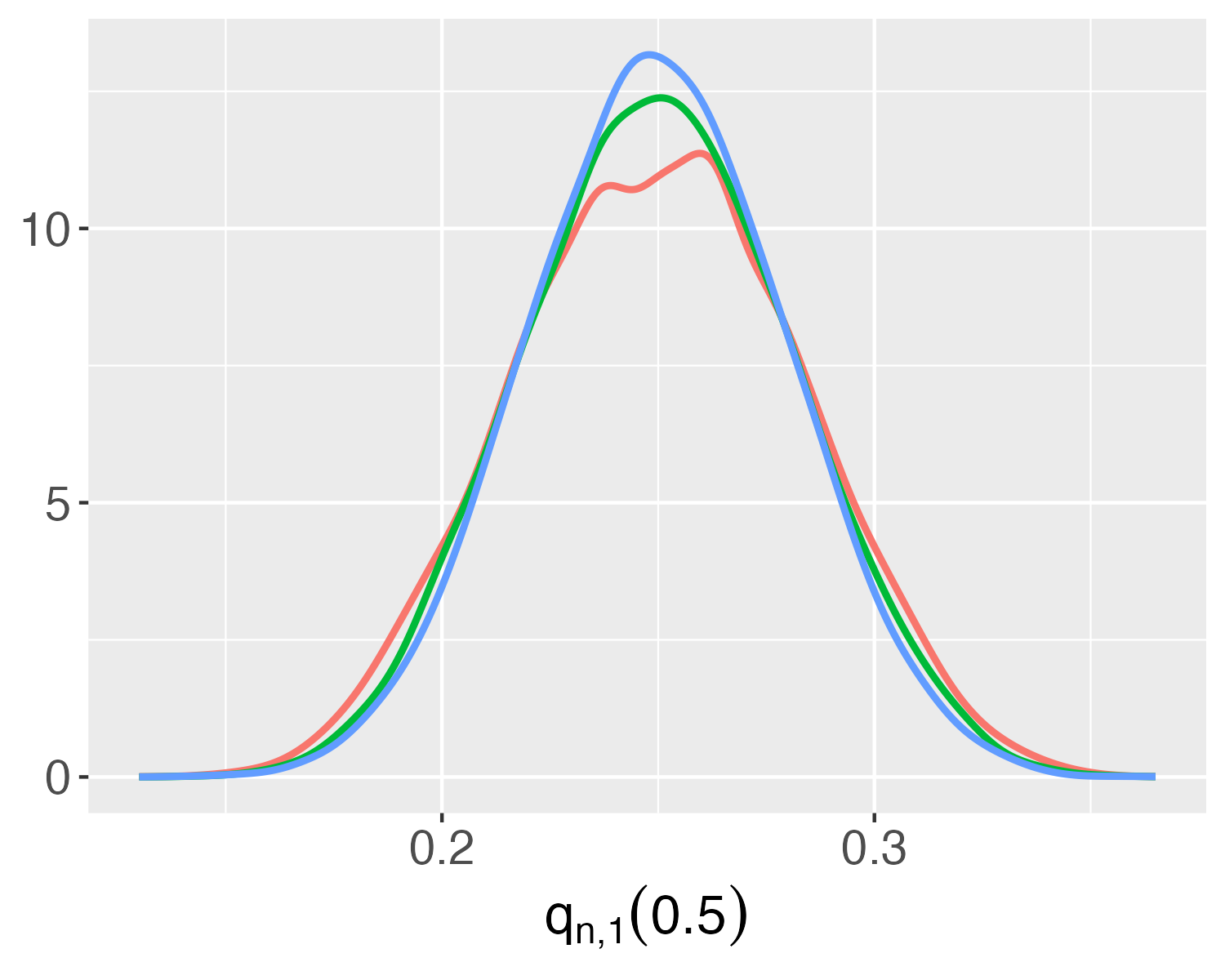}~\includegraphics[totalheight=5cm,height=4cm]{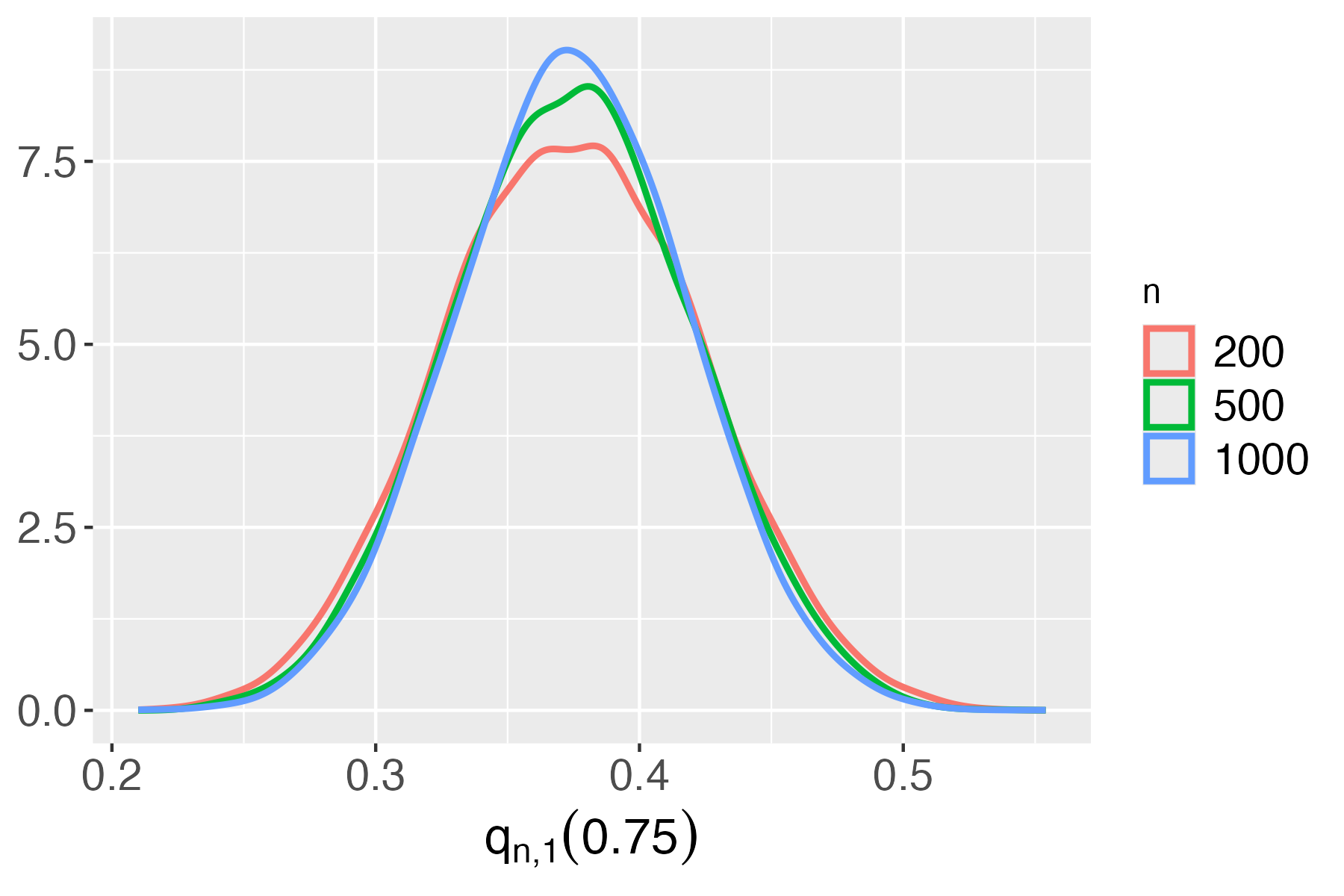}

{\scriptsize Note: Results from 2-armed bandit experiment with $Y^{(a)}\sim\textrm{Bernoulli}(\theta^{(a)})$
and $\theta^{(1)}=\theta^{(0)}=0.1$. }{\scriptsize\par}

\caption{Distribution of $q_{n,1}(t)$ under UCB\label{fig:q1_plots_UCB}}
\end{figure}

Suppose the two variants are identical, i.e., $\theta^{(0)}=\theta^{(1)}$.
Figure \ref{fig:q1_plots_TS} plots the sampling distribution of $q_{n,1}(t)$
at three time points---$t=0.25,0.5,0.75$---under TS, for various
values of $n$. As $n$ increases, the distributions converge, illustrating
a form of asymptotic stability. The same convergence also occurs under
UCB, as shown in Figure \ref{fig:q1_plots_UCB}, although the sampling
distributions differ quite substantially. The figures are plotted
for $\theta^{(0)}=\theta^{(1)}=0.1$, but changing these values would
not make much of a difference to the plots (as long as $\theta^{(0)}=\theta^{(1)}$).

This convergence behavior of the empirical allocation process $q_{n,1}(\cdot)$
is not specific to these algorithms. A central result of this article
is that under mild conditions, such asymptotic convergence holds for
almost every adaptive allocation rule. To this end, we start by describing
a general setup for adaptive experiments. 

\subsection{General setup }

Adaptive experiments involve multiple treatments, where the policy
rule---i.e., the probability of allocation to each treatment---can
continuously adjust over the course of the experiment. 

Let $K$ denote the number of treatments under consideration. To simplify
notation and proofs, we focus on the case $K=2$, though our results
extend to any fixed $K$. The outcome $Y^{(a)}$ under treatment $a\in\{0,1\}$
follows a parametric model $\{P_{\theta^{(a)}}^{(a)}\}$, where $\theta^{(a)}\in\mathbb{R}^{d}$
is an unknown parameter. For simplicity, we assume $\theta^{(0)}$
and $\theta^{(1)}$ have the same dimension, though this is not required
for our results. 

Since only ever observe one potential outcome $Y^{(a)}$ per unit,
we can assume $\{P_{\theta^{(a)}}^{(a)}\}_{a}$ are independent across
$a$, conditional on $(\theta^{(1)},\theta^{(0)})$. Let $j=1,\dots,n$
index the experimental periods. We define time $t$ as the scaled
number of periods, $t=j/n$, representing the fraction of the experiment
completed.

The decision-maker (DM) employs a policy rule $\{\pi_{n,j}\}_{j}\equiv\{\pi_{n,\lfloor nt\rfloor}\}_{t}$,
which prescribes the probability of assigning observation $j$ to
treatment 1 based on past information. The treatment assignment follows
$A_{j}\sim\text{Bernoulli}(\pi_{n,j})$. For the outcomes, it is useful
to conceptualize a stack of observations $\ensuremath{\bm{y}_{a}:=\{Y_{i}^{(a)}\}_{i=1}^{n}}$
for each treatment, generated at the outset as i.i.d draws from $\{P_{\theta^{(a)}}^{(a)}\}$,
but unobserved initially by the DM. This is the so-called stack-of-rewards
model (\citealp[Section 4.6]{lattimore2020bandit}). Each time a treatment
is sampled, it can be imagined that the DM observes the top element
of the corresponding treatment stack; this element is then taken out
of consideration. 

\subsubsection{Empirical allocation processes\label{subsec:Empirical-allocation-processes}}

As in the illustrative example, let
\[
q_{n,a}(t):=\frac{1}{n}\sum_{j=1}^{\lfloor nt\rfloor}\mathbb{I}\{A_{j}=a\}
\]
denote the fraction of (total) observations assigned to treatment
$a$ up to time $t$. We term $\{q_{n,a}(\cdot)\}_{a}$ the empirical
allocation process. One can interpret the policy rule $\{\pi_{n,j}\}_{j}$
as a function mapping the stacks $\ensuremath{(\bm{y}_{1},\bm{y}_{0})}$
and an exogenous random variable $U\sim\text{Uniform}[0,1]$ to the
observed trajectory of $\{q_{n,a}(\cdot)\}_{a}$. The exogenous randomness
accounts for probabilistic policy rules; specifically, we subsume
the policy randomizations at all $n$ stages of the adaptive experiment
into a single $U$.\footnote{Indeed, a single uniform random variable can be mapped into countably
infinite independent uniform random variables.} The key informational constraint is that the event $\{q_{n,1}(t)\leq\gamma_{1},q_{n,0}(t)\leq\gamma_{0}\}$
depends only on the first $\ensuremath{\lfloor n\gamma_{1}\rfloor},\ensuremath{\lfloor n\gamma_{0}\rfloor}$
observations from $(\bm{y}_{1},\bm{y}_{0})$ and the exogenous randomization
$U$. Formally, 
\[
\{q_{n,1}(t)\leq\gamma_{1},q_{n,0}(t)\leq\gamma_{0}\}\text{ is }\mathcal{G}_{n,\gamma_{0},\gamma_{1}}:=\mathcal{F}_{n,\gamma_{1}}^{(1)}\lor\mathcal{F}_{n,\gamma_{0}}^{(0)}\lor\sigma(U)\text{-measurable}
\]
for each $t,\gamma_{1},\gamma_{0}\in[0,1]$, where
\[
\mathcal{F}_{n,\gamma}^{(a)}:=\sigma\left(Y_{1}^{(a)},\dots,Y_{\lfloor n\gamma\rfloor}^{(a)}\right)
\]
is the filtration (i.e., information set) generated by the first $\lfloor n\gamma\rfloor$
observations from stack $\bm{y}_{a}$. Thus, each policy rule $\{\pi_{n,j}\}_{j}$
can be associated with monotonic empirical allocation processes $\{q_{n,a}(\cdot)\}_{a}$
satisfying the above informational constraint.

\subsubsection{Local asymptotics}

We are interested in the behavior of various statistics under local
perturbations of the form ${\theta_{0}^{(a)}+h^{(a)}/\sqrt{n}:h^{(a)}\in\mathbb{R}^{d}}$,
where $\theta_{0}^{(a)}$ denotes a reference parameter. This focus
is motivated by the fact that many decisions in adaptive experiments
hinge on distinguishing between parameter values that are close to
one another. For example, in the website optimization setting discussed
earlier, bandit algorithms are employed precisely because the differences
between website variants tend to be small. \citet{deng2013improving}
survey industry practices and report that typical differences in click-through
rates are often around 1\% or less. In such cases, bandit algorithms
are employed for detecting subtle effects using as little data as
possible. This naturally places us in the domain of local asymptotics.

In this article, we analyze the behavior of adaptive algorithms under
local asymptotics for parametric classes of families $\{P_{\theta}^{(a)}\}_{\theta}$.
Let $\nu$ denote a dominating measure for $\{P_{\theta}^{(a)}:\theta\in\mathbb{R}^{d},a\in\{0,1\}\}$,
and set $p_{\theta}^{(a)}:=dP_{\theta}^{(a)}/d\nu$. We require $\{P_{\theta}^{(a)}\}_{\theta}$
to be quadratic mean differentiable (qmd): 

\begin{asm1} The class $\{P_{\theta}^{(a)}:\theta\in\mathbb{R}^{d}\}$
is qmd around $\theta_{0}^{(a)}$ for each $a\in\{0,1\}$, i.e., there
exists a score function $\psi_{a}(\cdot)$ such that for each $h^{(a)}\in\mathbb{R}^{d},$
\[
\int\left[\sqrt{p_{\theta_{0}^{(a)}+h^{(a)}}^{(a)}}-\sqrt{p_{\theta_{0}^{(a)}}^{(a)}}-\frac{1}{2}h^{(a)\intercal}\psi_{a}\sqrt{p_{\theta_{0}^{(a)}}^{(a)}}\right]^{2}d\nu=o(\vert h^{(a)}\vert^{2}).
\]
Furthermore, the information matrix $I_{a}:=\mathbb{E}_{0}[\psi_{a}\psi_{a}^{\intercal}]$
is invertible for $a\in\{0,1\}$. \end{asm1}

In the illustrative example, the outcomes are Bernoulli, so Assumption
1 holds with $\psi_{a}(y)=\left(\theta_{0}^{(a)}(1-\theta_{0}^{(a)})\right)^{-1}(y-\theta_{0}^{(a)})$.
More broadly, this assumption is rather mild and satisfied for almost
all commonly used distributions, including the Normal, Cauchy, Exponential,
and Poisson distributions.

\subsubsection{Score processes}

For each $q\in[0,1]$, define $z_{n,a}(q)$ as the partial sum process
\[
z_{n,a}(q):=\frac{I_{a}^{-1/2}}{\sqrt{n}}\sum_{i=1}^{\left\lfloor nq\right\rfloor }\psi_{a}(Y_{i,j}^{(a)}).
\]
Knowledge of the process, $z_{n,a}(\cdot)$, on the domain $[0,q]$
is equivalent to knowledge of the scores from the first $\left\lfloor nq\right\rfloor $
observations of the stack ${\bf y}_{a}$. We then define the score
process for treatment $a$ as
\[
x_{n,a}(t):=z_{n,a}(q_{n,a}(t));\ a\in\{0,1\}.
\]
As we will show, the sample paths of this process serve as an asymptotically
sufficient statistic for the adaptive experiment.

\subsection{The limit experiment}

The primary result of this article establishes that any adaptive experiment
is asymptotically Blackwell equivalent to a limit experiment characterized
by Gaussian diffusions.

In this limit experiment, the decision maker observes a Gaussian process
signal $z_{a}(\cdot)$ associated with each treatment $a$, given
by
\begin{equation}
z_{a}(q)=I_{a}^{1/2}h^{(a)}q+W_{a}(q),\label{eq:definition of z_a()}
\end{equation}
where $\{\ensuremath{W_{a}(\cdot)\}_{a}}$ are independent $d$-dimensional
Wiener processes, and the drifts $\{h^{(a)}\}_{a}$ are unknown. Intuitively,
$z_{a}(\cdot)$ serves as the limiting counterpart of $z_{n,a}(\cdot)$
in the original experiment, while the index $q$ represents the amount
of `attention' devoted to that particular treatment. 

Define the natural filtration generated by the Gaussian process $z_{a}(\cdot)$
up to a given attention-value $\gamma$ as $\mathcal{F}_{\gamma}^{(a)}:=\sigma\{z_{a}(s):s\le\gamma\}$.
We can and do take $\mathcal{F}_{\gamma}^{(a)}$ to be right-continuous,
i.e., $\mathcal{F}_{\gamma}^{(a)}\equiv\bigcap_{\epsilon\downarrow0}\mathcal{F}_{\gamma+\epsilon}^{(a)}.$
Similar to the stack of rewards in the original experiment, the entire
process $z_{a}(\cdot)$ is not immediately observed. Instead, at time
$t$, the DM observes the sample paths of $z_{1}(\cdot)$ and $z_{0}(\cdot)$
over the intervals $[0,q_{1}(t)]$ and $[0,q_{0}(t)]$, respectively,
where $q_{a}(t)$ represents the amount of attention devoted to treatment
$a$ up to time $t$. The quantities $\{q_{a}(t)\}_{a}$, termed allocation
processes, are continuous-time analogues of $\{q_{n,a}(t)\}_{a}$,
and formally defined as follows:

\begin{definition}\label{Allocation_process_definition}Let $(z_{1}(\cdot),z_{0}(\cdot),U)$
represent a collection of independent stochastic processes and random
variables defined on a common probability space $(\Omega,\mathcal{F},\mathbb{P})$,
where $\{z_{a}(\cdot)\}_{a}$ are defined as in (\ref{eq:definition of z_a()})
and $U$ is an exogenous Uniform $[0,1]$ random variable. A collection
of non-negative stochastic processes, $\{q_{a}(\cdot)\}_{a}$, indexed
by $t\in[0,1]$, is termed an \textbf{allocation process} if:\\
(i) With probability 1, $q_{1}(t)+q_{0}(t)=t\ \forall\ t$;\\
(ii) Each $q_{a}(\cdot)$ is almost surely non-decreasing; and \\
(iii) For any $\gamma_{1},\gamma_{0},t\in[0,1]$ such that $\gamma_{1}+\gamma_{0}\ge t$,
the event $\{q_{1}(t)\le\gamma_{1},q_{0}(t)\le\gamma_{0}\}$ is measurable
with respect to $\mathcal{G}_{\gamma_{1},\gamma_{0}}$, the augmented
version of the filtration $\mathcal{\mathcal{F}}_{\gamma_{1}}^{(1)}\lor\mathcal{\mathcal{F}}_{\gamma_{0}}^{(0)}\lor\sigma(U)$.\footnote{The augmented version of $\mathcal{\mathcal{F}}_{\gamma_{1}}^{(1)}\lor\mathcal{\mathcal{F}}_{\gamma_{0}}^{(0)}\lor\sigma(U)$
is the smallest filtration containing $\mathcal{\mathcal{F}}_{\gamma_{1}}^{(1)}\lor\mathcal{\mathcal{F}}_{\gamma_{0}}^{(0)}\lor\sigma(U)$
that includes every null set of $(\Omega,\mathcal{F},\mathbb{P})$,
i.e., every $A\in\mathcal{F}$ such that $\mathbb{P}(A)=0$. } \end{definition}

Definition \ref{Allocation_process_definition} addresses the technical
challenge of defining adaptive experiments in continuous time without
recourse to policy rules, which are typically non-measurable (see
Section \ref{subsec:Allocation-processes-vs} below). Instead, it
defines allocation processes abstractly through their functional properties.
The first two conditions on $q_{a}(\cdot)$ are straightforward. The
third condition ensures that, aside from the exogenous randomization
$U$, whether or not $q_{1}(t)\le\gamma_{1}$ and $q_{0}(t)\le\gamma_{0}$
hold can be determined based only on the sample paths, $\{z_{a}(s):s\le\gamma_{a}\}$,
of the signal processes.\footnote{Although the informational constraint is only explicitly required
for $\gamma_{1}+\gamma_{0}\ge t$, the requirement $q_{1}(t)+q_{0}(t)=t$
almost surely implies that the event $\{q_{1}(t)\le\gamma_{1},q_{0}(t)\le\gamma_{0}\}$
is a measure-zero set when $\gamma_{1}+\gamma_{0}<t$. Consequently,
because $\mathcal{G}_{\gamma_{1},\gamma_{0}}$ is an augmented filtration,
this event is measurable with respect to $\mathcal{G}_{\gamma_{1},\gamma_{0}}$
for any $\gamma_{1}+\gamma_{0}<t$ as well.} It is the continuous time counterpart of the information constraint
on empirical allocation processes, described in Section \ref{subsec:Empirical-allocation-processes}.
The condition is analogous to the usual definition of a stopping time,
but extended to a multi-dimensional setting.

The quantities $\{z_{a}(\cdot),q_{a}(\cdot),U\}_{a}$ characterize
the limit adaptive experiment. Importantly, the inputs to the processes
$z_{a}(\cdot),q_{a}(\cdot)$ are different: $\{q_{a}(\cdot)\}_{a}$
are indexed by time, while $\{z_{a}(\cdot)\}_{a}$ are indexed by
the attention devoted to each treatment.

\subsubsection{Sufficient statistics}

For each $a\in\{0,1\}$, define $x_{a}(t):=z_{a}(q_{a}(t))$ as the
limit counterpart of $x_{n,a}(t)$. It is straightforward to verify
that the sample paths of $\{x_{a}(\cdot)\}_{a}$ constitute sufficient
statistics for the limit experiment up to time $t$. 

Let $\mathcal{I}_{t}:=\mathcal{G}_{q_{1}(t),q_{0}(t)}$ denote the
information accrued from the experiment until time $t$.\footnote{Formally, $\mathcal{G}_{q_{1}(t),q_{0}(t)}\equiv\left[A\in\mathcal{F}:A\cap\left\{ q_{1}(t)\le\gamma_{1},q_{0}(t)\le\gamma_{0}\right\} \in\mathcal{G}_{\gamma_{1},\gamma_{0}}\ \forall\ \gamma_{1},\gamma_{0}\in[0,1]\right]$. }
An important property of $x_{1}(\cdot),x_{0}(\cdot)$ is that they
are $\mathcal{I}_{t}$-martingales when $\bm{h}:=(h^{(1)},h^{(0)})=(0,0)$.
Furthermore, $q_{a}(t)\textup{I}_{d}$ is the quadratic variation
of $x_{a}(t)$, in that it captures the accumulated variability of
$x_{a}(t)$.

\begin{lem} \label{Lemma: Martingale} Under $\bm{h}=(0,0)$, the
processes $x_{1}(\cdot),x_{0}(\cdot)$ are $\mathcal{I}_{t}$-martingales
with quadratic variations $q_{1}(t)\textup{I}_{d},q_{0}(t)\textup{I}_{d}$,
where $\textup{I}_{d}$ is the $d$-dimensional identity matrix.\end{lem} 

\subsubsection{Allocation processes vs policy rules\label{subsec:Allocation-processes-vs}}

Theorem \ref{Thm: ART} in Section \ref{sec:Equivalence-of-experiments:}
states that we can take $q_{a}(\cdot)$ to be almost surely Lipschitz
continuous with a Lipschitz constant of $1$. By the fundamental theorem
of Lebesgue integral calculus, almost every sample path of $q_{a}(\cdot)$
is then differentiable almost everywhere, with a Lebesgue integrable
derivative $\pi_{a}(t):=\frac{dq_{a}(t)}{dt}.$ Consequently, $q_{a}(t)=\int_{0}^{t}\pi_{a}(s)ds$.
Moreover, Definition \ref{Allocation_process_definition} and the
right continuity of $\mathcal{G}_{q_{1},q_{0}}$ implies that $\pi_{a}(t)$
is $\mathcal{I}_{t}$-measurable coordinate-wise, for each $t$.

However, $\pi(\cdot)\equiv\pi_{1}(\cdot)$ cannot, in general, be
interpreted as a valid policy rule, since it need not be measurable
as a random process indexed by $t$. Indeed, in many continuous-time
optimal control problems, the optimal policy does not belong to a
separable space, and is therefore not measurable.\footnote{The measurability of sample paths---often referred to as strong or
Bochner measurability---differs from the weaker notion of coordinate-wise
measurability. The Pettis measurability theorem states that a stochastic
process is Bochner measurable if and only if its sample paths lie
in a separable subspace with probability one. See \citet[Section 5.1.1]{adusumilli2023optimal}
for further discussion.} For this reason, it is also generally impossible to define weak convergence
for a sequence of policies $\pi_{n,\lfloor nt\rfloor}$, as such sequences
are typically not asymptotically equicontinuous.

These observations indicate that the policy rule is not a particularly
suitable object for characterizing sequential strategies in the limit
experiment. In contrast, as will be established in Theorem \ref{Thm: ART}
below, any sequence of empirical allocation processes converge weakly
to an allocation process in the limit experiment. In this sense, the
allocation process serves as the more fundamental representation of
sequential decision-making.

\section{Asymptotic equivalence of experiments\label{sec:Equivalence-of-experiments:}}

We now establish the asymptotic equivalence between the original sequence
of adaptive experiments and the limit experiment. This equivalence
follows from two key results. The first, previously demonstrated in
\citet{adusumilli2025optimal}, states that the likelihood ratio processes
in the original experiment converge uniformly, at all possible `attention'
values, to their counterparts in the limit experiment. We restate
this result here for completeness. The second result, which is novel
to this article, asserts that any sequence of score and allocation
processes, $\{x_{n,a}(\cdot),q_{n,a}(\cdot)\}_{a}$, admits a corresponding
representation in the limit experiment. 

\subsection{Convergence of likelihood ratio processes\label{subsec:Convergence-of-likelihood-ratios}}

Let $\mathbb{P}_{n,\bm{h}}$ denote the induced probability over the
stacked rewards and the exogenous randomization, i.e., over $({\bf y}^{(1)},{\bf y}^{(0)},U)$
when $Y^{(a)}\sim P_{\theta_{0}+h^{(a)}/\sqrt{n}}^{(a)}$. For each
$a$, denote ${\bf y}_{\left\lfloor nq\right\rfloor }^{(a)}:=\{Y_{i}^{(a)}\}_{i=1}^{\left\lfloor nq\right\rfloor }$.
Suppose that we observe $U$ and $\left\lfloor n\gamma_{1}\right\rfloor ,\left\lfloor n\gamma_{0}\right\rfloor $
units from each treatment, i.e., we observe $U$ and ${\bf y}_{\gamma_{1},\gamma_{0}}:=\left({\bf y}_{\left\lfloor n\gamma_{1}\right\rfloor }^{(1)},{\bf y}_{\left\lfloor n\gamma_{0}\right\rfloor }^{(0)}\right)$.
Given this set of observations, the log-likelihood ratio process under
the local alternative $\bm{h}:=(h^{(1)},h^{(0)})$ is:
\begin{align*}
\hat{\varphi}(\bm{h};\gamma_{1},\gamma_{0}) & =\ln\frac{dP_{\theta_{0}^{(1)}+h^{(1)}/\sqrt{n}}^{(1)}}{dP_{\theta_{0}^{(1)}}^{(1)}}\left({\bf y}_{\left\lfloor n\gamma_{1}\right\rfloor }^{(1)}\right)+\ln\frac{dP_{\theta_{0}^{(0)}+h^{(0)}/\sqrt{n}}^{(0)}}{dP_{\theta_{0}^{(0)}}^{(0)}}\left({\bf y}_{\left\lfloor n\gamma_{0}\right\rfloor }^{(0)}\right)\\
 & :=\hat{\varphi}^{(1)}(h^{(1)};\gamma_{1})+\hat{\varphi}^{(0)}(h^{(0)};\gamma_{0}),
\end{align*}
where, for any $a\in\{0,1\}$ and $\gamma\in[0,1]$,
\[
\ln\frac{dP_{\theta_{0}^{(a)}+h^{(a)}/\sqrt{n}}^{(a)}}{dP_{\theta_{0}^{(a)}}^{(a)}}\left({\bf y}_{\left\lfloor n\gamma\right\rfloor }^{(a)}\right):=\sum_{i=1}^{\left\lfloor n\gamma\right\rfloor }\ln\frac{dP_{\theta_{0}^{(a)}+h^{(a)}/\sqrt{n}}^{(a)}}{dP_{\theta_{0}^{(a)}}^{(a)}}\left(Y_{i}^{(a)}\right).
\]
In \citet{adusumilli2025optimal}, this author showed that under Assumption
1, 
\begin{equation}
\hat{\varphi}^{(a)}(h^{(a)};\gamma)=h^{(a)\intercal}I_{a}^{1/2}z_{n,a}(\gamma)-\frac{\gamma}{2}h^{(a)\intercal}I_{a}h^{(a)}+o_{\mathbb{P}_{n,0}}(1)\ \textrm{uniformly over }\gamma\in[0,1].\label{eq:SLAN property}
\end{equation}

Analogously, in the limit experiment, the relevant probability measure
is $\mathbb{P}_{\bm{h}}:=\mathbb{P}_{h^{(1)}}^{(1)}\otimes\mathbb{P}_{h^{(0)}}^{(0)}\otimes P_{U}$,
where $\mathbb{P}_{h^{(a)}}^{(a)}$ is the induced probability over
the sample paths of $\{z_{a}(s);0\le s\le1\}$ when the local parameter
is $h^{(a)}$, and $P_{U}$ is the probability measure induced by
$U\sim\textrm{Uniform}[0,1]$. Also, for some fixed $\{\gamma_{a}\}_{a}$,
let
\[
\varphi^{(a)}(h^{(a)};\gamma_{a})=\mathbb{E}_{\mathbb{P}_{0}^{(a)}}\left[\left.\ln\frac{d\mathbb{P}_{h^{(a)}}^{(a)}}{d\mathbb{P}_{0}^{(a)}}\right|\bar{\mathcal{F}}_{\gamma_{a}}^{(a)}\lor\sigma(U)\right]
\]
denote the log-likelihood ratio under the local alternative $h^{(a)}$
given the sample path $\{z_{a}(s);s\le\gamma_{a}\}$ and $U$. Similarly,
$\varphi(\bm{h};\gamma_{1},\gamma_{0})$ denotes the likelihood ratio
under $\bm{h}=(h^{(1)},h^{(0)})$ given the sample paths $\{z_{a}(s);s\le\gamma_{a}\}_{a}$
and $U$. Since $z_{1}(\cdot),z_{0}(\cdot)$ are Wiener processes
under $\mathbb{P}_{\bm{0}}$, the Girsanov theorem implies 
\begin{equation}
\varphi^{(a)}(h^{(a)};\gamma_{a})=h^{(a)\intercal}I_{a}^{1/2}z_{a}(\gamma_{a})-\frac{\gamma_{a}}{2}h^{(a)\intercal}I_{a}h^{(a)}.\label{eq:Girsanov theorem}
\end{equation}
Furthermore, as the Wiener processes are independent, $\varphi(\bm{h};\gamma_{1},\gamma_{0})=\varphi^{(1)}(h^{(1)};\gamma_{1})+\varphi^{(0)}(h^{(0)};\gamma_{0}).$

Equations (\ref{eq:SLAN property}), (\ref{eq:Girsanov theorem})
imply $\hat{\varphi}^{(a)}(h^{(a)};\cdot)\xrightarrow[\mathbb{P}_{n,0}]{d}\varphi^{(a)}(h^{(a)};\cdot)$
for each $a,h^{(a)}$ and therefore, 
\begin{equation}
\hat{\varphi}(\bm{h};\cdot,\cdot)\xrightarrow[\mathbb{P}_{n,0}]{d}\varphi(\bm{h};\cdot,\cdot)\ \textrm{for each }\text{\ensuremath{\bm{h}}}.\label{eq:equivalence of LRs}
\end{equation}
By itself, (\ref{eq:equivalence of LRs}) does not reference a specific
allocation process. While it describes convergence of likelihood ratio
processes across the space of all possible `attention' values $(\gamma_{1},\gamma_{0})$,
it does not immediately guarantee convergence at all possible time
points $t$. Indeed, the relationship between the observed attention
values $(\gamma_{1},\gamma_{0})$ and time $t$ is determined by the
specific allocation process being employed. 

To demonstrate that a risk function in the finite-sample experiment
admits a corresponding representation or lower bound in the limit
experiment, we need to go further and match the joint distribution
of the asymptotically sufficient statistics $\{x_{n,a}(\cdot),q_{n,a}(\cdot)\}_{a}$
as well. Combining this joint convergence with (\ref{eq:equivalence of LRs})
would yield uniform convergence of likelihood ratios across all time
points, and thereby imply asymptotic equivalence between the actual
and limit experiments in the sense of \citet{le1986asymptotic}. The
following key result establishes this joint convergence.

\subsection{The main result}

\begin{thm} \label{Thm: ART}Suppose Assumption 1 holds. Let $\{x_{n,a}(\cdot),q_{n,a}(\cdot)\}_{a}$
be any sequence of score and allocation processes induced by a sequence
of policies $\{\pi_{n,j}\}_{j}$ in the actual experiment. Then, there
exists a further subsequence, $\{n_{k}\}_{k=1}^{\infty}$, and a random
collection $\{z_{a}(\cdot),q_{a}(\cdot),U\}_{a}$ defined on a probability
space $(\Omega,\mathcal{F},\mathbb{P})$ such that:\\
(i) $\{z_{a}(\cdot)\}_{a}$ are independent standard $d$-dimensional
Wiener processes and $U\sim\textrm{Uniform}[0,1]$ is independent
of $\{z_{a}(\cdot)\}_{a}$;\\
(ii) $\{q_{a}(\cdot)\}_{a}$ is an allocation process in the sense
of Definition \ref{Allocation_process_definition};\\
(iii) $\left\{ x_{n_{k},a}(\cdot),q_{n_{k},a}(\cdot)\right\} _{a}\xrightarrow[\mathbb{P}_{n,0}]{d}\left\{ x_{a}(\cdot),q_{a}(\cdot)\right\} _{a}$,
where $x_{a}(t):=z_{a}(q_{a}(t))$; and\\
(iv) $\{q_{a}(\cdot)\}_{a}$ is almost surely Lipschitz continuous,
with a Lipschitz constant of 1.\end{thm}

Theorem \ref{Thm: ART} establishes that the distribution of $\left\{ x_{n_{k},a}(\cdot),q_{n_{k},a}(\cdot)\right\} _{a}$
in the original experiment can be matched with that of $\left\{ x_{a}(\cdot),q_{a}(\cdot)\right\} _{a}$
in the limit experiment, where $\{q_{a}(\cdot)\}_{a}$ is a suitably
defined allocation process. Although these statistics are path-valued
processes, the convergence is uniform over time. In concert with (\ref{eq:equivalence of LRs}),
Theorem \ref{Thm: ART} enables us to derive lower bounds on losses
or statistical risk in various applications by employing change of
measure arguments. 

While the proof of Theorem \ref{Thm: ART} is somewhat involved, the
underlying intuition is relatively straightforward. The processes
$\{z_{n,a}(\cdot)\}_{a}$ are asymptotically tight (being standard
partial sum processes), and likewise, the processes $\{q_{n,a}(\cdot)\}_{a}$
are also tight since, by definition, 
\[
\sup_{t}\vert q_{n,a}(t+\delta)-q_{n,a}(t)\vert\le\delta+n^{-1},\quad\forall\ \delta>0.
\]
This ensures that $\{z_{n,a}(\cdot),q_{n,a}(\cdot)\}_{a}$ converges
to some weak limit $\{z_{a}(\cdot),q_{a}(\cdot)\}_{a}$. Moreover,
the measurability of the events $\{q_{n,1}(t)\le\gamma_{1},q_{n,0}(t)\le\gamma_{0}\}$
with respect to $\mathcal{G}_{n,\gamma_{0},\gamma_{1}}:=\mathcal{F}_{n,\gamma_{1}}^{(1)}\lor\mathcal{F}_{n,\gamma_{0}}^{(0)}\lor\sigma(U)$
suggests that $q_{a}(\cdot)$ can be constructed to inherit the appropriate
adaptedness properties required by Definition \ref{Allocation_process_definition}.
The construction requires some care and utilizes some results from
the theory of stable convergence \citep{hausler2015stable}; it is
perhaps the most intricate part of the proof. Setting $x_{n,a}(\cdot):=z_{n,a}(q_{n,a}(\cdot))$
and $x_{a}(\cdot):=z_{a}(q_{a}(\cdot))$ then gives the desired result. 

Although Theorem \ref{Thm: ART} is stated for sub-sequences, most
applications require weak convergence of the full sequence $\{x_{n,a}(\cdot),q_{n,a}(\cdot)\}_{a}$.
This motivates:

\begin{asm2} The sequence of policy rules $\{\pi_{n,j}\}_{j}$ is
such that $\{x_{n,a}(\cdot),q_{n,a}(\cdot)\}_{a}$ has a weak limit
under $\mathbb{P}_{n,0}$.\end{asm2}

Theorem \ref{Thm: ART} already ensures $\{x_{n,a}(\cdot),q_{n,a}(\cdot)\}_{a}$
is tight under $\mathbb{P}_{n,0}$. Assumption 2 strengthens this
to weak convergence. The assumption is needed to rule out pathological
sequences of policy rules, e.g., sequences where the policy rules
differ for even and odd $n$. It is therefore rather mild: if it does
not hold, one should extract convergent subsequences and treat each
as arising from a distinct protocol.

\subsection{Behavior under local alternatives}

Theorem \ref{Thm: ART} describes the behavior of $\left\{ x_{n_{k},a}(\cdot),q_{n_{k},a}(\cdot)\right\} _{a}$
under the reference distribution $\mathbb{P}_{n,0}$. Under local
alternatives of the form $\mathbb{P}_{n,\bm{h}}$, the partial sum
processes, $z_{n,a}(\cdot)$, acquire an asymptotic drift, converging
weakly to $z_{a}(\cdot)\sim I_{a}^{1/2}h_{a}\cdot+W_{a}(\cdot)$. 

Given that $\left\{ x_{a}(\cdot),q_{a}(\cdot)\right\} _{a}$ is adapted
to the filtrations generated by $\{z_{a}(\cdot)\}_{a}$ and the exogenous
randomization $U\sim\textrm{Uniform}[0,1]$, it follows that $\left\{ x_{n,a}(\cdot),q_{n,a}(\cdot)\right\} _{a}$
should converge weakly to $\{x_{a}(\cdot),q_{a}(\cdot)\}_{a}$, where
the only difference is that the underlying processes $\{z_{a}(\cdot)\}_{a}$
now exhibit a linear drift. This is formalized in the following corollary.

\begin{cor} \label{Cor: ART}Suppose Assumptions 1 and 2 hold. Let
$\{x_{n,a}(\cdot),q_{n,a}(\cdot)\}_{a}$ be any sequence of score
and allocation processes induced by a sequence of policies $\{\pi_{n,j}\}_{j}$
in the actual experiment. Then, there exists a random collection $\{z_{a}(\cdot),q_{a}(\cdot),U\}_{a}$
defined on a probability space $(\Omega,\mathcal{F},\mathbb{P})$
such that:\\
(i) $z_{a}(\cdot)\sim I_{a}^{1/2}h^{(a)}\cdot+W_{a}(\cdot)$ are independent
Gaussian processes and $U\sim\textrm{Uniform}[0,1]$ is independent
of $\{z_{a}(\cdot)\}_{a}$;\\
(ii) $\{q_{a}(\cdot)\}_{a}$---which is invariant across $\bm{h}$
as a function of $\left(U,\{z_{a}(s):0\le s\le1\}_{a}\right)$---is
an allocation process in the sense of Definition \ref{Allocation_process_definition};
and\\
(iii) $\left\{ x_{n,a}(\cdot),q_{n,a}(\cdot)\right\} _{a}\xrightarrow[\mathbb{P}_{n,\bm{h}}]{d}\left\{ x_{a}(\cdot),q_{a}(\cdot)\right\} _{a}$,
where $x_{a}(t):=z_{a}(q_{a}(t))$.\end{cor}

Corollary \ref{Cor: ART} is established in Appendix A using Theorem
\ref{Thm: ART} and standard change-of-measure arguments analogous
to Le Cam's third lemma.

Theorem \ref{Thm: ART} and Corollary \ref{Cor: ART} are existence
results: they establish that $\{q_{n,a}(\cdot)\}_{a}$ converge weakly
to an allocation process $\{q_{a}(\cdot)\}_{a}$ in the limit experiment.
While these results do not characterize the explicit form of $\{q_{a}(\cdot)\}_{a}$,
this is generally not a limitation in practice. As the applications
below illustrate, it is often possible to determine the form of optimal
decisions in the limit experiment without knowing the specific structure
of $\{q_{a}(\cdot)\}_{a}$. Theorem \ref{Thm: ART}, combined with
change-of-measure arguments, then allows us to transfer these decisions
back to the finite-sample setting and show that they are asymptotically
optimal.

\section{Application 1: Point estimation\label{subsec:Point-estimation}}

In this section, we illustrate how Theorem \ref{Thm: ART} can be
used to analyze estimation problems following an adaptive experiment. 

Suppose that, upon completion of the experiment, we aim to estimate
the unknown parameter vector $\bm{\theta}:=(\theta^{(1)},\theta^{(0)})$.
Let $T_{n}$ denote a proposed estimator based on the entire data
collected before the terminal time $t=1$. By construction, $T_{n}$
must be $\mathcal{I}_{n,1}:=\mathcal{G}_{n,q_{n,1}(1),q_{n,0}(1)}$
measurable. 

Let $l(\cdot)$ be a non-negative convex loss function. Following
the setup introduced earlier in this paper, we fix a reference parameter
$\bm{\theta}_{0}$ and evaluate estimator loss under local alternatives
of the form $\bm{\theta}_{0}+\bm{h}/\sqrt{n}$, where $\bm{h}\in\mathbb{R}^{d}$.
Unlike classical settings, however, the choice of $\bm{\theta}_{0}$
is subject to important restrictions. In many adaptive experiments,
only certain reference points yield non-degenerate diffusion asymptotics.
Let $\bm{\Theta}_{0}$ denote the equivalence class of such admissible
reference parameters. For example, in two-armed bandit experiments,
this class consists of all parameter vectors satisfying $\theta^{(1)}=\theta^{(0)}$;
otherwise, the resulting allocation processes become asymptotically
degenerate, collapsing to either 0 or 1. More generally, we define
$\bm{\Theta}_{0}$ to be the set of all parameter values $\bm{\theta}$
for which the weak limit of $q_{n,a}(\cdot)$ is not trivial (i.e.,
not identically 0) for any arm $a$. The reference parameter $\bm{\theta}_{0}$
must therefore lie in this set; otherwise some components of $\bm{\theta}$
will not be consistently estimable.\footnote{The set $\bm{\Theta}_{0}$ can be enlarged, however, if we are only
interested in estimating some sub-components of $\bm{\theta}$. } 

Given such a choice of $\bm{\theta}_{0}$, the frequentist risk of
$T_{n}$, evaluated at the local parameter $\bm{h}$, is defined as
\[
R_{n}(T_{n},\bm{h})=\mathbb{E}_{n,\bm{h}}\left[l\left(\sqrt{n}(T_{n}-\bm{\theta}(\bm{h}))\right)\right],
\]
where $\mathbb{E}_{n,\bm{h}}[\cdot]$ is the expectation under $\mathbb{P}_{n,\hm{h}}$,
and $\bm{\theta}(\bm{h}):=\bm{\theta}_{0}+\bm{h}/\sqrt{n}$. 

The estimation problem in the limit experiment is defined analogously.
Given access to the sample paths of $\left\{ x_{a}(\cdot),q_{a}(\cdot)\right\} _{a}$
over $t\in[0,1]$, we seek an estimate of the local parameter $\bm{h}$.
Let $T$ denote a candidate estimator, required to be $\mathcal{I}_{1}\equiv\mathcal{G}_{q_{1}(1),q_{0}(1)}$
measurable. The frequentist risk of this estimator is 
\[
R(T,\bm{h}):=\mathbb{E}_{\bm{h}}\left[l(T-\bm{h})\right].
\]

We term a sequence of estimators, $\{T_{n}\}_{n}$, \textit{tight}
at $\bm{\theta}_{0}$ if $\sqrt{n}(T_{n}-\bm{\theta}_{0})$ is asymptotically
tight, i.e., bounded in probability, under $\mathbb{P}_{n,0}$. Tightness
at $\bm{\theta}_{0}$ is a substantial relaxation of the usual notion
of regularity---which requires the limit distribution of $\sqrt{n}(T_{n}-\bm{\theta}(\bm{h}))$
under $\mathbb{P}_{n,\bm{h}}$ to be the same for all $\bm{h}$. 

The following theorem states that the asymptotic performance of any
tight sequence of estimators $\{T_{n}\}_{n}$ is lower bounded, along
subsequences, by the performance of some estimator $T$ in the limit
experiment, and that this limit estimator depends only on the terminal
values $\{x_{a}(1),q_{a}(1)\}_{a}$. 

\begin{thm} \label{Thm: Point estimation}Under Assumptions 1-2,
for any tight sequence of estimators, $\{T_{n}\}_{n}$, there exists
a further sub-sequence, $\{T_{n_{k}}\}_{k}$, and an estimator $T$
in the limit experiment depending only on $\{x_{a}(1),q_{a}(1)\}_{a}$
such that $\liminf_{k\to\infty}R_{n_{k}}(T_{n_{k}},\bm{h})\ge R(T,\bm{h})$
for each $\bm{h}$. \end{thm} 

The surprising implication of Theorem \ref{Thm: Point estimation}
is that knowledge of the just the terminal values of $\{x_{a}(\cdot),q_{a}(\cdot)\}_{a}$
is sufficient to characterize optimal estimators. The paths taken
by these processes are not directly informative for estimation. 

\subsection{Bayes risk\label{subsec:Bayes-risk}}

Let $m_{\bm{\theta}}(\cdot)$ denote a given prior over $\bm{\theta}$.
This induces a local prior, $m(\cdot)$, over $\bm{h}$ through the
transformation $\bm{h}=\sqrt{n}(\bm{\theta}-\bm{\theta}_{0})$. We
consider an asymptotic regime wherein $m(\cdot)$ is assumed to be
independent of $n$. The influence of the prior thus remains asymptotically
non-negligible and the Bernstein-von Mises theorem does not apply.
As discussed in \citet{adusumilli2025optimal}, local priors offer
a better approximation to finite-sample behavior because their influence
does not diminish with sample size.

Theorem \ref{Thm: Point estimation} implies a lower bound on the
Bayes risk corresponding to $m(\cdot)$: 

\begin{cor} \label{Cor: Bayes risk}Under Assumptions 1-2, for any
tight sequence of estimators, $\{T_{n}\}_{n}$, there exists a further
subsequence, $\{T_{n_{k}}\}_{k}$, and an estimator $T$ in the limit
experiment depending only on $\{x_{a}(1),q_{a}(1)\}_{a}$ such that
the Bayes risk, $\int R_{n_{k}}(T_{n_{k}},\bm{h})dm(\bm{h})$, of
$\{T_{n_{k}}\}_{k}$ is asymptotically lower bounded by the Bayes
risk, $\int R(T,\bm{h})dm(\bm{h})$, of $T$ in the limit experiment.
\end{cor} 

Let $T^{*}$ denote the optimal Bayes estimator in the limit experiment.
Then, Corollary \ref{Cor: Bayes risk} implies that $R^{*}(m):=R(T^{*},m)$
is an asymptotic lower bound on the Bayes risk of any tight sequence
of estimators $T_{n}$, i.e.,
\[
\liminf_{n\to\infty}R_{n}(T_{n},m)\ge R^{*}(m)\ \forall\ T_{n}.
\]
This lower bound does not require the use of subsequences. 

By the likelihood principle, the optimal Bayes estimator in the limit
experiment is algorithm independent and depends only on $\{x_{a}(1),q_{a}(1)\}_{a}$.
For example, consider a prior $m_{0}(\cdot)$ on $\bm{h}\equiv(h^{(1)},h^{(0)})\in\mathbb{R}^{2}$
with independent Gaussian components: $\mathcal{N}(\mu_{0}^{(1)},\nu_{(1)}^{2})\times\mathcal{N}(\mu_{0}^{(0)},\nu_{(0)}^{2})$.
Then, by standard results in stochastic filtering, the posterior distribution
of $h^{(a)}$ at the end of the experiment is
\[
h^{(a)}\vert q_{a}(1),x_{a}(1)\sim\mathcal{N}\left(\frac{I_{a}^{1/2}x_{a}(1)+\nu_{(a)}^{-2}\mu_{0}^{(a)}}{I_{a}q_{a}(1)+\nu_{(a)}^{-2}},\frac{1}{I_{a}q_{a}(1)+\nu_{(a)}^{-2}}\right),
\]
for each $a$. The optimal Bayes estimator of $h^{(a)}$ in the limit
experiment, under the squared error loss, $l(\delta)=\delta^{2}$,
is therefore 
\[
T^{*}=\frac{I_{a}^{1/2}x_{a}(1)+\nu_{(a)}^{-2}\mu_{0}^{(a)}}{I_{a}q_{a}(1)+\nu_{(a)}^{-2}}.
\]
Notably, this estimator is invariant to the choice of the sampling
algorithm. As $\nu^{(a)}\to\infty$, we get the MLE estimator $T_{\textrm{mle}}=I_{a}^{-1/2}x_{a}(1)/q_{a}(1)$. 

Corollary \ref{Cor: Bayes risk} implies that the set of estimators
depending only on $\{x_{a}(1),q_{a}(1)\}_{a}$ constitute a complete
class in the limit experiment. Furthermore, we can lower bound the
Bayes risk of any tight sequence of estimators using the Bayes risk
of estimators in the limit experiment. These results are useful because
determining the optimal estimator is a lot easier in the limit experiment.
As seen above, Gaussian priors are particularly straightforward to
analyze due to conjugacy. For general priors, computing the posterior
is more involved, but one can employ approximate methods such as MCMC.

\subsection{Attaining the bound\label{subsec:Attaining-the-bound}}

Given an optimal Bayes estimator, $T^{*}\left(\{x_{a}(1),q_{a}(1)\}_{a}\right)$,
in the limit experiment, we can construct a finite sample version,
\[
T_{n}^{*}:=\bm{\theta}_{0}+n^{-1/2}T^{*}\left(\{x_{n,a}(1),q_{n,a}(1)\}_{a}\right),
\]
by replacing $x_{a}(\cdot),q_{a}(\cdot)$ with the sample counterparts
$x_{n,a}(\cdot),q_{n,a}(\cdot)$. Since $T^{*}$ is an estimator of
$\bm{h}$, the transformation above converts it into an estimator
of $\bm{\theta}$. 

In practice, because $x_{n,a}(\cdot)$ depends on the information
matrix, $I_{a}$, one would need to replace it with a consistent estimate.
This can be supplied by the standard variance estimator, which remains
consistent under general conditions, even if only at slower-than $n^{-1/2}$
rates. The construction also requires knowledge of the reference parameter
$\bm{\theta}_{0}$. We suggest choosing this as the element from the
equivalence class, $\bm{\Theta}_{0}$, that is closest to the prior
median under $m_{\bm{\theta}}(\cdot)$.

Suppose that $T^{*}(\cdot)$ satisfies the conditions for a continuous
mapping theorem. Together with (\ref{eq:SLAN property}) and Theorem
\ref{Thm: ART}, this implies
\[
\left(\begin{array}{c}
\sqrt{n}\left(T_{n}^{*}-\bm{\theta}(\bm{h})\right)\\
\hat{\varphi}(\bm{h};q_{n,1}(1),q_{n,0}(1))
\end{array}\right)\xrightarrow[P_{nT,0}]{d}\left(\begin{array}{c}
T^{*}-h\\
\sum_{a}\left\{ h^{(a)\intercal}I_{a}^{1/2}x_{a}(1)-\frac{q_{a}(1)}{2}h^{(a)\intercal}I_{a}h^{(a)}\right\} 
\end{array}\right),
\]
for any $\bm{h}$. Then, a similar argument as in the proof of Theorem
\ref{Thm: Point estimation} shows that the frequentist risk of $T_{n}^{*}$
converges to that of $T^{*}$ in the limit experiment, as long as
the loss function is bounded. But $T^{*}$ is the optimal Bayes estimator
in the limit experiment, so the above implies that $T_{n}^{*}$ is
asymptotically Bayes optimal as well, in the sense that its Bayes
risk is arbitrarily close to $R^{*}(m)$ as $n\to\infty$.

\subsection{Minimax risk}

Minimax risk is defined as $\inf_{T_{n}}\sup_{m(\cdot)}\int R_{n}(T_{n},\bm{h})dm(\bm{h})$,
where $T_{n}$ ranges over all tight sequences of estimators, and
$m(\cdot)$ ranges over all possible priors. The estimator $T_{n}^{*}$
that solves this problem is referred to as the minimax estimator.
This estimator can also be interpreted as the equilibrium outcome
of a zero-sum game between a decision-maker and nature: nature selects
a prior $m(\cdot)$ to maximize the Bayes risk, while the decision-maker
selects an estimator to minimize it. 

In contrast to classical experiments, the minimax risk in adaptive
experiments is often infinite. To see why, consider the two-armed
bandit experiment from the illustrative example, and suppose the objective
is to estimate $h^{(1)}$ in the limit experiment. Nature can make
the problem arbitrarily hard by choosing a flat prior over $h^{(1)}$
and taking $h^{(0)}\to\infty$. In this case, because $h^{(0)}/h^{(1)}\to\infty$
with probability one, the bandit algorithm devotes negligible attention
to arm 1, effectively yielding no data from which to estimate $h^{(1)}$.
This leads to an infinite risk.

\subsection{Illustrative example (contd.)}

Continuing with the illustrative example from Section \ref{subsec:An-illustrative-example},
suppose we aim to estimate the parameter $\theta^{(1)}$ after conducting
the experiment using a bandit algorithm (we show results under both
UCB and Thompson Sampling). We assume independent and identical Gaussian
priors over $\theta^{(0)}$ and $\theta^{(1)}$, given by $\mathcal{N}(\bar{\theta}_{},\bar{\sigma}^{2})$.

To apply our asymptotic framework, we reformulate the problem as a
local estimation problem. Specifically, we treat $\bm{\theta}_{0}=(\bar{\theta},\bar{\theta})$---the
vector of prior medians---as the reference value and define the local
parameter $h^{(a)}:=\sqrt{n}(\theta^{(a)}-\bar{\theta}_{})$. This
transformation induces a prior over $h^{(a)}$ of the form $\mathcal{N}(0,\nu^{2})$,
where $\nu^{2}:=n\bar{\sigma}^{2}$. As detailed in Section \ref{subsec:Bayes-risk},
$\nu^{2}$ is held fixed in our asymptotic regime even as $n$ increases,
implying that the prior over $\theta^{(a)}$ increasingly concentrates
around $\bar{\theta}$. 

In the Bernoulli case, the score function is $\psi(y)=\left[\bar{\theta}_{}(1-\bar{\theta}_{})\right]^{-1}(y-\bar{\theta}_{})$,
and the Fisher information is $I=\left[\bar{\theta}_{}(1-\bar{\theta}_{})\right]^{-1}$.
Then,
\[
x_{n,a}(t):=\frac{1}{\sqrt{n\bar{\theta}_{}(1-\bar{\theta}_{})}}\sum_{j=1}^{\lfloor nq_{n,a}(t)\rfloor}(Y_{j}^{(a)}-\bar{\theta}_{}).
\]
As shown in Section \ref{subsec:Attaining-the-bound}, the asymptotically
optimal Bayes estimator of $h^{(1)}$ is
\[
\hat{h}^{(1)}=\left(Iq_{n,1}(1)+\nu^{-2}\right)^{-1}I^{1/2}x_{n,1}(1).
\]
This leads to the corresponding estimator for $\theta^{(1)}$:
\[
\hat{\theta}^{(1)}=\frac{I^{-1}\nu^{-2}}{q_{n,1}(1)+I^{-1}\nu^{-2}}\bar{\theta}_{}+\frac{1}{q_{n,1}(1)+I^{-1}\nu^{-2}}\left(\frac{1}{n}\sum_{j=1}^{\lfloor nq_{n,1}(1)\rfloor}Y_{j}^{(1)}\right).
\]
Clearly, $\hat{\theta}^{(1)}$ has the same form as the usual shrinkage
estimator of the population mean under a Gaussian prior. As $\nu\to\infty$,
$\hat{\theta}^{(1)}$ becomes the MLE estimator $\frac{1}{nq_{n,1}(1)}\sum_{j=1}^{\lfloor nq_{n,1}(1)\rfloor}Y_{j}^{(1)}.$
Both estimators are independent of the adaptive sampling algorithm
used. 

To evaluate the finite-sample performance of $\hat{\theta}^{(1)}$,
we conduct simulations based on the illustrative example from Section
\ref{subsec:An-illustrative-example}, fixing $\theta^{(0)}=\bar{\theta}_{}$
and setting $\theta^{(1)}=\bar{\theta}+h^{(1)}/\sqrt{n}$ for values
of $h^{(1)}$ ranging from $-0.5$ to $0.5$. We also take $\bar{\theta}_{}=0.1$
and set the prior standard deviation over $h^{(1)}$ to be $\nu=0.2$.
This implies that the 95\% prior credible interval for $\theta^{(1)}$
is approximately $[0.04,0.14]$ when $n=100$. 

\begin{figure}
\includegraphics[height=5cm]{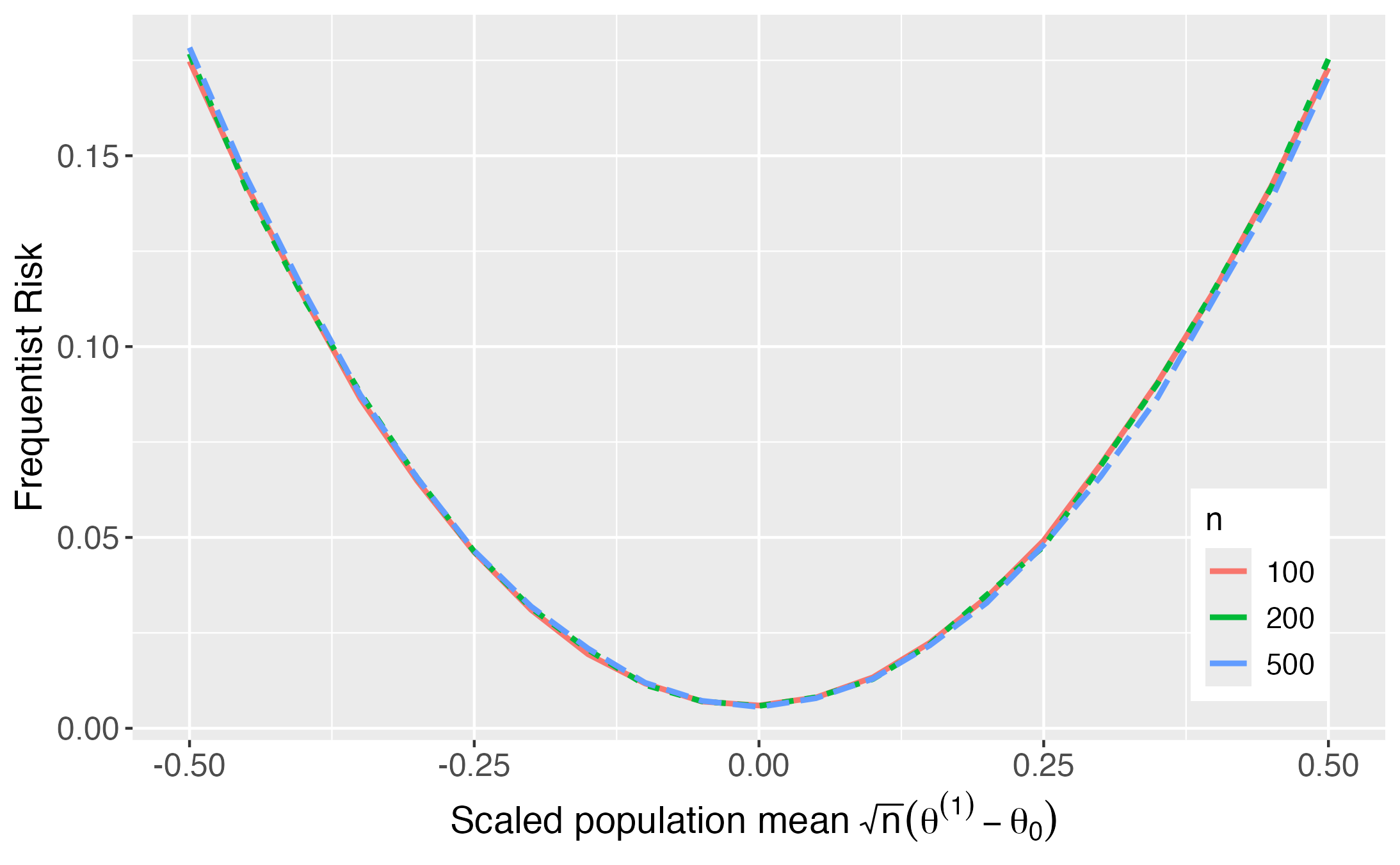}~\includegraphics[height=5cm]{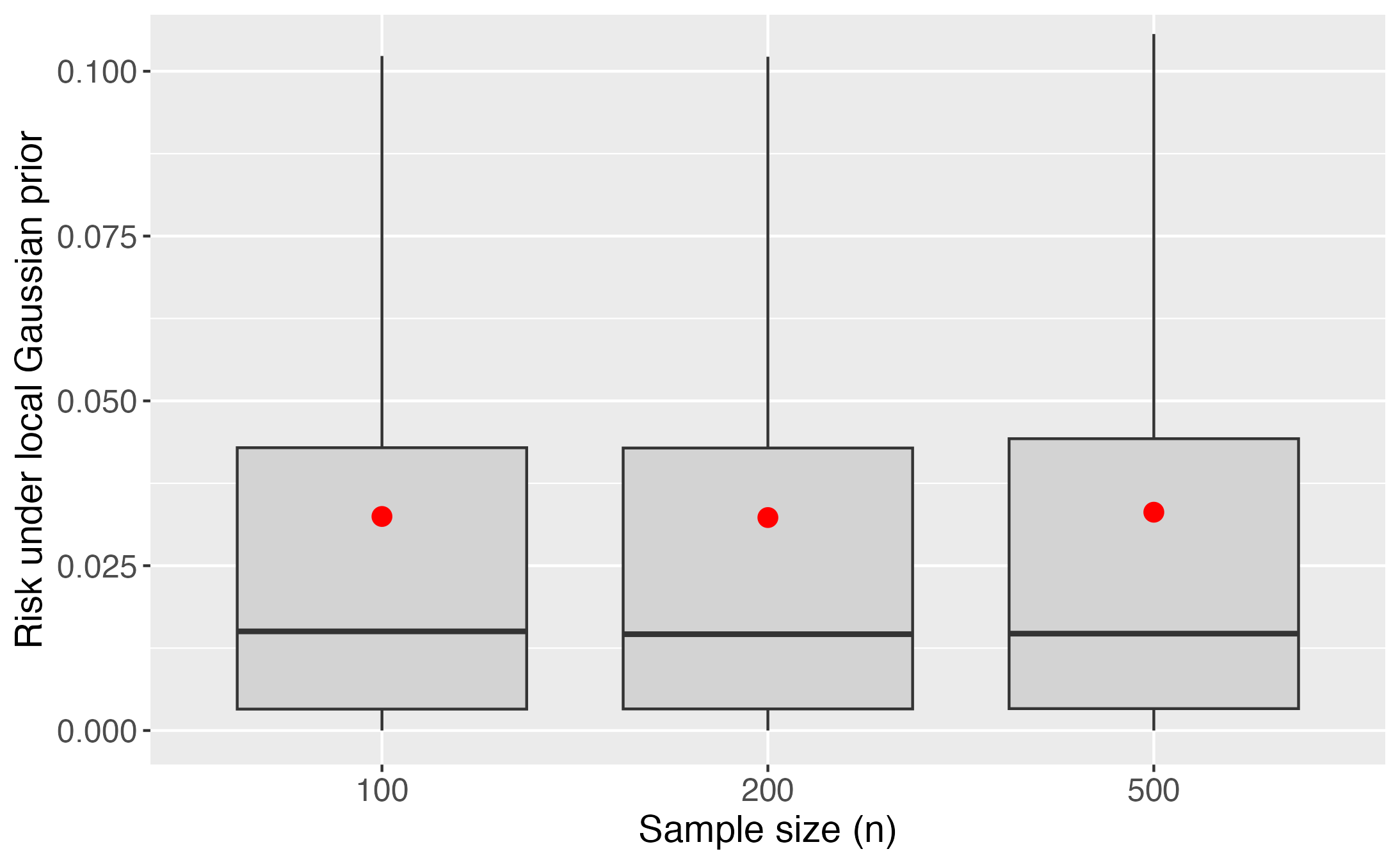}~

\begin{tabular}{>{\centering}p{7.5cm}>{\centering}p{7.5cm}}
{\scriptsize$\ $A: Frequentist risk profile} & {\scriptsize$\ \ $B: Bayes risk vs n}\tabularnewline
\end{tabular}
\begin{raggedright}
{\scriptsize Note: Panel A shows the frequentist risk profiles for
different values of $n$, when the data is obtained through UCB. Panel
B shows how the realized risk changes with $n$ under the prior $(h^{(1)},h^{(0)})\sim\textrm{i.i.d}\ \mathcal{N}(0,0.04)$.
The red points denote the Bayes risk under that prior. }{\scriptsize\par}
\par\end{raggedright}
\caption{Point estimation: two-armed UCB\label{fig:Point_estimation_example}}
\end{figure}

Panel A of Figure \ref{fig:Point_estimation_example} displays the
corresponding frequentist risk profiles of the estimator for different
values of $n$, when the data is obtained through UCB. Notably, the
risk profiles are nearly identical across sample sizes, even when
$n$ is as small as 100, highlighting the robustness of the estimator's
performance in small samples. Panel B of the same figure plots the
resulting Bayes risk under the local prior $(h^{(1)},h^{(0)})\sim\textrm{i.i.d}\ \mathcal{N}(0,\nu^{2})$.
The distributions of risk are again almost identical across $n$. 

\begin{figure}
\includegraphics[height=5cm]{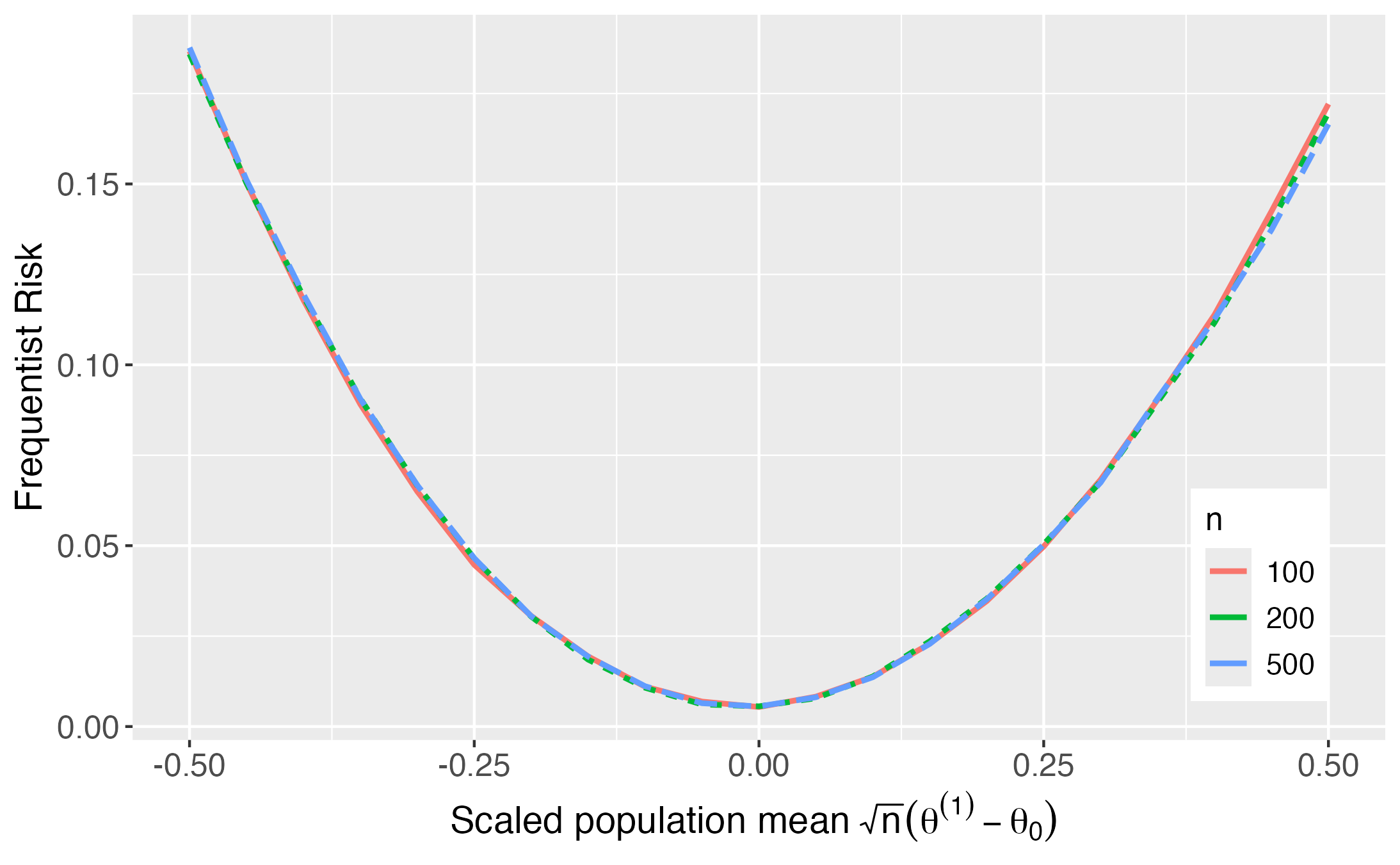}~\includegraphics[height=5cm]{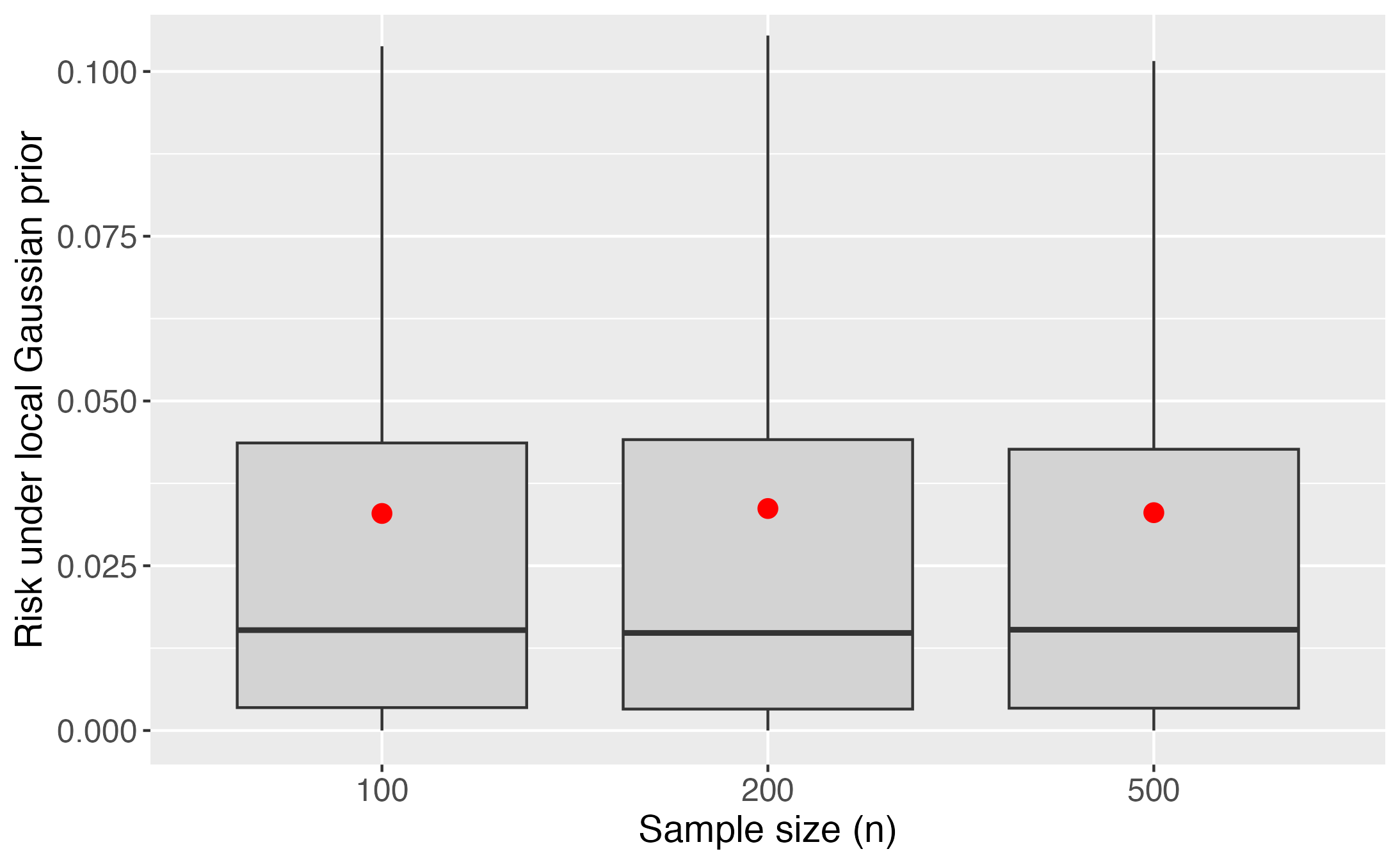}~

\begin{tabular}{>{\centering}p{7.5cm}>{\centering}p{7.5cm}}
{\scriptsize$\ $A: Frequentist risk profile} & {\scriptsize$\ \ $B: Bayes risk vs n}\tabularnewline
\end{tabular}
\begin{raggedright}
{\scriptsize Note: Panel A shows the frequentist risk profiles for
different values of $n$, when the data is obtained through Thompson
Sampling with a Beta$(1,1)$ prior. Panel B shows how the realized
risk changes with $n$ under the prior $(h^{(1)},h^{(0)})\sim\textrm{i.i.d}\ \mathcal{N}(0,0.04)$.
The red points denote the Bayes risk under that prior.}{\scriptsize\par}
\par\end{raggedright}
\caption{Point estimation: two-armed Thompson Sampling\label{fig:Point_estimation_example-TS}}
\end{figure}

Figure \ref{fig:Point_estimation_example-TS} displays equivalent
results when the data is obtained through Thompson Sampling. Surprisingly,
the estimator attains very similar values of Bayes risk under both
algorithms.

\section{Application 2: Equivalence of in-sample regret\label{sec:Application-2:-Equivalence}}

In-sample regret measures the difference between the (welfare) performance
of the best possible action---which is unknown---and a chosen policy,
evaluated on the same dataset used to select the policy. As we show
below, it follows from Theorem \ref{Thm: ART} that the in-sample
regret from any sequence of policy rules $\{\pi_{n,j}\}_{j}$ can
be asymptotically matched by that in the Gaussian diffusion limit
experiment.\footnote{See Appendix \ref{sec:Representations-for-out-of-sample} for extensions
to out-of-sample regret, also known as simple regret.} 

Let $\mu_{n,a}(\bm{h}):=\mathbb{E}_{n,\bm{h}}[Y_{i}^{(a)}]$ denote
the average reward corresponding to treatment $a$ when the local
parameter is $\bm{h}$. Following \citet{hirano2023asymptotic} and
\citet{adusumilli2025optimal}, the reference parameter is chosen
such that $\mu_{n,a}(\bm{0})=0$. The frequentist regret of any $\pi_{n}\equiv\{\pi_{n,j}\}_{j}$
is given by 
\[
W_{n}(\bm{h})=\sqrt{n}\left\{ \max_{a}\mu_{n,a}(\bm{h})-\sum_{a}\mu_{n,a}(\bm{h})\mathbb{E}_{n,\bm{h}}[q_{n,a}(1)]\right\} .
\]
Analogously, in the limit experiment, the frequentist regret is given
by 
\[
W(\bm{h})=\max_{a}\dot{\mu}_{a}^{\intercal}\bm{h}-\sum_{a}\dot{\mu}_{a}^{\intercal}\bm{h}\mathbb{E}_{\bm{h}}[q_{a}(1)],
\]
where $\dot{\mu}_{a}(\cdot)$ is defined in the assumption below:

\begin{asm3} There exists $\dot{\mu}_{a}\in\mathbb{R}^{d}$ and $\delta_{n}\to0$
such that $\sqrt{n}\mu_{n,a}(\bm{h})=\dot{\mu}_{a}^{\intercal}\bm{h}+\delta_{n}\vert\bm{h}\vert^{2}\ \forall\ \bm{h}$.
\end{asm3}

\begin{thm} \label{Thm: Welfare}Suppose Assumptions 1-3 hold. Then,
for any sequence of policies $\{\pi_{n,j}\}_{j}$ inducing the regret
function $W_{n}(\bm{h})$, there exists a limit experiment $\{x_{a}(\cdot),q_{a}(\cdot)\}_{a}$
with regret function $W(\bm{h})$ such that $W_{n}(\bm{h})\to W(\bm{h})$
for each $\bm{h}$.\end{thm} 

Recall from the measurability requirement on $q_{a}(\cdot)$ that
the allocation process at any time $t$ needs to be adapted to the
sample paths of $\{x_{a}(s);s\le t\}_{a}$. Theorem \ref{Thm: Welfare}
thus implies that the regret profile of any policy can be asymptotically
matched by one that depends only on the sample paths of $\{x_{a}(s);s\le t\}_{a}$. 

Previous work by this author (\citealp{adusumilli2025optimal}) characterized
the optimal Bayes and minimax risks in this setting, and showed that
they can be attained by a sequence of policy rules that depend on
just $\left\{ x_{n,a}(t),q_{n,a}(t)\right\} _{a}$, i.e., the past
values of these variables are not relevant. Theorem \ref{Thm: Welfare}
is more general, in that it applies to arbitrary sequences of policy
rules, but it allows for a larger information set that includes the
entire sample paths of $\{x_{n,a}(s);s\le t\}_{a}$ until time $t$. 

\section{Application 3: E-processes and anytime-valid inference\label{sec:Application-3:-E-processes}}

Anytime-valid inference refers to statistical procedures that maintain
valid error control (e.g., Type I error or confidence coverage) uniformly
over time, without compromising inference guarantees. A central tool
in this framework is the e-process, a nonnegative stochastic process
that starts at 1 and is a super-martingale (i.e., its expectations
is always less than or equal to 1) under the null hypothesis. Much
of the existing literature on anytime-valid inference has focused
on sequential experiments with fixed sampling strategies, where the
only adaptive element is the stopping rule. Here, we extend the notion
of an e-process to the more general setting of multi-treatment adaptive
experiments. 

Formally, we analyze e-processes and anytime-valid inference for tests
of the form $H_{0}:\bm{\theta}\in\Theta_{0}$ vs $H_{1}:\bm{\theta}\in\Theta_{1}$.
We assume that the reference parameter $\bm{\theta}_{0}:=(\theta_{0}^{(1)},\theta_{0}^{(0)})$
from Section \ref{sec:Diffusion-asymptotics-and} always lies in the
null region $\Theta_{0}$. As in Section \ref{subsec:Point-estimation},
we index each $\bm{\theta}$ by $\bm{h}\in\mathbb{R}^{d}\times\mathbb{R}^{d}$
such that $\bm{\theta}=\bm{\theta}_{0}+\bm{h}/\sqrt{n}$. Denote the
set of all $\bm{h}$ representing $\bm{\theta}\in\Theta_{0}$ by $\mathcal{H}_{0}$
and those representing $\bm{\theta}\in\Theta_{1}$ by $\mathcal{H}_{1}$.
We restrict attention to `asymptotically stable' hypothesis testing
problems, wherein the regions $\mathcal{H}_{0},\mathcal{H}_{1}$ do
not change with $n$. 

\subsection{E-processes in multi-treatment adaptive experiments}

Let $\mathcal{G}_{n,\gamma_{1},\gamma_{0}}$ denote the filtration
introduced in Section \ref{subsec:Empirical-allocation-processes}.
The e-process for multi-treatment adaptive experiments is formally
defined as follows:

\begin{definition}\label{Definition-e-process} An e-process, $\varepsilon_{n}(q_{1},q_{0})$,
for testing $H_{0}:\bm{h}\in\mathcal{H}_{0}$ is a non-negative stochastic
process indexed by $q_{1},q_{0}\in[0,1]^{2}$, such that:\\
(i) It is $\mathcal{G}_{n,q_{1},q_{0}}$-adapted at any given $(q_{1},q_{0})$;
and\\
(ii) For any empirical allocation process $\{q_{n,a}(\cdot)\}_{a}$,
\begin{equation}
\mathbb{E}_{n,\bm{h}}\left[\varepsilon_{n}\left(q_{n,1}(t),q_{n,0}(t)\right)\right]\le1\ \forall\ \bm{h}\in\mathcal{H}_{0},\forall\ t\in[0,1].\label{eq:e-process definition}
\end{equation}
\end{definition}

Definition \ref{Definition-e-process} generalizes the usual notion
of an e-process to a multi-indexed super-martingale, where the indices
are the treatment allocation proportions. In this framework, the value
of $\varepsilon_{n}(\cdot)$ depends on the trajectory of the empirical
allocation process, and the super-martingale property holds across
all possible empirical allocation processes, i.e., all possible adaptive
experiments. A central feature of the e-process is that it is algorithm-free:
it remains a valid supermartingale at any time point of any adaptive
experiment. 

Notably, Definition \ref{Definition-e-process} does not require $t$
to be a stopping time. Optimal stopping can be incorporated by introducing
a designated ``default'' treatment, such that assigning units to this
treatment is equivalent to halting the experiment. For notational
simplicity, we focus on the two-treatment case, though the extension
to more arms is conceptually straightforward. If there were optimal
stopping, the interpretation of $n$ would, however, change; it then
no longer denotes a fixed sample size, but instead serves as a normalization
constant. In particular, it associates the time index $t$ to the
closeness of alternatives being considered: if the aim is to analyze
performance against local alternatives of the form $\bm{\theta}_{0}+\bm{h}/\sqrt{n}$,
then $t$ should represent the number of observations collected in
units of $n$.

E-processes serve as dynamic measures of evidence against the null
hypothesis $H_{0}$. At any point $(q_{0},q_{1})$, the value $\varepsilon_{n}(q_{0},q_{1})$
can be interpreted as the current payoff from wagering one unit against
$H_{0}$. This interpretation holds uniformly over time and across
different adaptive experiments. Moreover, taking the inverse defines
a p-process $p_{n}(q_{1},q_{0}):=1/\varepsilon_{n}(q_{1},q_{0})$,
which yields an anytime-valid p-value. Specifically, for any $\bm{h}\in\mathcal{H}_{0}$,
empirical allocation $\{q_{n,a}(\cdot)\}_{a}$, and time $t$,
\begin{align*}
\mathbb{P}_{n,\bm{h}}\left(p_{n}\left(q_{n,1}(t),q_{n,0}(t)\right)\le\alpha\right) & =\mathbb{P}_{n,\bm{h}}\left(\varepsilon_{n}(q_{n,1}(t),q_{n,0}(t))\ge1/\alpha\right)\\
 & \le\alpha\mathbb{E}_{n,\bm{h}}[\varepsilon_{n}(q_{n,1}(t),q_{n,0}(t))]\le\alpha.
\end{align*}
Thus, $p_{n}\left(q_{n,1}(t),q_{n,0}(t)\right)$ is a valid classical
$p$-value at any time-point of any adaptive experiment. 

\subsection{GRO, mGRO and REGROW}

Following \citet{ramdas2023game} and \citet{grunwald2024safe}, a
common approach to evaluating e-processes in stopping-time experiments
is through the Growth Rate Optimality (GRO) criterion. GRO assesses
the quality of an e-process $\varepsilon_{n}(\cdot)$ based on its
expected log-growth under an alternative hypothesis $\bm{h}\in\mathcal{H}_{1}$.
We extend this criterion to the multi-treatment adaptive setting by
defining the GRO score
\begin{equation}
R_{n}\left(\varepsilon_{n};\bm{h},\{q_{n,a}(t)\}_{a}\right)=\mathbb{E}_{n,\bm{h}}\left[\ln\varepsilon_{n}(q_{n,1}(t),q_{n,0}(t))\right];\ \bm{h}\in\mathcal{H}_{1},\label{eq:GRO-criterion}
\end{equation}
where $\{q_{n,a}(t)\}_{a}$ denotes the empirical allocation process
at time $t$. The GRO score thus depends jointly on the e-process,
the alternative $\bm{h}$, and the adaptive experiment employed (as
indexed by the empirical allocation process).

We say that an e-process $\varepsilon_{n}(\cdot)$ uniformly dominates
another process $\varepsilon_{n}'(\cdot)$ in terms of GRO if
\[
R_{n}\left(\varepsilon_{n};\bm{h},\{q_{n,a}(t)\}_{a}\right)\ge R_{n}\left(\varepsilon_{n}';\bm{h},\{q_{n,a}(t)\}_{a}\right)
\]
for all $\bm{h}\in\mathcal{H}_{1}$, at all time points $t$, and
for all possible experiments, i.e., all possible empirical allocation
processes $\{q_{n,a}(t)\}_{a}$. This notion of uniform GRO dominance
is quite strong and, in general, no single e-process may achieve it.
One way to relax this requirement is to instead use the mixture-GRO
(mGRO) criterion, which averages the GRO score using a prior, or weight
function, $w(\cdot)$ over $\mathcal{H}_{1}$:
\[
R_{n}\left(\varepsilon_{n};w(\cdot),\{q_{n,a}(t)\}_{a}\right)=\int\mathbb{E}_{n,\bm{h}}\left[\ln\varepsilon_{n}(q_{n,1}(t),q_{n,0}(t))\right]dw(\bm{h}).
\]
The mGRO criterion ranks e-processes in terms of their average performance
over plausible alternatives, rather than requiring dominance for every
possible alternative.

An alternative criterion, following \citet{grunwald2024safe}, is
the REGROW (RElative GRowth Optimality in Worst case) score: 
\begin{align*}
 & \mathcal{R}_{n}\left(\varepsilon_{n};\{q_{n,a}(t)\}_{a}\right)\\
 & =\inf_{\bm{h}\in\mathcal{H}_{1}}\left\{ \mathbb{E}_{n,\bm{h}_{}}\left[\ln\varepsilon_{n}(q_{n,1}(t),q_{n,0}(t))\right]-\mathbb{E}_{n,\bm{h}}\left[\ln\frac{d\mathbb{P}_{n,\hm{h}}}{d\mathbb{P}_{n,0}}(q_{n,1}(t),q_{n,0}(t))\right]\right\} .
\end{align*}
REGROW measures the (negative of the) worst-case GRO-regret, where
GRO-regret is defined as the difference between the GRO value of $\varepsilon_{n}(\cdot)$
and the GRO value of the log-likelihood ratio process corresponding
to a specific alternative $\bm{h}_{1}\in\mathcal{H}_{1}$. The latter
is the ideal, i.e., uniformly GRO optimal, e-process for testing $H_{0}:\bm{h}=0$
versus $H_{1}:\bm{h}=\bm{h}_{1}$. REGROW thus benchmarks the performance
of $\varepsilon_{n}(\cdot)$ against the optimal e-process that would
be achievable if $\bm{h}_{1}$ were known in advance. A higher REGROW
score indicates more robust performance across alternatives.

\subsection{Local asymptotics and e-processes in the limit experiment\label{subsec:Local-asymptotics-and-e-processes}}

GRO and its variants, mGRO and REGROW, offer an alternative paradigm
to the classical power criterion for hypothesis testing. However,
determining and computing optimal e-processes becomes considerably
more challenging when $\mathcal{H}_{0}$ is composite or when the
REGROW criterion is used. Closed-form expressions are only known for
certain parametric families $\mathbb{P}_{n,h}$.

In order to circumvent this complexity in the fixed $n$ setting,
we propose employing a local-asymptotic criterion. Accordingly, we
relax the definition of e-processes given previously, and call a sequence
of non-negative, $\mathcal{G}_{n,q_{1},q_{0}}$-adapted stochastic
processes, $\varepsilon_{n}(\cdot)$, \textit{asymptotic e-processes}
if 
\[
\limsup_{n}\mathbb{E}_{n,\bm{h}}\left[\varepsilon_{n}\left(q_{n,1}(t),q_{n,0}(t)\right)\right]\le1\ \forall\ \bm{h}\in\mathcal{H}_{0},\ \forall\ t\in[0,1],
\]
and for all possible empirical allocation processes $\{q_{n,a}(\cdot)\}_{a}$.

We then define an equivalent notion of an e-process in the limit experiment.
Recall that the limit experiment is characterized by Gaussian process
signals $z_{a}(q):=I_{a}^{1/2}h^{(a)}q+W_{a}(q)$. The e-process in
the limit experiment is a $\mathcal{G}_{q_{1},q_{0}}$-adapted stochastic
process designed for testing $H_{0}:\bm{h}\in\mathcal{H}_{0}$ versus
$H_{1}:\bm{h}\in\mathcal{H}_{1}$. The formal definition is given
below. First, a bit of terminology: let $\bm{\mathcal{Q}}$ denote
the collection of all allocation processes in the limit experiment
that are the weak limits of some possible sequence of empirical allocation
processes $\{q_{n,a}(\cdot)\}_{a}$.

\begin{definition}\label{Def:e_process_limit_experiment}An e-process,
$\varepsilon(\cdot,\cdot)$, for testing $H_{0}:\bm{h}\in\mathcal{H}_{0}$
in the limit experiment is a non-negative stochastic process indexed
by $q_{1},q_{0}\in[0,1]^{2}$, such that:\\
(i) It is $\mathcal{G}_{q_{1},q_{0}}$-adapted at any given $(q_{1},q_{0})$;
and\\
(ii) For any allocation process $\{q_{a}(\cdot)\}_{a}\in\bm{\mathcal{Q}}$,
\begin{equation}
\mathbb{E}_{\bm{h}}[\varepsilon(q_{1}(t),q_{0}(t))]\le1\ \forall\ \bm{h}\in\mathcal{H}_{0},\forall\ t\in[0,1].\label{eq:e-processes limit experiment}
\end{equation}
\end{definition}

Relative to Definition \ref{Definition-e-process}, the above definition
restricts the set of allocation processes to $\bm{\mathcal{Q}}$.
The rationale behind this is mainly technical: it allows us to avoid
dealing with $\{q_{a}(\cdot)\}_{a}$ that cannot be generated by any
sequence of experimental protocols in the actual experiment. Whether
$\bm{\mathcal{Q}}$, in fact, includes all possible allocation processes
is currently unknown (to this author).

The GRO, mGRO and REGROW criteria in the limit experiment retain the
same form as (\ref{eq:GRO-criterion}), except that $\ensuremath{\mathbb{E}_{n,\bm{h}}[\cdot]}$
is replaced by $\mathbb{E}_{\bm{h}}[\cdot]$, e.g., the GRO criterion
in the limit experiment becomes
\[
R\left(\varepsilon_{};\bm{h},\{q_{a}(t)\}_{a}\right)=\mathbb{E}_{\bm{h}}\left[\ln\varepsilon_{}(q_{1}(t),q_{0}(t))\right];\ \bm{h}\in\mathcal{H}_{1}.
\]

\subsection{Representation theorems for e-processes}

We derive an asymptotic representation theorem for e-processes, which
establishes that for any sequence of asymptotic e-processes, there
exists a dominating e-process in the limit experiment with respect
to the GRO criterion and its variants. This is based on the following
regularity conditions:

\begin{asm4} The sequence $\{\varepsilon_{n}(\cdot,\cdot),z_{n,a}(\cdot),U\}_{a}$,
with $\varepsilon_{n}(\cdot,\cdot)$ regarded as a stochastic process
indexed by $(q_{1},q_{0})\in[0,1]^{2}$, converges weakly under $\mathbb{P}_{n,\bm{0}}$
to a limit whose sample paths are almost surely continuous. Furthermore,
for each $t$ and each possible weakly convergent sequence of empirical
processes $\{q_{n,a}(\cdot)\}_{a}$, the sequence $\left\{ \ln\varepsilon_{n}\left(q_{n,1}(t),q_{n,0}(t)\right)\right\} _{n}$
is uniformly integrable with respect to each element in $\{\mathbb{P}_{n,\bm{h}}\}_{\bm{h}\in\mathcal{H}_{1}}$.\end{asm4}

The first part of Assumption 4 restricts the class of asymptotic e-processes
by requiring them to be asymptotically equicontinuous (so that they
have a weak limit together with the underlying signal processes).
The second part of Assumption 4 is an additional regularity condition
ensuring that the GRO scores are asymptotically convergent under the
alternative hypotheses. Both properties will need to be verified on
a case-by-case basis. For an example, see Appendix \ref{subsec:Verification-of-the-conditions}.

\begin{thm} \label{Thm:e-processes}Suppose Assumptions 1 and 4 hold.
Then, for any sequence of asymptotic e-processes $\varepsilon_{n}(\cdot)$,
there exists an e-process $\varepsilon(\cdot)$ in the limit experiment
such that
\[
\limsup_{n}R_{n}\left(\varepsilon_{n};\bm{h},\{q_{n,a}(t)\}_{a}\right)\le R\left(\varepsilon;\bm{h},\{q_{a}(t)\}_{a}\right)
\]
 for all $\bm{h}\in\mathcal{H}_{1}$, all $t\in[0,1]$, and all sequences
of policy rules satisfying Assumption 2 so that $\{q_{n,a}(\cdot)\}_{a}$
converges to some $\{q_{a}(\cdot)\}_{a}\in\bm{\mathcal{Q}}$ in the
limit experiment. \end{thm} 

The extension to the mGRO criterion is a straightforward consequence
of Theorem \ref{Thm:e-processes} and the monotone convergence theorem.

\begin{cor} \label{Cor:mGRO}Suppose Assumptions 1 and 4 hold. Assume
further that there exists $g\left(\bm{h}\right)\in[0,\infty)$ satisfying
$\int g(\bm{h})dw(\bm{h})<\infty$ and $\left|R_{n}\left(\varepsilon_{n};\bm{h},\{q_{n,a}(t)\}_{a}\right)\right|\le g\left(\bm{h}\right)$
for all $\bm{h}$, all allocation processes $\{q_{n,a}(t)\}_{n}$,
and all sufficiently large $n$. Then, for any sequence of asymptotic
e-processes $\varepsilon_{n}(\cdot)$, there exists an e-process $\varepsilon(\cdot)$
in the limit experiment such that 
\[
\limsup_{n}R_{n}\left(\varepsilon_{n};w,\{q_{n,a}(t)\}_{a}\right)\le R\left(\varepsilon;w,\{q_{a}(t)\}_{a}\right)
\]
for all $t\in[0,1]$, and and all sequences of policy rules satisfying
Assumption 2 so that $\{q_{n,a}(\cdot)\}_{a}$ converges to some $\{q_{a}(\cdot)\}_{a}\in\bm{\mathcal{Q}}$
in the limit experiment.\end{cor} 

In most cases, $R_{n}\left(\varepsilon_{n};\bm{h},\{q_{n,a}(t)\}_{a}\right)\ge0$,
while an upper bound has to verified on a case-by-case basis; see
Appendix \ref{subsec:Verification-of-the-conditions}, for instance.

Analysis of the REGROW criterion requires an additional assumption:

\begin{asm5} For any time $t$, $\bm{h}_{1}\in\mathcal{H}_{1}$,
and any sequence of empirical allocation processes $\{q_{n,a}(\cdot)\}_{a}$
weakly converging to some $\{q_{n,a}(\cdot)\}_{a}\in\bm{\mathcal{Q}}$,
we have that $\mathbb{E}_{n,\bm{h}_{1}}\left[\ln\frac{d\mathbb{P}_{n,\hm{h}_{1}}}{d\mathbb{P}_{n,\bm{0}}}(q_{n,1}(t),q_{n,0}(t))\right]$
converges to $\mathbb{E}_{\bm{h}_{1}}\left[\ln\frac{d\mathbb{P}_{\hm{h}_{1}}}{d\mathbb{P}_{\bm{0}}}(q_{1}(t),q_{0}(t))\right]$.\end{asm5}

The assumption states that the KL-divergences, $\textrm{KL}\left(\mathbb{P}_{n,\bm{h}_{1}}\mid\mid\mathbb{P}_{n,\bm{0}}\right)$,
in the original experiment converge asymptotically to KL-divergences,
$\textrm{KL}\left(\mathbb{P}_{\bm{h}_{1}}\mid\mid\mathbb{P}_{\bm{0}}\right)$,
in the limit experiment. This involves restrictions on the parametric
models allowed. 

\begin{cor} \label{Cor:REGROW}Suppose Assumptions 1, 4 and 5 hold.
Then, for any sequence of asymptotic e-processes $\varepsilon_{n}(\cdot)$,
there exists an e-process $\varepsilon(\cdot)$ in the limit experiment
such that 
\[
\limsup_{n}\mathcal{R}_{n}\left(\varepsilon_{n};\{q_{n,a}(t)\}_{a}\right)\le\mathcal{R}\left(\varepsilon;\{q_{a}(t)\}_{a}\right)
\]
for all $t\in[0,1]$, and and all sequences of policy rules satisfying
Assumption 2 so that $\{q_{n,a}(\cdot)\}_{a}$ converges to some $\{q_{a}(\cdot)\}_{a}\in\bm{\mathcal{Q}}$
in the limit experiment.\end{cor} 

\subsection{Applying the representation theorems\label{subsec:Applying-the-representation-Any-time}}

In what follows, we simplify matters by assuming that the null hypothesis
$\mathcal{H}_{0}$ is a singleton, consisting solely of the reference
parameter $\bm{\theta}_{0}$.\footnote{When there is only adaptive stopping, compound nulls are typically
addressed in anytime-valid inference using the method of reverse information
projection (see \citealp{ramdas2023game}, for a survey). Extending
this approach to the adaptive sampling setting is more involved and
left for future work. See, however, Appendix \ref{sec:Additional-results-on-anytime-inference}
for the simpler case of testing parameters corresponding to a single
treatment arm.} Theorem \ref{Thm:e-processes} and Corollaries \ref{Cor:mGRO}, \ref{Cor:REGROW}
establish asymptotic upper bounds on the GRO, mGRO and REGROW criteria.
These bounds are obtained by optimizing the respective criteria within
the limit experiment. 

A natural question is whether the restriction $\{q_{a}(\cdot)\}_{a}\in\bm{\mathcal{Q}}$,
imposed in the definition of the limiting e-process (Definition \ref{Def:e_process_limit_experiment}),
affects the outcome of the optimization. In practice, it does not.
The general approach to optimizing these criteria involves constructing,
for each fixed allocation process $\{q_{a}(\cdot)\}_{a}$ and time
$t$, an e-value: an $\mathcal{I}_{t}$-measurable random variable
$\varepsilon$ satisfying
\[
\mathbb{E}_{\bm{h}}[\varepsilon]\leq1\quad\text{for all }\bm{h}\in\mathcal{H}_{0}.
\]
We then identify the point-wise optimal e-value, $\varepsilon_{q_{1}(t),q_{0}(t)}^{*}$,
that maximizes the desired criterion (GRO, mGRO, or REGROW) at the
given allocation point $\{q_{a}(t)\}_{a}$. The GRO, mGRO or REGROW
value of $\varepsilon_{q_{1}(t),q_{0}(t)}^{*}$ furnishes a sharp
upper bound---uniformly over all possible e-processes---for the
corresponding criterion evaluated at that allocation. 

The final step is to determine whether these pointwise-optimal e-values
can be coherently combined, or ``strung together'', into a full e-process
satisfying Definition \ref{Def:e_process_limit_experiment}. If such
a construction is not possible, it usually implies that a globally
optimal e-process---one that simultaneously achieves pointwise optimality
at all allocation points---does not exist. 

\subsubsection{mGRO optimality}

As a first illustration of this approach, consider the mGRO criterion.
For a given $\{q_{a}(t)\}_{a}$, the mGRO optimal e-value in the limit
experiment is
\[
\varepsilon_{q_{1}(t),q_{0}(t)}^{*}=\int\exp\sum_{a}\left\{ h^{(a)\intercal}I_{a}^{1/2}z_{a}(q_{a}(t))-\frac{q_{a}(t)}{2}h^{(a)\intercal}I_{a}h^{(a)}\right\} dw(\bm{h}).
\]
Importantly, the form of $\varepsilon_{q_{1}(t),q_{0}(t)}^{*}$ does
not change with $\{q_{a}(t)\}_{a}$, implying that the optimal e-process
$\varepsilon^{*}(\cdot,\cdot)$ can be constructed as $\varepsilon^{*}(q_{1},q_{0})=\varepsilon_{q_{1},q_{0}}^{*}$.
It is straightforward to verify that the resulting e-process satisfies
(\ref{eq:e-processes limit experiment}) using Lemma \ref{Lemma: Martingale}
and standard martingale arguments.

Replacing $\{z_{a}(\cdot)\}_{a}$ with $\{z_{n,a}(\cdot)\}_{a}$ yields
the asymptotically mGRO-optimal e-process: 
\begin{equation}
\varepsilon_{n}^{*}(q_{1},q_{0})=\int\exp\sum_{a}\left\{ h^{(a)\intercal}I_{a}^{1/2}z_{n,a}(q_{a})-\frac{q_{a}}{2}h^{(a)\intercal}I_{a}h^{(a)}\right\} dw(\bm{h}).\label{eq:asymptotic_mGRO}
\end{equation}
Appendix \ref{subsec:Verification-of-the-conditions} describes primitive
conditions under which $\varepsilon_{n}^{*}(q_{1},q_{0})$ satisfies
the requirements for Corollary \ref{Cor:mGRO}. Essentially, we need
$w(\cdot)$ to be a sub-Gaussian distribution and $\psi_{a}(Y_{i}^{(a)})$
to have finite $2+p$ moments, for some $p>0$.

\subsubsection{REGROW optimality}

\citet{grunwald2024safe} show that the REGROW-optimal e-value, $\bar{\varepsilon}_{q_{1}(t),q_{0}(t)}^{*}$,
coincides with the mGRO-optimal e-value for a specific, least-favorable
weighting function $w_{q_{1}(t),q_{0}(t)}^{*}(\cdot)$. The authors
also show that this weighting function is obtained as the solution
to
\begin{align}
w_{q_{1}(t),q_{0}(t)}^{*} & =\argmax_{w\in\Delta(\mathcal{H}_{1})}\mathbb{E}_{\bm{h}\sim w}\left[\textrm{KL}_{q_{1}(t),q_{0}(t)}\left(\mathbb{P}_{\bm{h}}\mid\mid\mathbb{P}_{w}\right)\right],\label{eq:least-favorable distribution - REGROW}
\end{align}
where $\Delta(\mathcal{H}_{1})$ denotes the set of all probability
measures supported on $\mathcal{H}_{1}$, 
\[
\mathbb{P}_{w}(\cdot):=\int\mathbb{P}_{\bm{h}}(\cdot)dw(\bm{h}),
\]
and $\textrm{KL}_{q_{1}(t),q_{0}(t)}\left(\mathbb{P}_{\bm{h}}\mid\mid\mathbb{P}_{w}\right)$
represents the KL divergence between $\mathbb{P}_{\bm{h}},\mathbb{P}_{w}$
when these probability measures are restricted to the filtration $\mathcal{I}_{t}\equiv\mathcal{G}_{q_{1}(t),q_{0}(t)}$.\footnote{For instance, the restriction of $\mathbb{P}_{\bm{h}}$ to $\mathcal{I}_{t}$
is the probability measure induced by the sample paths of $z_{a}(s)=I_{a}^{1/2}h^{(a)}s+W_{a}(s)$
between $0$ and $q_{a}(t)$.} 

\citet{grunwald2024safe} further demonstrate that the optimized value
of the objective in (\ref{eq:least-favorable distribution - REGROW})
provides an upper bound on the REGROW criterion $\mathcal{R}\left(\varepsilon;\{q_{a}(t)\}_{a}\right)$:
\[
\sup_{\varepsilon}\mathcal{R}\left(\varepsilon;\{q_{a}(t)\}_{a}\right)\le\mathbb{E}_{\bm{h}\sim w_{q_{1}(t),q_{0}(t)}^{*}}\left[\textrm{KL}_{q_{1}(t),q_{0}(t)}\left(\mathbb{P}_{\bm{h}}\mid\mid\mathbb{P}_{w}\right)\right].
\]

Replacing the $q_{1}(t),q_{0}(t)$ subscripts with $(\cdot)$ for
ease of notation, note that $w_{(\cdot)}^{*}$ can be alternatively
characterized as: 
\begin{equation}
w_{(\cdot)}^{*}=\argmax_{w\in\Delta(\mathcal{H}_{1})}\textrm{KL}_{(\cdot)}\left(p_{\bm{h}}\cdot w\mid\mid p_{w}\cdot w\right)=\argmax_{w\in\Delta(\mathcal{H}_{1})}I_{(\cdot)}\left(w;p_{w}\right),\label{eq:l.f distribution - REGROW - dual}
\end{equation}
where $p_{\bm{h}},p_{w}$ denote the densities of $\mathbb{P}_{\bm{h}},\mathbb{P}_{w}$
with respect to some dominating measure, and $I_{(\cdot)}\left(\cdot;\cdot\right)$
represents mutual information under the restricted filtration. The
optimization problem (\ref{eq:l.f distribution - REGROW - dual})
has a natural information-theoretic interpretation. Consider an information
transmission channel with input $\hm{h}$ and output $\{z_{a}(s);s\le q_{a}(t)\}_{a}$.
Then, the quantity $\sup_{w}I_{(\cdot)}\left(w;p_{w}\right)$ corresponds
to the channel capacity, and $w_{(\cdot)}^{*}$ represents the optimal
signal distribution. 

Crucially, the least-favorable distribution $w_{(\cdot)}^{*}$ depends
on both the chosen allocation process and the structure of the alternative
hypothesis set $\mathcal{H}_{1}$. It is essential that $\mathcal{H}_{1}$
be compact---without it, the channel capacity (and hence the REGROW
value) becomes infinite. Because of the dependence on the allocation
process, a globally optimal REGROW e-process does not exist. 

As an alternative, one could fix a specific allocation process $\{q_{a}(t)\}_{a}$
and an alternative region $\mathcal{H}_{1}$, from which an optimal
weighting function $w_{(\cdot)}^{*}$ can be derived, e.g., by employing
the Blahut-Arimoto algorithm (see, \citealp{arimoto1972algorithm,blahut1972computation}).
The resulting e-process is thus \textit{locally REGROW optimal} relative
to the chosen design and alternative set (it would also be globally
mGRO optimal relative to the least favorable distribution $w^{*}(\cdot)$).
Appendix \ref{sec:Additional-results-on-anytime-inference} provides
an example of such a locally REGROW optimal e-process, constructed
to be optimal against fixed values of $\{q_{a}\}_{a}$. 

\subsection{Illustrative example (contd.)\label{subsec:Illustrative-example-(contd.)-anytime}}

We revisit again the illustrative example from Section \ref{subsec:An-illustrative-example},
now focusing on constructing an anytime-valid test of the null hypothesis
$H_{0}:\theta^{(1)}=0.1$ against the two-sided alternative $H_{1}:\theta^{(1)}\neq0.1$.
We take the reference parameter vector to be $\bm{\theta}_{0}=(0.1,0.1)$---as
it is the only reference parameter that induces non-trivial asymptotic
limits---and consider local alternatives of the form $\bm{\theta}=\bm{\theta}_{0}+\bm{h}/\sqrt{n}$.

Although the null is composite due to the unrestricted nature of $h^{(0)}$,
Appendix \ref{sec:Additional-results-on-anytime-inference} shows
that it is without loss of generality to ignore observations from
arm 0. This dimensionality reduction allows us to construct optimal
e-processes using only the data from arm 1, leveraging the techniques
developed in Section \ref{subsec:Applying-the-representation-Any-time}.

We employ the mGRO criterion with the weighting function $w_{1}(\cdot)\sim\mathcal{N}(0,\nu^{2})$,
where $\nu^{2}=1$. This corresponds to a $\mathcal{N}(0.1,1/n)$
prior over $\theta^{(1)}$. We examine sample sizes $n\in\{500,1000,1500,2000\}$.
For instance, when $n=1000$, this prior places 95\% of the mass within
the credible interval $[0.06,0.14]$ for $\theta^{(1)}$. As noted
earlier, in real-world scenarios, the difference in click-through
rates between various variants of a website is typically less than
1\%. Hence, our chosen weighting aligns well with realistic alternative
values of $\theta^{(1)}$.

Under this setup, the asymptotically optimal mGRO e-process (from
equation \ref{eq:asymptotic_mGRO}) takes the explicit form:
\begin{equation}
\varepsilon_{n}^{*}(q_{1})=\frac{1}{\sqrt{1+q_{1}I_{1}\nu^{2}}}\exp\left[\frac{I_{1}^{2}\nu^{2}\left\{ \sum_{j=1}^{\lfloor nq_{1}\rfloor}\left(Y_{j}^{(1)}-0.1\right)\right\} ^{2}}{2n(1+q_{1}I_{1}\nu^{2})}\right],\label{eq:e-process_application}
\end{equation}
where $I_{1}:=[\theta_{0}^{(1)}(1-\theta_{0}^{(1)})]^{-1}$ denotes
the Fisher information.

\begin{figure}
\includegraphics[height=5cm]{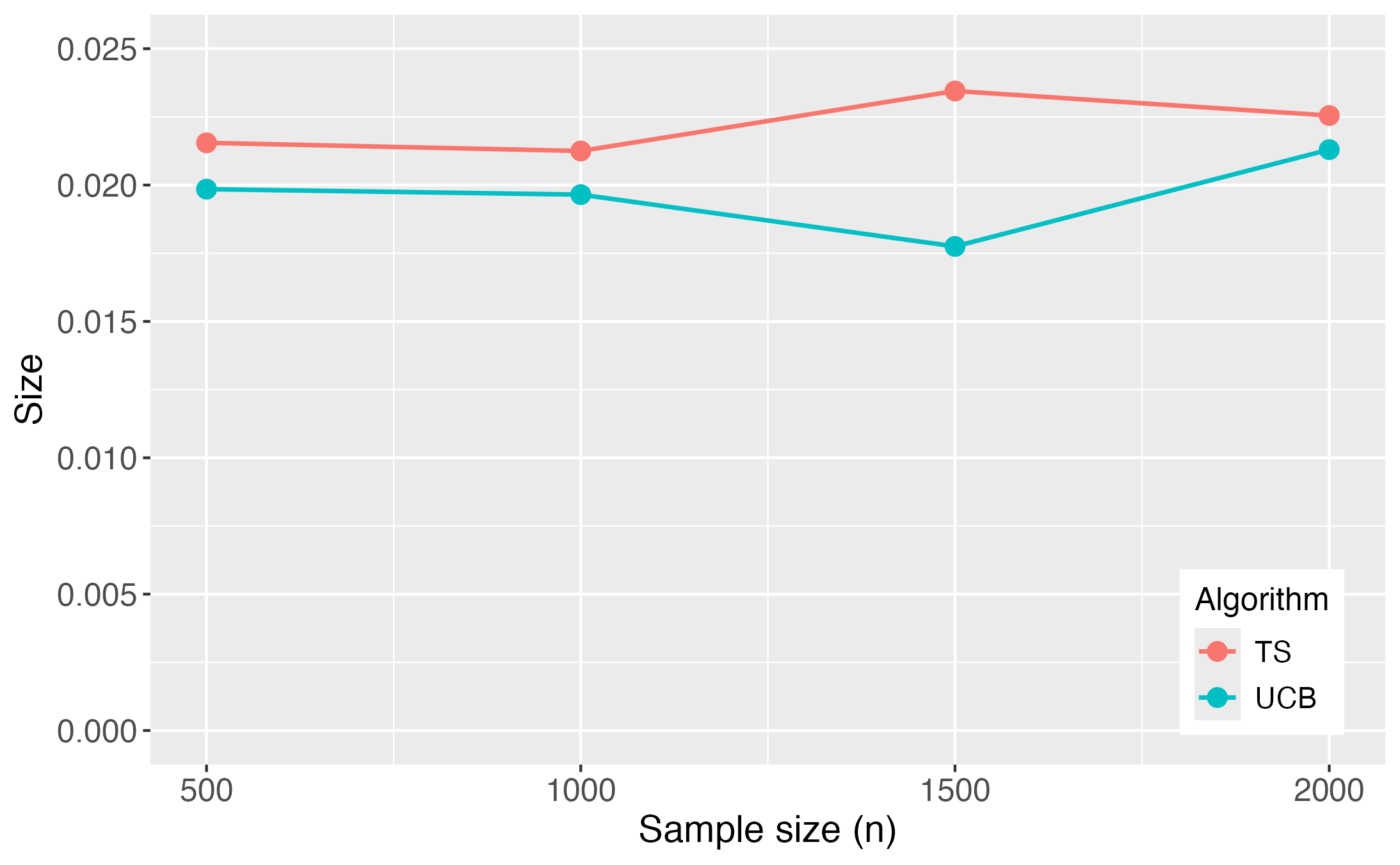}~\includegraphics[height=5cm]{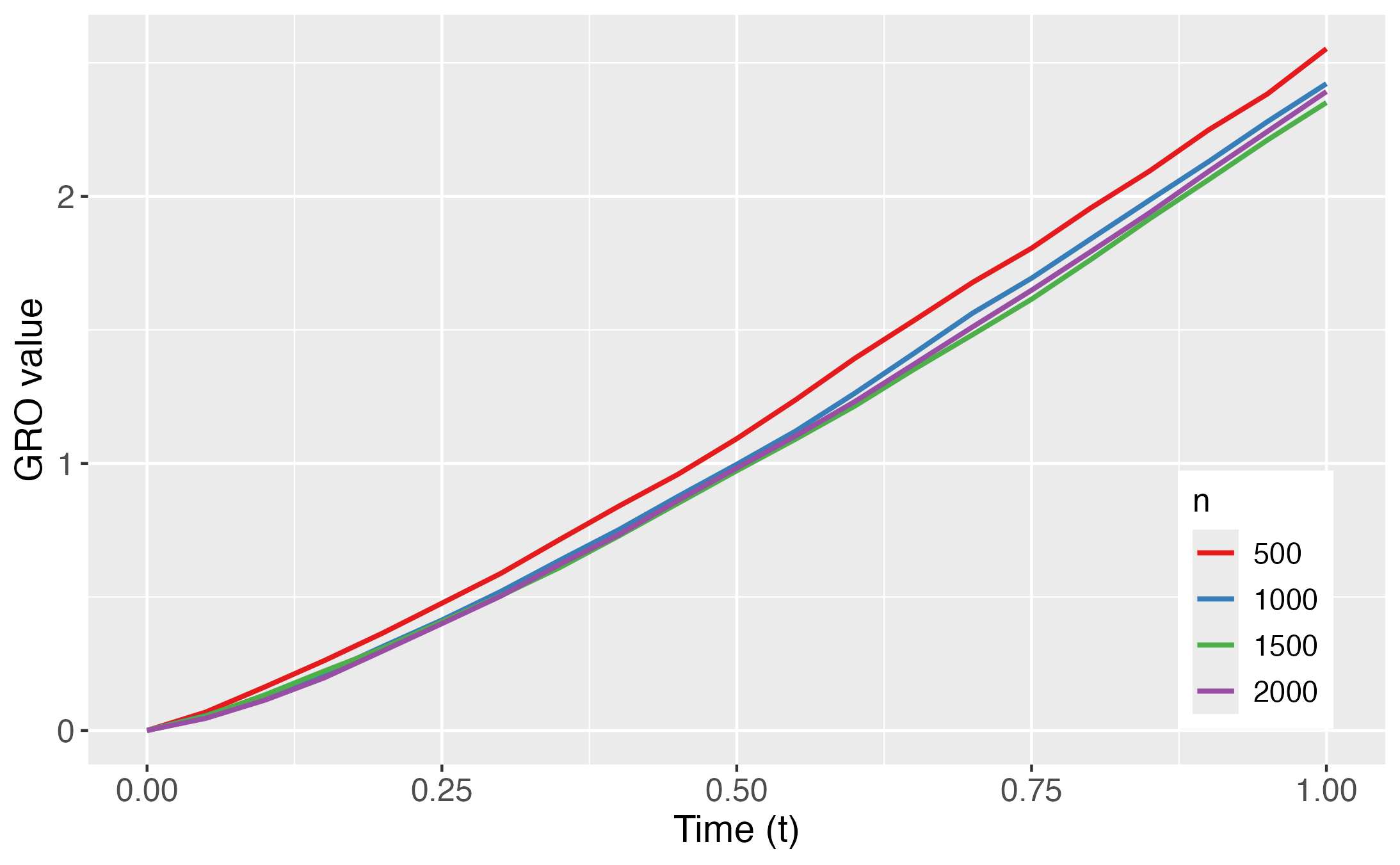}~

\begin{tabular}{>{\centering}p{7.5cm}>{\centering}p{7.5cm}}
{\scriptsize A: Size} & {\scriptsize$\ \ $B: GRO vs $t$ (UCB)}\tabularnewline
\end{tabular}
\begin{raggedright}
{\scriptsize Note: Panel A displays the uniform over time finite sample
size of the e-process in equation (\ref{eq:e-process_application})
at different values of $n$. Panel B plots the evolution of GRO value
of this test over time under UCB, at the local alternative $(h^{(1)},h^{(0)})=(1/\sqrt{n},0)$.}{\scriptsize\par}
\par\end{raggedright}
\caption{Anytime-valid inference\label{fig:Any_time_valid_inference}}
\end{figure}

Panel A of Figure \ref{fig:Any_time_valid_inference} reports the
uniform-over-time finite-sample size of this e-process under Thompson
Sampling and UCB allocation rules, calculated as\footnote{The least-favorable configuration for $h^{(0)}$ appears to be $-\infty$,
corresponding to $\theta^{(0)}=0$.} 
\[
\sup_{h^{(0)}}P_{n,(0,h^{(0)})}\left(\max_{t}\varepsilon_{n}^{*}(q_{n,1}(t))\ge20\right),
\]
The critical value of $20$ implies that the p-process conversion
(from the e-process) targets an anytime valid size of $5$\%. However,
the size under specific policies can be smaller and it indeed turns
out that the test is conservative for the policies considered---a
behavior that is expected, since the validity of the e-process applies
to all adaptive algorithms and not just Thompson Sampling or UCB.

Panel B plots the evolution of the GRO value over time for this e-process
under a local alternative $\bm{h}\equiv(h^{(1)},h^{(0)})=(1/\sqrt{n},0)$,
for varying sample sizes, all under UCB (see Appendix \ref{sec:Additional-results-on-anytime-inference}
for equivalent results under Thompson Sampling). The resulting curves
exhibit remarkable stability across $n$, indicating that the asymptotic
approximation is already accurate at such small sample sizes as $n=500$.

\section{Conclusion}

This article introduces a continuous-time formalism for analyzing
fully adaptive experiments. The formalism is based on the notion of
allocation processes, also introduced in this article. We show that
any empirical allocation process, as induced by some policy rule,
converges weakly to a corresponding allocation process in a limit
experiment governed by Gaussian diffusions. The various applications,
ranging from point-estimation to anytime-valid inference illustrate
the utility of this framework. 

Beyond these applications, the continuous-time formulation offers
a powerful tool for addressing design problems in adaptive experimentation.
Though not based on the results reported in this article, prior work
by this author has applied the continuous-time framework to derive
optimal algorithms for bandit experiments and costly sampling problems
(\citealt{adusumilli2025optimal,adusumilli2025sample}). The current
results provide an easy-to-use and generalizable template for transferring
optimal designs from the limit experiment back to the finite-sample
setting. Looking ahead, we expect this approach to be broadly useful
in addressing a range of open questions in adaptive experimentation---e.g.,
in deriving optimal strategies for best-arm identification with multiple
treatments, or in designing adaptive experiments in strategic environments
involving interactions between an experimenter and a regulator.

\bibliographystyle{IEEEtranSN}
\bibliography{Continuous_time_ARTs}

\appendix

\section{Proofs of Lemma \ref{Lemma: Martingale} and Theorem \ref{Thm: ART}\label{sec:Appendix:A}}

\subsection{Proof of Lemma \ref{Lemma: Martingale}}

We start by showing that $x_{1}(t)$ is an $\mathcal{I}_{t}$-martingale;
the claim for $x_{0}(t)$ follows analogously. 

Fix any $t_{1},t_{2}\in[0,1]$ such that $t_{2}\ge t_{1}$, and define
$\tau_{1}:=q_{1}(t_{1}),\tau_{2}:=q_{1}(t_{2})$. Also, for each $\gamma_{1}\ge0$,
define $\mathcal{H}_{\gamma_{1}}:=\mathcal{G}_{\gamma_{1},1}$. By
the definition of the allocation process, the event $\{q_{1}(t)\le\gamma_{1}\}$
is $\mathcal{H}_{\gamma_{1}}$ measurable for all $\gamma_{1}$, under
any given $t$. Hence, $\tau_{1},\tau_{2}$ are both $\{\mathcal{H}_{\gamma_{1}}\}_{\gamma_{1}\ge0}$-adapted
stopping times. It is easily verified from the definition of $\mathcal{G}_{\gamma_{1},\gamma_{0}}$
that $z_{1}(\gamma_{1})$ is a Wiener process with respect to $\mathcal{H}_{\gamma_{1}}$.
Since $\tau_{2}\ge\tau_{1}$ almost surely (a.s.,) due to the almost
sure monotonicity of $q_{1}(\cdot)$, it follows by Doob's optional
sampling theorem that $\mathbb{E}\left[z_{1}(\tau_{2})\vert\mathcal{H}_{\tau_{1}}\right]=z_{1}(\tau_{1})$
a.s. In other words, $\mathbb{E}\left[z_{1}(q_{1}(t_{2}))\vert\mathcal{H}_{q_{1}(t_{1})}\right]=z_{1}(q_{1}(t_{1}))$
a.s., i.e., $\mathbb{E}\left[x_{1}(t_{2})\vert\mathcal{H}_{q_{1}(t_{1})}\right]=x_{1}(t_{1})$.
Since $\mathcal{I}_{t}\equiv\mathcal{G}_{q_{1}(t),q_{0}(t)}\subseteq\mathcal{G}_{q_{1}(t),1}\equiv\mathcal{H}_{q_{1}(t)}$
for any $t$ and $x_{1}(t)$ is $\mathcal{I}_{t}$-adapted, the tower
property of conditional expectations then implies $\mathbb{E}\left[x_{1}(t_{2})\vert\mathcal{I}_{t_{1}}\right]=x_{1}(t_{1})$
a.s. But $t_{1},t_{2}$ were arbitrary, so the claim follows. 

To show that the quadratic variation of $x_{1}(t)$ is $q_{1}(t)$,
we simplify notation by supposing that $z_{1}(t)$ is 1-dimensional;
the general case follows the same logic. Observe that $G(\gamma_{1}):=z_{1}(\gamma_{1}){}^{2}-\gamma_{1}$
is a $\mathcal{H}_{\gamma_{1}}$-martingale by the well known properties
of Wiener processes. Then, by similar arguments as above, we have
that $\mathbb{E}\left[G(\tau_{2})\vert\mathcal{H}_{\tau_{1}}\right]=G(\tau_{1})$
almost surely, and it thereby follows, also by similar arguments as
before, that $\mathbb{E}\left[x_{1}^{2}(t_{2})-q_{1}(t_{2})\vert\mathcal{I}_{t_{1}}\right]=x_{1}^{2}(t_{1})-q_{1}(t_{1})$
almost surely. Hence, $x_{1}^{2}(t)-q_{1}(t)$ is an $\mathcal{I}_{t}$-martingale.
This proves that $q_{1}(t)$ is the quadratic variation of $x_{1}(t)$.
The argument for $x_{0}(t)$ is analogous.

\subsection{Proof of Theorem \ref{Thm: ART}}

An informal outline of the proof is as follows: We employ dyadic approximations
to discretize the processes $\{q_{n,a}(\cdot)\}_{a}$ across both
their support (i.e., time) and range. The resulting discrete processes
are shown to converge in distribution under $\mathbb{P}_{n,0}$ to
a limit (discrete) process that is a function of $(z_{1}(\cdot),z_{0}(\cdot),U)$,
where $z_{1}(\cdot),z_{0}(\cdot)$ are independent $d$-dimensional
Wiener processes, and $U\sim\textrm{Uniform}[0,1]$ represents exogenous
randomization. 

Next, we demonstrate that as the discretization becomes arbitrarily
fine, the intermediate limit processes converge to a continuous-time
allocation process satisfying the criteria in Definition \ref{Allocation_process_definition}.
The theory of stable convergence \citep{hausler2015stable} plays
a key role here in ensuring that the informational/measurability constraints
satisfied by the empirical allocation processes are preserved during
the transition to the continuous-time (and range) limit. 

The proof proceeds in the following steps:

\subsubsection*{\textbf{Step 1: Convergence under dyadic approximations.}}

\textbf{\\}We discretize time into dyadic sets $\mathcal{D}_{m}=\{t_{k}:k=0,\dots,2^{m}\}$,
where $t_{k}=k2^{-m}$. Denote
\begin{equation}
q_{n,a,k}^{(m)}:=q_{n,a}(t_{k}).\label{eq:pf:Thm1:0.05}
\end{equation}
We then employ a further dyadic discretization $\mathcal{D}_{l}\equiv\{\eta_{0}\}\cup\{\eta_{l}:l=1,\dots,2^{L}\}$
of the range, $[0,1]$, of the empirical allocation process, $q_{n,1}(\cdot)$,
where $\eta_{0}=0$, $\eta_{l}:=l2^{-L}$ and $2^{L}:=\bar{c}2^{m}$
for some natural number $\bar{c}$ that is a power of $2$. Subsequently,
we approximate each of the random variables $q_{n,1,k}^{(m)}$ by
\begin{equation}
q_{n,1,k}^{(m,L)}=\sum_{l=1}^{2^{L}}\eta_{l}\mathbb{I}\{\eta_{l-1}<q_{n,1,k}^{(m)}\le\eta_{l}\}.\label{eq:pf:Thm1:0.07}
\end{equation}

Let $\phi_{n,k,l}$ denote the indicator functions $\mathbb{I}\{q_{n,1,k}^{(m,L)}=\eta_{l}\}$.
The random variables $\{\phi_{n,k,l}\}_{k,l}$ are tight, as are the
processes $z_{n,a}(\cdot)$, the latter due to standard results in
empirical process theory. Hence, by Prohorov's theorem, there exists
a subsequence, $\{n_{k}\}_{k}$, represented as $n$ for simplicity,
such that
\begin{equation}
\left(\begin{array}{c}
\{\phi_{n,k,l}\}_{k,l}\\
z_{n,1}(\cdot)\\
z_{n,0}(\cdot)
\end{array}\right)\xrightarrow[\mathbb{P}_{n,0}]{d}\left(\begin{array}{c}
\{\tilde{\phi}_{k,l}\}_{k,l}\\
z_{1}(\cdot)\\
z_{0}(\cdot)
\end{array}\right),\label{eq:pf:Thm1:0.1}
\end{equation}
where $z_{1}(\cdot),z_{0}(\cdot)$ are independent $d$-dimensional
Wiener processes. Denote
\begin{align*}
\tilde{q}_{1,k}^{(m,L)} & :=\sum_{l=1}^{2^{L}}\eta_{l}\tilde{\phi}_{k,l},\ \ \tilde{q}_{0,k}^{(m,L)}:=t_{k}-\tilde{q}_{1,k}^{(m,L)},\\
s_{n,a,l} & :=z_{n,a}(\eta_{l})-z_{n,a}(\eta_{l-1}),\ \textrm{and}\\
s_{a,l} & :=z_{a}(\eta_{l})-z_{a}(\eta_{l-1}).
\end{align*}

Lemma \ref{Lemma:Supporting_result_1} in Appendix C shows that we
can construct versions of $\tilde{q}_{a,k}^{(m,L)}$ , denoted $q_{a,k}^{(m,L)}$,
that satisfy the following conditions:
\begin{description}
\item [{C1}] $q_{1,k}^{(m,L)}+q_{0,k}^{(m,L)}=t_{k}$ for all $k$.
\item [{C2}] For each $k$, $\left\{ q_{1,k}^{(m,L)}\le\eta_{l},q_{0,k}^{(m,L)}\le\eta_{l^{\prime}}\right\} $
is
\[
\sigma\left\{ U_{1},\dots,U_{k},\{s_{1,j}\}_{j\le l+\bar{c}+1},\{s_{0,j}\}_{j\le l^{\prime}+\bar{c}+1}\right\} 
\]
measurable for all $l+l^{\prime}\ge k\bar{c}$, where $U_{1},\dots,U_{k}$
are uniform random variables independent of $z_{1}(\cdot),z_{0}(\cdot)$.\\
The random variables $U_{1},\dots,U_{2^{m}}$ can be subsumed into
a single $U\sim\textrm{Uniform}[0,1]$. Also, define $l(\gamma)=\inf\{l:\eta_{l}>\gamma\}$
and note that $\eta_{l(\gamma)+\bar{c}+1}\le\gamma+2^{-m}+2^{-L}$.
Then, we can rewrite the first part of this condition as: $\left\{ q_{1,k}^{(m,L)}\le\gamma_{1},q_{0,k}^{(m,L)}\le\gamma_{0}\right\} $
is
\[
\sigma\left\{ \{s_{1,j}\}_{j\le l(\gamma_{1})+\bar{c}+1},\{s_{0,j}\}_{j\le l(\gamma_{0})+\bar{c}+1},U\right\} \subseteq\bar{\mathcal{F}}_{\gamma_{1}+2^{-m}+2^{-L}}^{(1)}\lor\bar{\mathcal{F}}_{\gamma_{0}+2^{-m}+2^{-L}}^{(0)}\lor\sigma(U)
\]
measurable for any $\gamma_{1},\gamma_{0}\in[0,1]$ such that $\gamma_{1}+\gamma_{0}\ge t_{k}$. 
\item [{C3}] Letting $\sim$ denote equivalence in distributions,
\begin{equation}
\left(\{s_{a,l}\}_{a,l},\left\{ \tilde{q}_{1,\tilde{k}}^{(m,L)}\right\} _{\tilde{k}=1}^{k}\right)\sim\left(\{s_{a,l}\}_{a,l},\left\{ q_{1,\tilde{k}}^{(m,L)}\right\} _{\tilde{k}=1}^{k}\right)\ \textrm{for each }k.\label{eq:pf:Thm1:0.2}
\end{equation}
\end{description}
Equations (\ref{eq:pf:Thm1:0.1}) and (\ref{eq:pf:Thm1:0.2}) imply
\[
\left(\{s_{n,a,l}\}_{a,l},\{q_{n,a,k}^{(m,L)}\}_{a,k}\right)\xrightarrow[\mathbb{P}_{n,0}]{d}\left(\{s_{a,l}\}_{a,l},\{q_{a,k}^{(m,L)}\}_{a,k}\right).
\]
As a particular consequence of the above and the continuous mapping
theorem, if we define $x_{n,a,k}^{(m,L)}:=z_{n,a}(q_{n,a,k}^{(m,L)})$
and $x_{a,k}^{(m,L)}:=z_{a}(q_{a,k}^{(m,L)})$, then 
\begin{equation}
\left\{ x_{n,a,k}^{(m,L)},q_{n,a,k}^{(m,L)}\right\} _{a,k}\xrightarrow[\mathbb{P}_{n,0}]{d}\left\{ x_{a,k}^{(m,L)},q_{a,k}^{(m,L)}\right\} _{a,k}.\label{eq:pf:Thm1:0.3}
\end{equation}

Note that, by construction, $q_{n,a,k}^{(m,L)}\le t_{k}$ and $q_{n,a,k}^{(m,L)}\le q_{n,a,k^{\prime}}^{(m,L)}$
for all $k$ and $k^{\prime}\ge k$. Hence (\ref{eq:pf:Thm1:0.3})
implies---by the properties of weak convergence---that we also have
$q_{a,k}^{(m,L)}\le t_{k}$ and $q_{a,k}^{(m,L)}\le q_{a,k^{\prime}}^{(m,L)}$
almost surely, for all $k$ and $k^{\prime}\ge k$. 

\subsubsection*{\textbf{Step 2: Taking $L\to\infty$.}}

\textbf{\\}Equations (\ref{eq:pf:Thm1:0.1})-(\ref{eq:pf:Thm1:0.3})
apply under any fixed $L$. In fact, as $z_{1}(\cdot),z_{0}(\cdot),U$
do not depend on $L$, and $\{q_{a,k}^{(m,L)}\}_{a,k}$ is a measurable
function of these quantities by Condition C2, we can construct versions
of $\left\{ z_{1}(\cdot),z_{0}(\cdot),\{q_{a,k}^{(m,L)}\}_{a,k}\right\} $
that lie in the same probability space and where $z_{1}(\cdot),z_{0}(\cdot),U$
are the same quantities across $L$. Since $q_{a,k}^{(m,L)}$ is tight
(it takes values in $[0,1]$), by Prohorov's theorem, there exists
a sequence $L_{j}\to\infty$ and some random variables $\{q_{a,k}^{(m)}\}_{k}$
such that 
\begin{equation}
\left(U,z_{1}(\cdot),z_{0}(\cdot),\left\{ q_{a,k}^{(m,L_{j})}\right\} _{a,k}\right)\xrightarrow{d}\left(U,z_{1}(\cdot),z_{0}(\cdot),\{q_{a,k}^{(m)}\}_{a,k}\right)\ \textrm{as }j\to\infty.\label{eq:pf:Thm1:0.35}
\end{equation}

Recall from the end of Step 1 that $q_{a,k}^{(m,L_{j})}\le t_{k}$,
$q_{a,k}^{(m,L_{j})}\le q_{a,k^{\prime}}^{(m,L_{j})}$ and $q_{1,k}^{(m,L_{j})}+q_{0,k}^{(m,L_{j})}=t_{k}$
almost surely, for all $k$ and $k^{\prime}\ge k$. Equation (\ref{eq:pf:Thm1:0.35})
then implies $q_{a,k}^{(m)}\le t_{k}$, $q_{a,k}^{(m)}\le q_{a,k^{\prime}}^{(m)}$
and $q_{1,k}^{(m)}+q_{0,k}^{(m)}=t_{k}$ almost surely, for all $k$
and $k^{\prime}\ge k$. 

Define
\begin{align}
q_{a}^{(m,L)}(t) & =\sum_{k=0}^{2^{m}-1}q_{a,k}^{(m,L)}\mathbb{I}\{t_{k}\le t<t_{k+1}\},\ \textrm{and}\nonumber \\
q_{a}^{(m)}(t) & =\sum_{k=0}^{2^{m}-1}q_{a,k}^{(m)}\mathbb{I}\{t_{k}\le t<t_{k+1}\}.\label{eq:pf:Thm1:0.37}
\end{align}
It then follows from (\ref{eq:pf:Thm1:0.35}) that 
\begin{equation}
\left\{ U,z_{a}(\cdot),q_{a}^{(m,L_{j})}(\cdot)\right\} _{a}\xrightarrow{d}\left\{ U,z_{a}(\cdot),q_{a}^{(m)}(\cdot)\right\} _{a}\ \textrm{as }j\to\infty.\label{eq:pf:Thm1:0.4}
\end{equation}

Another implication of (\ref{eq:pf:Thm1:0.35}) is that
\begin{equation}
\left\{ x_{a,k}^{(m,L_{j})},q_{a,k}^{(m,L_{j})}\right\} _{a,k}\xrightarrow{d}\left\{ x_{a,k}^{(m)},q_{a,k}^{(m)}\right\} _{a,k}\ \textrm{as }j\to\infty,\label{eq:pf:Thm1:0.5}
\end{equation}
where
\[
x_{a,k}^{(m)}:=z_{a}(q_{a,k}^{(m)}).
\]

We conclude this step by deriving some limit approximations for $q_{n,a,k}^{(m)}$
(defined in \ref{eq:pf:Thm1:0.07}) and $x_{n,a,k}^{(m)}:=z_{n,a}(q_{n,a,k}^{(m)})$.
The dyadic discretization $q_{n,a,k}^{(m,L_{j})}$ of $q_{n,a,k}^{(m)}$
is such that $q_{n,a,k}^{(m,L_{j})}\to q_{n,a,k}^{(m)}$ and 
\begin{equation}
\sup_{n,k}\left|q_{n,a,k}^{(m)}-q_{n,a,k}^{(m,L_{j})}\right|\le2^{-L_{j}}\to0\ \textrm{as }j\to\infty.\label{eq:pf:Thm1:0.6}
\end{equation}
Recall the definition $x_{n,a,k}^{(m,L)}:=z_{n,a}(q_{n,a,k}^{(m,L)})$
from Step 1. For every $\epsilon>0$ and $a\in\{0,1\}$, (\ref{eq:pf:Thm1:0.6})
implies
\begin{align}
\limsup_{n\to\infty}\mathbb{P}_{n,0}\left(\sup_{k}\left|x_{n,a,k}^{(m)}-x_{n,a,k}^{(m,L_{j})}\right|>\epsilon\right) & \le\limsup_{n\to\infty}\mathbb{P}_{n,0}\left(\sup_{q\in[0,1],\delta\in[0,2^{-L_{j}}]}\left|z_{n,a}\left(q+\delta\right)-z_{n,a}(q)\right|>\epsilon\right)\nonumber \\
 & \to0\ \textrm{as }j\to\infty,\label{eq:pf:Thm1:0.7}
\end{align}
where the second step follows from \citet[Lemma 2.4.19]{karatzas2012brownian}.
Combining (\ref{eq:pf:Thm1:0.3}), (\ref{eq:pf:Thm1:0.5}) and (\ref{eq:pf:Thm1:0.7}),
we conclude 
\begin{equation}
\left\{ x_{n,a,k}^{(m)},q_{n,a,k}^{(m)}\right\} _{a,k}\xrightarrow[\mathbb{P}_{n,0}]{d}\left\{ x_{a,k}^{(m)},q_{a,k}^{(m)}\right\} _{a,k}.\label{eq:pf:Thm1:0.8}
\end{equation}

\subsubsection*{\textbf{Step 3: Taking $m\to\infty$.}}

\textbf{\\}Equation (\ref{eq:pf:Thm1:0.4}) applies for any fixed
$m$. Therefore the construction in Step 2 can be applied for each
$m$, giving rise to a sequence of processes $\{z_{a}(\cdot),q_{a}^{(m)}(\cdot),U\}_{a}$.

By construction, $\left|\sum_{a}q_{a}^{(m)}(t)-t\right|\le2^{-m}\ \forall\ t$
almost surely since, as shown in Step 2, $q_{1,k}^{(m)}+q_{0,k}^{(m)}=t_{k}\ \forall\ k$
almost surely. Furthermore, as $q_{a,k}^{(m)}\le q_{a,k+1}^{(m)}$
for all $k$ almost surely (as also shown in Step 2), it follows that
$q_{1}^{(m)}(\cdot),q_{0}^{(m)}(\cdot)$ are also almost surely monotone.
We now claim that the sequence $\{q_{a}^{(m)}(\cdot)\}_{m=1}^{\infty}$
is stochastically equicontinuous. Recall the definition of $\{q_{n,1,k}^{(m)}\}_{k}$
from (\ref{eq:pf:Thm1:0.05}) and observe that by the structure of
$q_{n,1}(\cdot)$, 
\[
\sup_{k}\left|q_{n,1,k+1}^{(m)}-q_{n,1,k}^{(m)}\right|\le2^{-m}+n^{-1}.
\]
In view of (\ref{eq:pf:Thm1:0.8}), the above implies 
\[
\mathbb{P}\left(\sup_{k}\left|q_{1,k+1}^{(m)}-q_{1,k}^{(m)}\right|>2^{-m}\right)=0
\]
for each $m$. Consequently, from the definition of $q_{1}^{(m)}(\cdot)$
in (\ref{eq:pf:Thm1:0.37}), it follows that for any $\delta>0$,
\begin{equation}
\mathbb{P}\left(\sup_{t}\left|q_{1}^{(m)}(t+\delta)-q_{1}^{(m)}(t)\right|>\delta+2^{-m}\right)=0.\label{eq:Lipschitz continuity of q_a^m}
\end{equation}
This implies $\{q_{1}^{(m)}(\cdot)\}_{m=1}^{\infty}$ is stochastically
equicontinuous. Stochastic equicontinuity of $\{q_{0}^{(m)}(\cdot)\}_{m}$
also follows since $q_{0}^{(m)}(t)=t-q_{1}^{(m)}(t)$ almost surely.

As it is stochastically equicontinuous, the sequence $\{q_{1}^{(m)}(\cdot),q_{0}^{(m)}(\cdot)\}_{m=1}^{\infty}$
is tight. Combined with the tightness of $\{z_{a}(\cdot)\}_{a},U$,
we conclude that the joint $\left\{ z_{a}(\cdot),U,q_{a}^{(m)}(\cdot)\right\} _{a}$
is also tight. Then, by Prohorov's theorem, there exists a subsequence
$\{m_{k}\}_{k=1}^{\infty}$, represented as $\{m\}$ without loss
of generality, such that
\begin{equation}
\left\{ z_{a}(\cdot),U,q_{a}^{(m)}(\cdot)\right\} _{a}\xrightarrow{d}\left\{ z_{a}(\cdot),U,q_{a}(\cdot)\right\} _{a}.\label{eq:pf:Thm1:0.85}
\end{equation}

\subsubsection*{\textbf{Step 4: Existence of an allocation process satisfying }(\ref{eq:pf:Thm1:0.85}\textbf{).}}

\textbf{\\}We now show there exists a version of $q_{a}(\cdot)$,
defined on a suitably constructed probability space $(\bar{\Omega},\bar{\mathcal{F}},\bar{\mathbb{P}})$,
that is a valid allocation process.

First, (\ref{eq:pf:Thm1:0.85}) and $\left|\sum_{a}q_{a}^{(m)}(t)-t\right|\le2^{-m}\ \forall\ t$
(as shown in Step 3) implies $\sum_{a}q_{a}(t)=t\ \forall\ t$ almost
surely. Second, $q_{a}(\cdot)$ is almost surely monotone as it is
the weak limit of almost surely monotone processes $\{q_{a}^{(m)}(\cdot)\}_{a}\in D[0,1]^{2}$
and the set of monotone functions is closed under the Skorokhod topology.
It thus remains to construct a version of $q_{a}(\cdot)$ that satisfies
the measurability requirement of Definition \ref{Allocation_process_definition}. 

By (\ref{eq:pf:Thm1:0.4}) and (\ref{eq:pf:Thm1:0.85}), there exists
a sequence $\{(m_{j},L_{j})\}_{j=1}^{\infty}$ with $(m_{j},L_{j})\to(\infty,\infty)$,
under which\footnote{It is possible to choose such a sequence since weak convergence can
be metrized, e.g., using the bounded Lipschitz metric.} 
\begin{equation}
\left\{ z_{a}(\cdot),U,q_{a}^{(m_{j},L_{j})}(\cdot)\right\} _{a}\xrightarrow{d}\left\{ z_{a}(\cdot),U,q_{a}(\cdot)\right\} _{a}\ \textrm{as }j\to\infty.\label{eq:pf:Thm1:0.9}
\end{equation}
Let $(\Omega,\mathcal{F},\mathbb{P})$ represent the canonical probability
space corresponding to $\{z_{a}(\cdot)\}_{a},U$. Specifically, we
set $\Omega\equiv C^{d}[0,1]^ {}\times C^{d}[0,1]\times[0,1]$, $\mathcal{F}\equiv\mathcal{B}(\Omega)$
and $\mathbb{P}=\mathcal{W}^{d}\otimes\mathcal{W}^{d}\otimes\lambda$,
where $C^{d}[0,1]^ {}$ represents the space of continuous functions
from $[0,1]$ to $\mathbb{R}^{d}$, $\mathcal{W}^{d}$ denotes the
$d$-dimensional Wiener measure on $C^{d}[0,1]$, and $\lambda$ is
the Lebesgue measure on $[0,1]$. Then, (\ref{eq:pf:Thm1:0.9}) implies,
by the properties of stable convergence, that\footnote{See \citet{hausler2015stable} for a textbook treatment of stable
convergence.} 
\begin{equation}
\left\{ q_{a}^{(m_{j},L_{j})}(\cdot)\right\} _{a}\xrightarrow[\mathcal{\mathcal{F}}]{\textrm{stably}}\left\{ q_{a}(\cdot)\right\} _{a}.\label{eq:pf:Thm1:0.92}
\end{equation}

Set $\bm{\gamma}:=(\gamma_{1},\gamma_{0})$, $\bm{Q}_{j}(t):=\left(q_{1}^{(m_{j},L_{j})}(t),q_{0}^{(m_{j},L_{j})}(t)\right)$
and $\bm{Q}(t):=(q_{1}(t),q_{0}(t))$. Since $\bm{Q}_{j}(\cdot)_{}$
takes values on the Skorokhod space $D[0,1]^{2}$---which is Polish---stable
convergence $\bm{Q}_{j}(\cdot)\xrightarrow[\mathcal{\mathcal{F}}]{\textrm{stably}}\bm{Q}(\cdot)$
implies there exists a Markov kernel, $K(\omega,dy)$, from $\Omega$
to $D[0,1]^{2}$ that acts as the limit version of the conditional
distribution of $\bm{Q}_{j}$ given $\mathcal{F}$.\footnote{Formally, $K(\omega,dy)$ is such that $\mathbb{E}[f(\bm{Q}_{j})\vert\mathcal{F}]\xrightarrow{p}\int f(y)K(\cdot,dy)$
for all bounded continuous $f(\cdot)$.} We can then construct a measurable representation of $\bm{Q}(\cdot)$
on the extended probability space, $(\bar{\Omega},\bar{\mathcal{F}},\bar{\mathbb{P}})\equiv(\Omega\times[0,1],\mathcal{F}\otimes\mathcal{B}[0,1],\mathbb{P}\otimes\lambda)$,
such that $K(\omega,dy)$ represents the conditional probability of
$\bm{Q}(\cdot)$ given $\mathcal{F}$. In essence, the extended probability
space augments the underlying set of variables $\{z_{a}(\cdot)\}_{a},U$
with another exogenous randomization $V\sim\textrm{Uniform}[0,1]$.
By the usual properties of stable convergence, (\ref{eq:pf:Thm1:0.9})
continues to hold for this representation of $\bm{Q}(\cdot)$. 

Let $\mathcal{\mathcal{F}}_{\gamma}^{(a)}\subseteq\mathcal{F}$ denote
the right-continuous filtration generated by the sample paths of $z_{a}(\cdot)$
between $0$ and $\gamma$, and take $\mathcal{G}_{\gamma_{1},\gamma_{0}}\subseteq\mathcal{F}$
to be the augmentation of $\mathcal{\mathcal{F}}_{\gamma_{1}}^{(1)}\lor\mathcal{\mathcal{F}}_{\gamma_{0}}^{(0)}\lor\sigma(U)$
with respect to $(\Omega,\mathcal{F},\mathbb{P})$. In the extended
probability space, this gives rise to the extended filtration $\bar{\mathcal{G}}_{\gamma_{1},\gamma_{0}}\equiv\mathcal{G}_{\gamma_{1},\gamma_{0}}\lor\sigma(V)\subseteq\bar{\mathcal{F}}$.
Note that the filtration $\bar{\mathcal{G}}_{\gamma_{1},\gamma_{0}}$
inherits the right-continuous and augmented nature of $\mathcal{G}_{\gamma_{1},\gamma_{0}}$.
We now argue that $\left\{ \bm{Q}(t)\le\bm{\gamma}\right\} :=\{q_{1}(t)\le\gamma_{1},q_{0}(t)\le\gamma_{0}\}$
is $\bar{\mathcal{G}}_{\gamma_{1},\gamma_{0}}$ measurable for each
$\gamma_{1},\gamma_{0},t\in[0,1]$ such that $\gamma_{1}+\gamma_{0}\ge t$. 

For any $\epsilon>0$ and $\bm{u}=(u_{1},u_{0})\in[0,1]^{2}$, let
$\phi_{\epsilon,\bm{\gamma}}(\bm{u})$ denote a smoothed version of
$\mathbb{I}\{u_{1}\le\gamma_{1},u_{0}\le\gamma_{0}\}$, defined as
\[
\phi_{\epsilon,\bm{\gamma}}(\bm{u})=\begin{cases}
1 & \textrm{if }u_{1}\le\gamma_{1},u_{0}\le\gamma_{0}\\
0 & \textrm{if }u_{1}\ge\gamma_{1}+\epsilon\textrm{ or }u_{0}\ge\gamma_{0}+\epsilon\\
\textrm{linear decay} & \textrm{otherwise}.
\end{cases}
\]
Given a fixed value of $j$, let $k\in\{0,\dots,2^{m_{j}}\}$, $t_{k}:=k2^{-m_{j}}$
and $k(t):=\max\{k:t\ge t_{k}\}$. By the definition of $q_{a}^{(m_{j},L_{j})}(\cdot)$,
along with Condition C2 from Step 1, $\{\bm{Q}_{j}(t)\le\bm{\gamma}+\epsilon^{\prime}\}\equiv\left\{ q_{1,k(t)}^{(m_{j},L_{j})}\le\gamma_{1}+\epsilon^{\prime},q_{0,k(t)}^{(m_{j},L_{j})}\le\gamma_{0}+\epsilon^{\prime}\right\} $
is 
\[
\mathcal{G}_{\gamma_{1}+\epsilon^{\prime}+2^{-m_{j}}+2^{-L_{j}},\gamma_{0}+\epsilon^{\prime}+2^{-m_{j}}+2^{-L_{j}}}\subseteq\mathcal{G}_{\gamma_{1}+2\epsilon^ {},\gamma_{0}+2\epsilon^ {}}
\]
measurable for all $\epsilon^{\prime}\in[0,\epsilon]$ and $j$ sufficiently
large. This implies, by the definition of $\phi_{\epsilon,\bm{\gamma}}(\cdot)$,
that $\phi_{\epsilon,\bm{\gamma}}(\bm{Q}_{j}(t))$ is $\mathcal{G}_{\gamma_{1}+2\epsilon,\gamma_{0}+2\epsilon}$-measurable.
Hence, for any bounded $\mathcal{F}$-measurable random variable $W$
that is independent of $\mathcal{G}_{\gamma_{1}+2\epsilon,\gamma_{0}+2\epsilon}$
(i.e., $W$ relies only on Wiener process increments ``after'' $\mathcal{G}_{\gamma_{1}+2\epsilon,\gamma_{0}+2\epsilon}$),
we must have 
\begin{equation}
\mathbb{E}\left[\phi_{\epsilon,\bm{\gamma}}(\bm{Q}_{j}(t))W\right]=\mathbb{E}\left[\phi_{\epsilon,\bm{\gamma}}(\bm{Q}_{j}(t))\right]\mathbb{E}[W].\label{eq:pf:Thm1:0.95}
\end{equation}

The definition of stable convergence states that 
\[
\mathbb{E}\left[f(\bm{Q}_{j}(t))Z\right]\to\bar{\mathbb{E}}\left[f(\bm{Q}(t))Z\right]\ \textrm{as }j\to\infty
\]
for any bounded continuous function $f(\cdot)$, and any bounded $\mathcal{F}$-measurable
random variable $Z$. Applying the above to the factorization (\ref{eq:pf:Thm1:0.95})
with $f=\phi_{\epsilon,\bm{\gamma}}$ and $Z=\{1,W\}$, we get 
\begin{equation}
\bar{\mathbb{E}}\left[\phi_{\epsilon,\bm{\gamma}}(\bm{Q}(t))W\right]=\bar{\mathbb{E}}\left[\phi_{\epsilon,\bm{\gamma}}(\bm{Q}(t))\right]\mathbb{E}[W].\label{eq:pf:Thm1:0.99}
\end{equation}
(\ref{eq:pf:Thm1:0.99}) holds for any bounded random variable $W$
independent of $\mathcal{G}_{\gamma_{1}+2\epsilon,\gamma_{0}+2\epsilon}$
in the original probability space. But space of all such $W$ is equivalent
to the space of all bounded random variables independent of $\bar{\mathcal{G}}_{\gamma_{1}+2\epsilon,\gamma_{0}+2\epsilon}$
in the extended probability space. Hence, (\ref{eq:pf:Thm1:0.99})
implies $\phi_{\epsilon,\bm{\gamma}}(\bm{Q}(t))$ is independent of
the ``future'' Wiener-process noise relative to $\bar{\mathcal{G}}_{\gamma_{1}+2\epsilon,\gamma_{0}+2\epsilon}$,
and is therefore $\bar{\mathcal{G}}_{\gamma_{1}+2\epsilon,\gamma_{0}+2\epsilon}$-measurable. 

We are interested in the event $E\equiv\{\bm{Q}_{}(t)\le\bm{\gamma}\}$.
Notice, from the definition of $\phi_{\epsilon,\bm{\gamma}}(\cdot)$,
that $\mathbb{I}_{E}=\lim_{\epsilon\downarrow0}\phi_{\epsilon,\bm{\gamma}}(\bm{Q}(t))$
point-wise for each $\omega\in\bar{\Omega}$. But as shown earlier,
$\phi_{\epsilon,\bm{\gamma}}(\bm{Q}(t))$ is $\bar{\mathcal{G}}_{\gamma_{1}+2\epsilon,\gamma_{0}+2\epsilon}$-measurable;
consequently, $\mathbb{I}_{E}$ must be $\cap_{\epsilon\downarrow0}\bar{\mathcal{G}}_{\gamma_{1}+2\epsilon,\gamma_{0}+2\epsilon}\equiv\bar{\mathcal{G}}_{\gamma_{1},\gamma_{0}}$-measurable,
where the equivalence is due to the right continuity of the filtrations. 

This concludes the existence of a valid allocation process $\{q_{a}(\cdot)\}_{a}$---defined
on an extended probability space $(\bar{\Omega},\bar{\mathcal{F}},\bar{\mathbb{P}})$---that
satisfies (\ref{eq:pf:Thm1:0.85}). But the additional $\textrm{V\ensuremath{\sim}Uniform[0,1]}$
randomization employed in the definition of $(\bar{\Omega},\bar{\mathcal{F}},\bar{\mathbb{P}})$
can be combined with $U$ to form a new $\textrm{Uniform}[0,1]$ random
variable $\bar{U}$. So the measurability requirement of Definition
\ref{Allocation_process_definition} is satisfied by taking the relevant
probability space to be $(\bar{\Omega},\bar{\mathcal{F}},\bar{\mathbb{P}})$. 

\subsubsection*{\textbf{Step 5: Completing the proof.}}

\textbf{\\}Define $x_{a}^{(m)}(t)=z_{a}(q_{a}^{(m)}(t))$ and $x_{a}(t)=z_{a}(q_{a}(t))$.
By construction, 
\[
x_{a}^{(m)}(t)=\sum_{k=0}^{2^{m}-1}x_{a,k}^{(m)}\{t_{k}\le t<t_{k+1}\},
\]
where $x_{a,k}^{(m)}=z_{a}(q_{a,k}^{(m)})$, as defined earlier in
Step 2. Since $q_{a}(\cdot)$ is the limit of stochastically equicontinuous
processes, it has almost surely continuous sample paths. Combined
with (\ref{eq:pf:Thm1:0.85}), this implies 
\begin{equation}
\left\{ x_{a}^{(m)}(\cdot),U,q_{a}^{(m)}(\cdot)\right\} _{a}\xrightarrow{d}\left\{ x_{a}(\cdot),U,q_{a}(\cdot)\right\} _{a}\ \textrm{as }m\to\infty.\label{eq:pf:Thm1:1.1}
\end{equation}

By the properties of weak convergence (see, e.g., \citealt[Chapter 1.12]{van1996weak}),
part (iii) of Theorem \ref{Thm: ART} follows if we show 
\begin{equation}
\mathbb{E}_{n,0}\left[f\left(x_{n,1}(\cdot),x_{n,0}(\cdot),q_{n,1}(\cdot),q_{n,0}(\cdot)\right)\right]\to\mathbb{E}\left[f\left(x_{1}(\cdot),x_{0}(\cdot),q_{1}(\cdot),q_{0}(\cdot)\right)\right]\label{eq:pf:Thm1:1.2}
\end{equation}
for all bounded Lipschitz continuous $f(\cdot)$. 

Fix a value of $m$ and construct dyadic approximations for $x_{n,a}(\cdot),q_{n,a}(\cdot)$
of the form
\begin{align*}
x_{n,a}^{(m)}(t) & =\sum_{k=0}^{2^{m}-1}x_{n,a}(t_{k})\mathbb{I}\left\{ t_{k}\le t<t_{k+1})\right\} \equiv\sum_{k=0}^{2^{m}-1}x_{n,a,k}^{(m)}\mathbb{I}\left\{ t_{k}\le t<t_{k+1})\right\} ,\\
q_{n,a}^{(m)}(t) & =\sum_{k=0}^{2^{m}-1}q_{n,a}(t_{k})\mathbb{I}\left\{ t_{k}\le t<t_{k+1})\right\} \equiv\sum_{k=0}^{2^{m}-1}q_{n,a,k}^{(m)}\mathbb{I}\left\{ t_{k}\le t<t_{k+1})\right\} .
\end{align*}
In what follows, let $S_{n}:=\{x_{n,a}(\cdot),q_{n,a}(\cdot)\}_{a}$,
$S_{n}^{(m)}:=\{x_{n,a}^{(m)}(\cdot),q_{n,a}^{(m)}(\cdot)\}_{a}$,
$S:=\{x_{a}(\cdot),q_{a}(\cdot)\}_{a}$ and $S^{(m)}:=\{x_{a}^{(m)}(\cdot),q_{a}^{(m)}(\cdot)\}_{a}$.
We can then decompose
\begin{align*}
\mathbb{E}_{n,0}\left[f\left(S_{n}\right)\right]-\mathbb{E}\left[f\left(S\right)\right] & =\text{\ensuremath{\left\{  \mathbb{E}\left[f\left(S^{(m)}\right)\right]-\mathbb{E}\left[f\left(S\right)\right]\right\} } }+\text{\ensuremath{\left\{  \mathbb{E}_{n,0}\left[f\left(S_{n}^{(m)}\right)\right]-\mathbb{E}\left[f\left(S^{(m)}\right)\right]\right\} } }\\
 & \quad+\text{\ensuremath{\left\{  \mathbb{E}_{n,0}\left[f\left(S_{n}\right)\right]-\mathbb{E}_{n,0}\left[f\left(S_{n}^{(m)}\right)\right]\right\} } }\\
 & :=T_{1}^{(m)}+T_{2n}^{(m)}+T_{3n}^{(m)}.
\end{align*}

By (\ref{eq:pf:Thm1:1.1}), $\vert T_{1}^{(m)}\vert\to0$ as $m\to\infty$.
Furthermore, (\ref{eq:pf:Thm1:0.8}) implies $\lim_{n\to\infty}T_{2n}^{(m)}=0$
for any given $m$. It remains to bound $T_{3n}^{(m)}$. Note that
by the definition of $q_{n,a}(\cdot)$, 
\[
\sup_{t\in[0,1]}\vert q_{n,a}(t+2^{-m})-q_{n,a}(t)\vert\le2^{-m}+n^{-1}:=\delta_{m,n}
\]
for any $m>0$. Then, letting $B$ denote the upper bound of $\vert f(\cdot)\vert$
and $C$ the Lipschitz constant of $f(\cdot)$, we observe that for
every $\epsilon>0$, and all $n$ sufficiently large so that $\delta_{m,n}<2^{-(m-1)}$,
\begin{align*}
\vert T_{3n}^{(m)}\vert & \le2C(2^{-m}+n^{-1})+\sum_{a\in\{0,1\}}\left\{ C\epsilon+2B\cdot\mathbb{P}_{n,0}\left(\sup_{q\in[0,1],\delta\in[0,2^{-(m-1)}]}\left|z_{n,a}\left(q+\delta\right)-z_{n,a}(q)\right|>\epsilon\right)\right\} .
\end{align*}
Define 
\[
r_{m}^{(a)}(\epsilon):=\limsup_{n\to\infty}\mathbb{P}_{n,0}\left(\sup_{q\in[0,1],\delta\in[0,2^{-(m-1)}]}\left|z_{n,a}\left(q+\delta\right)-z_{n,a}(q)\right|>\epsilon\right).
\]
Then, for any $\epsilon>0$, 
\[
\limsup_{n\to\infty}\vert T_{3n}^{(m)}\vert\le C2^{-m}+2C\epsilon+2B\left(r_{m}^{(1)}(\epsilon)+r_{m}^{(0)}(\epsilon)\right):=\bar{r}(m,\epsilon),
\]
By \citet[Lemma 2.4.19]{karatzas2012brownian}, $\lim_{m\to\infty}r_{m}^{(a)}(\epsilon)=0$
for any $\epsilon>0$, so $\lim_{m\to\infty}\bar{r}(m,\epsilon)=2C\epsilon$. 

To conclude, we have shown that for any given $m,\epsilon$,
\begin{equation}
\limsup_{n\to\infty}\left|\mathbb{E}_{n,0}\text{\ensuremath{\left[f\left(S_{n}\right)\right]}}-\mathbb{E}\left[f\left(S\right)\right]\right|\le\vert T_{1}^{(m)}\vert+\bar{r}(m,\epsilon).\label{eq:pf:Thm1:6}
\end{equation}
In view of the previous results, the right hand side of (\ref{eq:pf:Thm1:6})
can be made arbitrarily small by taking $m\to\infty$ and $\epsilon\to0$.
This proves (\ref{eq:pf:Thm1:1.2}). 

It remains to show that $q_{a}(\cdot)$ is Lipschitz continuous. But
this is a straightforward consequence of (\ref{eq:Lipschitz continuity of q_a^m})
and the fact that $q_{a}(\cdot)$ is the weak limit of $q_{a}^{(m)}(\cdot)$,
see (\ref{eq:pf:Thm1:0.9}). 

\newpage{}

\section*{\textbf{SUPPLEMENTARY APPENDIX (Online Appendix) }}

\section{Proofs of the remaining results}

\subsection{Proof of Corollary \ref{Cor: ART}}

Recall the quantity $\hat{\varphi}(\bm{h};\gamma_{1},\gamma_{0})$
from Section \ref{sec:Equivalence-of-experiments:}. By (\ref{eq:SLAN property}),
\[
\hat{\varphi}(\bm{h};q_{n,1}(1),q_{n,0}(1))=\sum_{a}\left\{ h^{(a)\intercal}I_{a}^{1/2}x_{n,a}(1)-\frac{q_{n,a}(1)}{2}h^{(a)\intercal}I_{a}h^{(a)}\right\} +o_{\mathbb{P}_{n,0}}(1).
\]
Combining the above with Theorem \ref{Thm: ART} and Assumption 2
gives
\begin{align}
\left(\begin{array}{c}
\{x_{n,a}(\cdot),q_{n,a}(\cdot)\}_{a}\\
\hat{\varphi}(\bm{h};q_{n,1}(1),q_{n,0}(1))
\end{array}\right) & \xrightarrow[\mathbb{P}_{n,0}]{d}\left(\begin{array}{c}
\{x_{a}(\cdot),q_{a}(\cdot)\}_{a}\\
\ln V
\end{array}\right);\ \textrm{where}\label{eq:pf:Thm1:weak convergence-1}\\
V\sim & \exp\sum_{a}\left\{ h^{(a)\intercal}I_{a}^{1/2}x_{a}(1)-\frac{q_{a}(1)}{2}h^{(a)\intercal}I_{a}h^{(a)}\right\} .\nonumber 
\end{align}
Denote 
\[
S(t):=\sum_{a}h^{(a)\intercal}I_{a}^{1/2}x_{a}(t)
\]
and 
\[
M(t):=\exp\sum_{a}\left\{ h^{(a)\intercal}I_{a}^{1/2}x_{a}(t)-\frac{q_{a}(t)}{2}h^{(a)\intercal}I_{a}h^{(a)}\right\} .
\]
By Lemma \ref{Lemma: Martingale}, $S(t)$ is an $\mathcal{I}_{t}$-martingale
and its quadratic variation is given by $\sum_{a}q_{a}(t)h^{(a)\intercal}I_{a}h^{(a)}$.
Hence, $M(t)$ is the stochastic/Doleans-Dade exponential of $S(t)$.
As $q_{a}(t)\le1$ almost surely, 
\[
E\left[\exp\left\{ \sum_{a}\frac{q_{a}(t)}{2}h^{(a)\intercal}I_{a}h^{(a)}\right\} \right]\le\exp\left\{ \sum_{a}\frac{1}{2}h^{(a)\intercal}I_{a}h^{(a)}\right\} <\infty.
\]
Thus, Novikov's condition is satisfied and $M(t)$ is also an $\mathcal{I}_{t}$-martingale.
Doob's optional sampling theorem then implies $E[V]\equiv E[M(1)]=E[M(0)]=1$. 

Since the processes, $\{x_{a}(\cdot),q_{a}(\cdot)\}_{a}$, are tight,
their sample paths lie in a separable metric space $\mathcal{D}$,
with an associated Borel sigma-algebra $\mathcal{B}(\mathcal{D})$.
This, together with the fact that $V\ge0$ and $\mathbb{E}[V]=1$,
implies, by a version of Le Cam's third lemma for processes (see,
e.g., \citealt[Theorem 3.10.7]{van1996weak}), that 
\[
\{x_{n,a}(\cdot),q_{n,a}(\cdot)\}_{a}\xrightarrow[\mathbb{P}_{n,\bm{h}}]{d}\mathcal{L};\ \textrm{where }\mathcal{L}(B):=E\left[\mathbb{I}\left(\{x_{a}(\cdot),q_{a}(\cdot)\}_{a}\in B\right)V\right]\ \forall\ B\in\mathcal{B}(\mathcal{D}).
\]
But $\{x_{a}(\cdot),q_{a}(\cdot)\}_{a}$ is adapted to $\mathcal{I}_{1}:=\mathcal{G}_{q_{1}(1),q_{0}(1)}$
due to Definition \ref{Allocation_process_definition}, so, by the
Girsanov theorem, 
\begin{align*}
\mathcal{L}(B) & =E\left[\mathbb{I}\left(\{x_{a}(\cdot),q_{a}(\cdot)\}_{a}\in B\right)\exp\sum_{a}\left\{ h^{(a)\intercal}I_{a}^{1/2}z_{a}(q_{a}(1))-\frac{q_{a}(1)}{2}h^{(a)\intercal}I_{a}h^{(a)}\right\} \right],\\
 & =\mathbb{P}_{\bm{h}}\left(\{x_{a}(\cdot),q_{a}(\cdot)\}_{a}\in B\right),
\end{align*}
where the probability $\mathbb{P}_{\bm{h}}$ (defined in Section \ref{subsec:Convergence-of-likelihood-ratios})
is the one induced by the sample paths of $z_{a}(\cdot)\sim I_{a}^{1/2}h^{(a)}\cdot+W_{a}(\cdot)$,
together with an exogenous randomization $U$. The claim therefore
follows. 

\subsection{Proof of Theorem \ref{Thm: Point estimation}}

Recall the definition $\hat{\varphi}(\bm{h};\gamma_{1},\gamma_{0})$
from Section \ref{sec:Equivalence-of-experiments:}. By (\ref{eq:SLAN property}),
\[
\hat{\varphi}(\bm{h};q_{n,1}(1),q_{n,0}(1))=\sum_{a}\left\{ h^{(a)\intercal}I_{a}^{1/2}x_{n,a}(1)-\frac{q_{n,a}(1)}{2}h^{(a)\intercal}I_{a}h^{(a)}\right\} +o_{\mathbb{P}_{n,0}}(1).
\]
Combining the above with Theorem \ref{Thm: ART} and Assumption 2
gives 
\begin{equation}
\hat{\varphi}(\bm{h};q_{n,1}(1),q_{n,0}(1))\xrightarrow[\mathbb{P}_{n,0}]{d}\sum_{a}\left\{ h^{(a)\intercal}I_{a}^{1/2}x_{a}(1)-\frac{q_{a}(1)}{2}h^{(a)\intercal}I_{a}h^{(a)}\right\} .\label{pf:Thm:ART:eq:1}
\end{equation}

By the definition, the sequence $\sqrt{n}(T_{n}-\bm{\theta}_{0})$
is tight, and consequently, so is the sequence $\sqrt{n}(T_{n}-\bm{\theta}(\bm{h}))=\sqrt{n}(T_{n}-\bm{\theta}_{0})-\bm{h}$.
Since $\left\{ x_{n,a}(1),q_{n,a}(1)\right\} _{a}$ converges in distribution
due to Theorem \ref{Thm: ART} and Assumption 2, it follows that the
joint $\left(\sqrt{n}(T_{n}-\bm{\theta}(\bm{h})),\left\{ x_{n,a}(1),q_{n,a}(1)\right\} _{a}\right)$
is also tight. Hence, by Prohorov's theorem, given any sequence $\{n_{j}\}$,
there exists a further sub-sequence $\{n_{j_{m}}\}$---represented
as $\{n\}$ for ease of notation---and a random variable $\bar{T}$
such that 
\[
\left(\sqrt{n}(T_{n}-\bm{\theta}(\bm{h})),\left\{ x_{n,a}(1),q_{n,a}(1)\right\} _{a}\right)\xrightarrow[\mathbb{P}_{n,0}]{d}\left(\bar{T}-\bm{h},\left\{ x_{a}(1),q_{a}(1)\right\} _{a}\right).
\]
Combined with (\ref{pf:Thm:ART:eq:1}), this implies
\begin{align}
\left(\begin{array}{c}
\sqrt{n}(T_{n}-\bm{\theta}(\bm{h}))\\
\hat{\varphi}(\bm{h};q_{n,1}(1),q_{n,0}(1))
\end{array}\right) & \xrightarrow[\mathbb{P}_{n,0}]{d}\left(\begin{array}{c}
\bar{T}-\bm{h}\\
\ln V
\end{array}\right);\ \textrm{where}\label{eq:pf:Thm1:weak convergence}\\
V\sim & \exp\sum_{a}\left\{ h^{(a)\intercal}I_{a}^{1/2}x_{a}(1)-\frac{q_{a}(1)}{2}h^{(a)\intercal}I_{a}h^{(a)}\right\} .\nonumber 
\end{align}
As in the proof of Corollary \ref{Cor: ART}, $E[V]=1$. 

We now claim that 
\begin{equation}
\sqrt{n}(T_{n}-\bm{\theta}(\bm{h}))\xrightarrow[\mathbb{P}_{n,\bm{h}}]{d}\mathcal{L};\ \textrm{where }\mathcal{L}(B):=E\left[\mathbb{I}\{\bar{T}-\bm{h}\in B\}V\right]\ \forall\ B\in\mathcal{B}(\mathbb{R}^{2d}).\label{eq:pf:Thm1:weak convergence 2}
\end{equation}
It is clear from $V\ge0$ and $E[V]=1$ that $\mathcal{L}$ is a probability
measure, and that for every bounded measurable function $f:\mathbb{R}\to\mathbb{R}$,
$\int fd\mathcal{L}=E[f(\bar{T}-\bm{h})V]$. Furthermore, for any
$f(\cdot)$ lower-semicontinuous and non-negative, 
\begin{align}
 & \lim\inf\mathbb{E}_{n,\bm{h}}\left[f\left(\sqrt{n}(T_{n}-\bm{\theta}(\bm{h}))\right)\right]\nonumber \\
 & \ge\lim\inf\mathbb{E}_{n,0}\left[f\left(\sqrt{n}(T_{n}-\bm{\theta}(\bm{h}))\right)\frac{d\mathbb{P}_{n,\bm{h}}}{d\mathbb{P}_{n,0}}\right]\nonumber \\
 & =\lim\inf\mathbb{E}_{n,0}\left[f\left(\sqrt{n}(T_{n}-\bm{\theta}(\bm{h}))\right)\exp\left\{ \hat{\varphi}(\bm{h};q_{n,1}(1),q_{n,0}(1))\right\} \right]\nonumber \\
 & \ge E[f\left(\bar{T}-\bm{h}\right)V].\label{eq:weak convergence 3}
\end{align}
The equality in (\ref{eq:weak convergence 3}) follows from the law
of iterated expectations since $T_{n}$ is $\mathcal{I}_{n,1}:=\mathcal{G}_{n,q_{n,1}(1),q_{n,0}(1)}$
measurable, 
\[
\frac{d\mathbb{P}_{n,\bm{h}}}{d\mathbb{P}_{n,0}}\equiv\frac{d\mathbb{P}_{n,\bm{h}}}{d\mathbb{P}_{n,0}}\left({\bf y}_{n}^{(1)},{\bf y}_{n}^{(0)}\right)=\exp\left\{ \hat{\varphi}(\bm{h};1,1)\right\} 
\]
by definition (see Section \ref{subsec:Convergence-of-likelihood-ratios}),
and 
\[
\mathbb{E}_{n,0}\left[\exp\left\{ \hat{\varphi}(\bm{h};1,1)\right\} \vert\mathcal{I}_{n,1}\right]=\exp\left\{ \hat{\varphi}(\bm{h};q_{n,1}(1),q_{n,0}(1))\right\} 
\]
as the observations are iid given $\bm{h}$. The last inequality in
(\ref{eq:weak convergence 3}) follows from applying the portmanteau
lemma on (\ref{eq:pf:Thm1:weak convergence}). Applying the portmanteau
lemma again, in the converse direction, on (\ref{eq:weak convergence 3}),
gives (\ref{eq:pf:Thm1:weak convergence 2}). 

Weak convergence, (\ref{eq:pf:Thm1:weak convergence 2}), implies
that for any non-negative loss function $l(\cdot)$,
\begin{align}
 & \liminf_{n\to\infty}\mathbb{E}_{n,\bm{h}}\left[l\left(\sqrt{n}(T_{n}-\bm{\theta}(\bm{h}))\right)\right]\nonumber \\
 & \ge E\left[l(\bar{T}-\bm{h})\exp\sum_{a}\left\{ h^{(a)\intercal}I_{a}^{1/2}x_{a}(1)-\frac{q_{a}(1)}{2}h^{(a)\intercal}I_{a}h^{(a)}\right\} \right].\label{eq:pf:Thm:ART:2}
\end{align}
Define $s:=\{x_{1}(1),x_{0}(1),q_{1}(1),q_{0}(1)\}$ and $T(s):=E[\bar{T}\vert s]$.
Since $l(\cdot)$ is convex, the conditional version of Jensen's inequality
implies 
\begin{align*}
 & E\left[l(\bar{T}-\bm{h})\exp\sum_{a}\left\{ h^{(a)\intercal}I_{a}^{1/2}x_{a}(1)-\frac{q_{a}(1)}{2}h^{(a)\intercal}I_{a}h^{(a)}\right\} \right]\\
 & =E\left[E\left[\left.l(\bar{T}-\bm{h})\right|s\right]\exp\sum_{a}\left\{ h^{(a)\intercal}I_{a}^{1/2}x_{a}(1)-\frac{q_{a}(1)}{2}h^{(a)\intercal}I_{a}h^{(a)}\right\} \right]\\
 & \ge E\left[l(T-\bm{h})\exp\sum_{a}\left\{ h^{(a)\intercal}I_{a}^{1/2}z_{a}(q_{a}(1))-\frac{q_{a}(1)}{2}h^{(a)\intercal}I_{a}h^{(a)}\right\} \right].
\end{align*}
But by the Girsanov theorem, applied on the processes $\{z_{a}(\cdot)\}_{a}$,
the last term is just the expectation, $\mathbb{E}_{h}[l(T-\bm{h})]$,
of $l(T-\bm{h})$ when $x_{a}(t):=z_{a}(q_{a}(t))$ and $z_{a}(\cdot)$
is distributed as a Gaussian process with drift $I_{a}^{1/2}h^{(a)}$,
i.e., when $z_{a}(q)\sim I_{a}^{1/2}h^{(a)}q+W_{a}(q)$. 

\subsection{Proof of Corollary \ref{Cor: Bayes risk}}

For any tight sequence of estimators $\{T_{n}\}_{n}$, there exists
a further subsequence $\{T_{n_{k}}\}_{k}$, and an estimator $T$
in the limit experiment such that
\[
\liminf_{k\to\infty}\int R_{n_{k}}(T_{n_{k}},\bm{h})dm(\bm{h})\ge\int\liminf_{k\to\infty}R_{n_{k}}(T_{n_{k}},\bm{h})dm(\bm{h})\ge\int R(T,\bm{h})dm(\bm{h}),
\]
where the first inequality follows by Fatou's lemma, and the second
inequality by Theorem \ref{Thm: Point estimation}. 

\subsection{Proof of Theorem \ref{Thm: Welfare}}

Due to Assumption 3, the claim follows if we show that $\mathbb{E}_{n,\bm{h}}[q_{n,a}(1)]\to\mathbb{E}_{\bm{h}}[q_{a}(1)]$
for each $\bm{h}$ and $a\in\{0,1\}$. Theorem \ref{Thm: ART} and
Assumption 2 gives 
\begin{align*}
\left(\begin{array}{c}
q_{n,a}(1)\\
\hat{\varphi}(\bm{h};q_{n,1}(1),q_{n,0}(1))
\end{array}\right) & \xrightarrow[\mathbb{P}_{n,0}]{d}\left(\begin{array}{c}
q_{a}(1)\\
\ln V
\end{array}\right);\ \textrm{where}\\
V\sim & \exp\sum_{a}\left\{ h^{(a)\intercal}I_{a}^{1/2}x_{a}(1)-\frac{q_{a}(1)}{2}h^{(a)\intercal}I_{a}h^{(a)}\right\} .
\end{align*}
By similar arguments as in the proof of Theorem \ref{Thm: Point estimation},
the above implies 
\[
q_{n,a}(1)\xrightarrow[\mathbb{P}_{n,\bm{h}}]{d}\mathcal{L};\ \textrm{where }\mathcal{L}(B):=E\left[\mathbb{I}\{q_{a}(1)\in B\}V\right]\ \forall\ B\in\mathcal{B}(\mathbb{R}).
\]
Consequently, 
\[
\mathbb{E}_{n,\bm{h}}[q_{n,a}(1)]\to\mathbb{E}\left[q_{a}(1)e^{\sum_{a}\left\{ h^{(a)\intercal}I_{a}^{1/2}x_{a}(1)-\frac{q_{a}(1)}{2}h^{(a)\intercal}I_{a}h^{(a)}\right\} }\right].
\]
But by the Girsanov theorem, the right hand side is just the expectation,
$\mathbb{E}_{\bm{h}}[q_{a}(1)]$, of $q_{a}(1)$ when $x_{a}(t):=z_{a}(q_{a}(t))$
and $z_{a}(\cdot)$ is distributed as a Gaussian process with drift
$I_{a}^{1/2}h^{(a)}$. 

\subsection{Proof of Theorem \ref{Thm:e-processes}}

Throughout the proof, let $\left(\bar{\varepsilon}(\cdot,\cdot),\{z_{a}(\cdot)\}_{a},U\right)$
denote the weak limit of $\left(\bar{\varepsilon}_{n}(\cdot,\cdot),\{z_{n,a}(\cdot)\}_{a},U\right)$
in Assumption 4, realized on the canonical probability space $(\Omega,\mathcal{F},\mathbb{P})$
from Step 4 of the proof of Theorem \ref{Thm: ART}, enlarged so as
to carry $\bar{\varepsilon}(\cdot)$. For any sequence of empirical
allocation processes satisfying Assumption 2, $\{z_{n,a}(\cdot),q_{n,a}(\cdot)\}_{a}$
converges weakly to some $\{z_{a}(\cdot),q_{a}(\cdot)\}_{a}$. Then,
by Prohorov's theorem, the sequence $\left(\varepsilon_{n}(\cdot,\cdot),\{z_{n,a}(\cdot),q_{n,a}(\cdot)\}_{a}\right)$
is tight and therefore converges to weak limits along subsequences.
In what follows, assume we are on such a subsequence, represented
as $\{n\}$ for ease of notation; this is without loss of generality,
as long as we take the subsequence to be one along which the $\limsup$
in formulation of the theorem is attained. Now, append $\varepsilon_{n}(\cdot,\cdot)$
to the tuple in (\ref{eq:pf:Thm1:0.1}) and repeat the construction
in the proof of Theorem \ref{Thm: ART} verbatim. The stable convergence
arguments of Step 4 then deliver, on the extension $(\bar{\Omega},\bar{\mathcal{F}},\bar{\mathbb{P}})\equiv(\Omega\times[0,1],\mathcal{F}\otimes\mathcal{B}[0,1],\mathbb{P}\otimes\lambda)$,
an allocation process $\{q_{a}(\cdot)\}_{a}$ such that
\begin{equation}
\left(\varepsilon_{n}(\cdot,\cdot),\{x_{n,a}(\cdot),q_{n,a}(\cdot)\}_{a}\right)\xrightarrow[\mathbb{P}_{n,\bm{0}}]{d}\left(\bar{\varepsilon}(\cdot,\cdot),\{x_{a}(\cdot),q_{a}(\cdot)\}_{a}\right),\label{eq:pf:e-processes-0.5}
\end{equation}
where, importantly, the auxiliary randomization $V\sim\textrm{Uniform}[0,1]$
employed in that construction is drawn independently of the entire
base space, including $\bar{\varepsilon}(\cdot)$. 

Now define $\varepsilon(q_{1},q_{0})$ as (a version of) $\mathbb{E}[\bar{\varepsilon}(q_{1},q_{0})\vert\mathcal{G}_{q_{1},q_{0}}]$.
Note that, as $V$ is independent of the base space,
\[
\mathbb{E}\left[\bar{\varepsilon}(q_{1},q_{0})\vert\mathcal{G}_{q_{1},q_{0}}\lor\sigma(V)\right]=\mathbb{E}\left[\bar{\varepsilon}(q_{1},q_{0})\vert\mathcal{G}_{q_{1},q_{0}}\right],
\]
so folding $V$ into the enlarged randomization $\bar{U}$, as is
done at the end of Step 4 of the proof of Theorem \ref{Thm: ART},
leaves $\varepsilon(\cdot)$ unchanged. Moreover, by the Doob-{}-Dynkin
lemma, $\varepsilon(q_{1},q_{0})$ is a fixed measurable functional
of $(\{z_{1}(s):s\le q_{1}\},\{z_{0}(s):s\le q_{0}\},U)$ determined
by the law in Assumption 4, so the same field $\varepsilon(\cdot)$
serves every allocation process in $\bm{\mathcal{Q}}$, as required
by Definition \ref{Def:e_process_limit_experiment}; indeed, in our
construction above, $\{q_{a}(\cdot)\}_{a}$ is a deterministic function
of $\{z_{a}(\cdot)\}_{a},\bar{U}$ only. Finally, we also observe
that by similar stochastic analysis arguments as those employed to
prove Doob's optional sampling theorem, we can derive an equivalent
version for allocation processes:
\begin{equation}
\mathbb{E}\left[\bar{\varepsilon}(\cdot,\cdot)\vert\mathcal{G}_{q_{1}(t),q_{0}(t)}\right]=\varepsilon(q_{1}(t),q_{0}(t)).\label{eq:Them-e_processes-0.75}
\end{equation}

Recall the definition of $\hat{\varphi}(\bm{h};\cdot,\cdot)$ in Section
\ref{sec:Equivalence-of-experiments:}. By (\ref{eq:SLAN property}),
\begin{equation}
\hat{\varphi}(\bm{h};q_{1},q_{0})=\sum_{a}\left\{ h^{(a)\intercal}I_{a}^{1/2}z_{n,a}(q_{a})-\frac{q_{a}}{2}h^{(a)\intercal}I_{a}h^{(a)}\right\} +o_{\mathbb{P}_{n,0}}(1),\label{eq:Thm-e_processes:1}
\end{equation}
uniformly over all $(q_{1},q_{0})\in[0,1]^{2}$. Combined with Assumption
4, this gives
\begin{align}
\left(\begin{array}{c}
\varepsilon_{n}(\cdot,\cdot)\\
\hat{\varphi}(\bm{h};\cdot,\cdot)
\end{array}\right) & \xrightarrow[\mathbb{P}_{n,0}]{d}\left(\begin{array}{c}
\bar{\varepsilon}(\cdot,\cdot)\\
\ln V(\cdot,\cdot)
\end{array}\right);\ \textrm{where}\label{eq:pf:Thm3:weak convergence}\\
V(q_{1},q_{0})\sim & \exp\sum_{a}\left\{ h^{(a)\intercal}I_{a}^{1/2}z_{a}(q_{a})-\frac{q_{a}}{2}h^{(a)\intercal}I_{a}h^{(a)}\right\} .\nonumber 
\end{align}

We now claim that $\varepsilon(\cdot,\cdot)$ is a valid e-process
in the limit experiment. By (\ref{eq:pf:e-processes-0.5}), (\ref{eq:pf:Thm3:weak convergence})
and almost surely continuity of the sample paths (Assumption 4),
\[
\left(\begin{array}{c}
\varepsilon_{n}\left(q_{n,1}(t),q_{n,0}(t)\right)\\
\hat{\varphi}\left(\bm{h};q_{n,1}(t),q_{n,0}(t)\right)
\end{array}\right)\xrightarrow[\mathbb{P}_{n,0}]{d}\left(\begin{array}{c}
\bar{\varepsilon}\left(q_{1}(t),q_{0}(t)\right)\\
\ln V\left(q_{1}(t),q_{0}(t)\right)
\end{array}\right).
\]
As in the proof of Theorem \ref{Thm: Point estimation}, $E[V(q_{1}(t),q_{0}(t))]=1$.
Furthermore, by the same arguments as in that proof, we also have
\begin{align}
 & \varepsilon_{n}\left(q_{n,1}(t),q_{n,0}(t)\right)\xrightarrow[\mathbb{P}_{n,\bm{h}}]{d}\mathcal{L};\ \textrm{where, }\nonumber \\
 & \mathcal{L}(B):=E\left[\mathbb{I}\left\{ \bar{\varepsilon}\left(q_{1}(t),q_{0}(t)\right)\in B\right\} V\left(q_{1}(t),q_{0}(t)\right)\right]\ \forall\ B\in\mathcal{B}(\mathbb{R}).\label{eq:pf:Thm3:2}
\end{align}
 As $\varepsilon_{n}\left(q_{n,1}(t),q_{n,0}(t)\right)$ is a sequence
of non-negative random variables, the portmanteau lemma and (\ref{eq:pf:Thm3:2})
imply
\begin{align*}
 & \liminf_{n}\mathbb{E}_{n,\bm{h}}\left[\varepsilon_{n}\left(q_{n,1}(t),q_{n,0}(t)\right)\right]\\
 & \ge E\left[\bar{\varepsilon}\left(q_{1}(t),q_{0}(t)\right)\exp\left\{ \sum_{a}\left\{ h^{(a)\intercal}I_{a}^{1/2}z_{a}(q_{a}(t))-\frac{q_{a}(t)}{2}h^{(a)\intercal}I_{a}h^{(a)}\right\} \right\} \right]
\end{align*}
for each $\bm{h}$. Furthermore, by the law of iterated expectations
and (\ref{eq:Them-e_processes-0.75}),
\begin{align*}
 & E\left[\bar{\varepsilon}\left(q_{1}(t),q_{0}(t)\right)\exp\left\{ \sum_{a}\left\{ h^{(a)\intercal}I_{a}^{1/2}z_{a}(q_{a}(t))-\frac{q_{a}(t)}{2}h^{(a)\intercal}I_{a}h^{(a)}\right\} \right\} \right]\\
 & =E\left[\varepsilon(q_{1}(t),q_{0}(t))\exp\left\{ \sum_{a}\left\{ h^{(a)\intercal}I_{a}^{1/2}z_{a}(q_{a}(t))-\frac{q_{a}(t)}{2}h^{(a)\intercal}I_{a}h^{(a)}\right\} \right\} \right]\\
 & =\mathbb{E}_{\bm{h}}[\varepsilon(q_{1}(t),q_{0}(t))],
\end{align*}
where the last step follows by the Girsanov theorem as in the proof
of Theorem \ref{Thm: Point estimation}. But $\liminf_{n}\mathbb{E}_{n,\bm{h}}\left[\varepsilon_{n}(q_{n,1}(t),q_{n,0}(t))\right]\le1$
for any $\bm{h}\in\mathcal{H}_{0}$ and $t\in[0,1]$ by the definition
of an asymptotic e-process, so we conclude by the above argument that
for any allocation process $\{q_{a}(\cdot)\}_{a}\in\bm{\mathcal{Q}}$,
\[
\mathbb{E}_{\bm{h}}\left[\varepsilon\left(q_{1}(t),q_{0}(t)\right)\right]\le1\ \forall\ \bm{h}\in\mathcal{H}_{0},\ t\in[0,1].
\]
Since $\varepsilon(\cdot,\cdot)$ is $\mathcal{G}_{q_{1},q_{0}}$-adapted
by definition, the above implies that $\varepsilon\left(\cdot,\cdot\right)$
is a valid e-process in the limit experiment (in the sense of Definition
\ref{Def:e_process_limit_experiment}). 

Equation (\ref{eq:pf:Thm3:2}) and Assumption 4 also imply that for
each $\bm{h}\in\mathcal{H}_{1}$ and allocation processes $\{q_{n,a}(\cdot)\}_{a}$
converging to an allocation process $\{q_{a}(\cdot)\}_{a}$ in the
limit experiment,
\[
\lim_{n\to\infty}\mathbb{E}_{n,\bm{h}}\left[\ln\varepsilon_{n}\left(q_{n,1}(t),q_{n,0}(t)\right)\right]=E\left[V\left(q_{1}(t),q_{0}(t)\right)\ln\bar{\varepsilon}\left(q_{1}(t),q_{0}(t)\right)\right].
\]
But by (\ref{eq:Them-e_processes-0.75}), the conditional Jensen's
inequality, and the Girsanov theorem, 
\begin{align*}
E\left[V\left(q_{1}(t),q_{0}(t)\right)\ln\bar{\varepsilon}\left(q_{1}(t),q_{0}(t)\right)\right] & =E\left[V\left(q_{1}(t),q_{0}(t)\right)E\left[\left.\ln\bar{\varepsilon}\left(q_{1}(t),q_{0}(t)\right)\right|\mathcal{G}_{q_{1}(t),q_{0}(t)}\right]\right]\\
 & \le E\left[V\left(q_{1}(t),q_{0}(t)\right)\ln\varepsilon(q_{1}(t),q_{0}(t))\right]\\
 & =\mathbb{E}_{\bm{h}}\left[\ln\varepsilon\left(q_{1}(t),q_{0}(t)\right)\right].
\end{align*}
We thus conclude that 
\begin{align*}
\limsup_{n}R_{n}\left(\varepsilon_{n};\bm{h},\{q_{n,a}(t)\}\right) & =\lim_{n\to\infty}\mathbb{E}_{n,\bm{h}}\left[\ln\varepsilon_{n}\left(q_{n,1}(t),q_{n,0}(t)\right)\right]\\
 & \le\mathbb{E}_{\bm{h}}\left[\ln\varepsilon\left(q_{1}(t),q_{0}(t)\right)\right]=R\left(\varepsilon;\bm{h},\{q_{a}(t)\}\right),
\end{align*}
for all $\bm{h}\in\mathcal{H}_{1}$. This proves the desired claim. 

\subsection{Proof of Corollary \ref{Cor:mGRO}}

By the statement of Corollary \ref{Cor:mGRO}, $R_{n}\left(\varepsilon_{n};\bm{h},\{q_{n,a}(t)\}\right)$
is dominated by $g(\bm{h})$ for all $\bm{h},\{q_{n,a}(t)\}_{n}$
and $n$ sufficiently large. Therefore, 
\begin{align*}
 & \limsup_{n\to\infty}\int R_{n}\left(\varepsilon_{n};\bm{h},\{q_{n,a}(t)\}\right))dw(\bm{h})\\
 & =\int\limsup_{n\to\infty}R_{n}\left(\varepsilon_{n};\bm{h},\{q_{n,a}(t)\}\right)dw(\bm{h})\\
 & \le\int R\left(\varepsilon;\bm{h},\{q_{a}(t)\}_{a}\right)dw(\bm{h}),
\end{align*}
for some asymptotic e-process $\varepsilon(\cdot)$, where the equality
follows from the dominated convergence theorem, and the inequality
follows from Theorem \ref{Thm:e-processes}. 

\subsection{Proof of Corollary \ref{Cor:REGROW}}

Observe that
\begin{align*}
 & \limsup_{n}\mathcal{R}_{n}\left(\varepsilon_{n};\{q_{n,a}(t)\}_{a}\right)\\
 & \le\inf_{\bm{h}\in\mathcal{H}_{1}}\limsup_{n}\left\{ \mathbb{E}_{n,\bm{h}_{1}}\left[\ln\varepsilon_{n}(q_{n,1}(t),q_{n,0}(t))\right]-\mathbb{E}_{n,\bm{h}_{1}}\left[\ln\frac{d\mathbb{P}_{n,\hm{h}_{1}}}{d\mathbb{P}_{n,0}}(q_{n,1}(t),q_{n,0}(t))\right]\right\} \\
 & \le\inf_{\bm{h}\in\mathcal{H}_{1}}\left\{ \mathbb{E}_{\bm{h}_{1}}\left[\ln\varepsilon(q_{1}(t),q_{0}(t))\right]-\mathbb{E}_{\bm{h}_{1}}\left[\ln\frac{d\mathbb{P}_{\hm{h}_{1}}}{d\mathbb{P}_{0}}(q_{1}(t),q_{0}(t))\right]\right\} =\mathcal{R}\left(\varepsilon;\{q_{a}(t)\}_{a}\right),
\end{align*}
where the second inequality follows from Theorem \ref{Thm:e-processes}
and Assumption 5. 

\section{Auxiliary results for the proof of Theorem \ref{Thm: ART}\label{sec:Supporting-results}}

\begin{lem} \label{Lemma:Supporting_result_1} Consider the setup
in Step 1 of the proof of Theorem \ref{Thm: ART}. For each $m,L$,
there exists a collection of random variables $\left\{ q_{a,k}^{(m,L)}\right\} _{a,k}$
satisfying the conditions C1-C3 laid out in that step.\end{lem} 
\begin{proof}
The construction is inductive. We start by recording two consequences
of Condition C1 and the dyadic discretization that will be used repeatedly.
Since $q_{0,k}^{(m,L)}=t_{k}-q_{1,k}^{(m,L)}$ and $t_{k}=\eta_{k\bar{c}}$,
the process $q_{1,k}^{(m,L)}$ takes values in $\{\eta_{0},\dots,\eta_{k\bar{c}}\}$,
and for any $l+l^{\prime}\ge k\bar{c}$, 
\begin{equation}
\left\{ q_{1,k}^{(m,L)}\le\eta_{l},q_{0,k}^{(m,L)}\le\eta_{l^{\prime}}\right\} \equiv\bigcup_{i=(k\bar{c}-l^{\prime})\vee0}^{l}\left\{ q_{1,k}^{(m,L)}=\eta_{i}\right\} .\label{eq:sup_lemma_1}
\end{equation}
In particular, taking $l^{\prime}=k\bar{c}-l$ shows that this `tight
pair' event coincides with the point event 
\[
\left\{ q_{1,k}^{(m,L)}\le\eta_{l},q_{0,k}^{(m,L)}\le\eta_{k\bar{c}-l}\right\} \equiv\left\{ q_{1,k}^{(m,L)}=\eta_{l}\right\} .
\]
The same identities also hold for $\tilde{q}_{a,k}^{(m,L)}$. In view
of (\ref{eq:sup_lemma_1}), to verify Condition C2 it suffices to
show that each event $\left\{ q_{1,k}^{(m,L)}=\eta_{i}\right\} $
is 
\[
\sigma\left\{ U_{1},\dots,U_{k},\{s_{1,j}\}_{j\le i+\bar{c}+1},\{s_{0,j}\}_{j\le k\bar{c}-i+\bar{c}+1}\right\} 
\]
measurable: indeed, for every term in the right hand side of (\ref{eq:sup_lemma_1}),
we have $i\le l$ and $k\bar{c}-i\le l^{\prime}$, so every term and
the union lies within the $\sigma$-algebra prescribed by Condition
C2. 

\begin{spacing}{2}
\noindent\textbf{\textit{Initialization}}
\end{spacing}

For $k=1$, denote $\psi_{1,l}:=\mathbb{E}[\tilde{\phi}_{1,l}\vert\{s_{a,l}\}_{a,l}]$
and 
\[
\varphi_{1,l}:=\mathbb{E}[\tilde{\phi}_{1,l}\vert\{s_{1,j}\}_{j\le l},\{s_{0,j}\}_{j\le\bar{c}-l+1}].
\]
It is straightforward to show that the sets of random variables 
\[
\left\{ \tilde{\phi}_{1,l},\{s_{1,j}\}_{j\le l},\{s_{0,j}\}_{j\le\bar{c}-l+1}\right\} \ \textrm{and }\left\{ \{s_{1,j}\}_{j>l},\{s_{0,j}\}_{j>\bar{c}-l+1}\right\} 
\]
are independent of each other for any $l$. Indeed, this is a consequence
of (\ref{eq:pf:Thm1:0.1}) and the fact that in the actual experiment
\[
\left\{ \phi_{n,1,l},\{s_{n,1,j}\}_{j\le l},\{s_{n,0,j}\}_{j\le\bar{c}-l+1}\right\} \ \textrm{and }\left\{ \{s_{n,1,j}\}_{j>l},\{s_{n,0,j}\}_{j>\bar{c}-l+1}\right\} 
\]
are independent. Then, we have that (almost surely): (i) $\varphi_{1,l}=\psi_{1,l}$,
and (ii) $\sum_{j\le l}\varphi_{1,j}\le\sum_{j\le l^{\prime}}\varphi_{1,j}$
for all $l^{\prime}\ge l$ with $\sum_{j}\varphi_{1,j}=1$. Property
(i) follows from well known properties of regular conditional probabilities
(see the proof of Proposition 3 in \citealp{LeCam1979}). Property
(ii) follows from (\ref{eq:pf:Thm1:0.1}) and the definition of $\varphi_{1,l}$
after noting that $\sum_{j\le l}\phi_{n,1,j}\le\sum_{j\le l^{\prime}}\phi_{n,1,j}$
for all $l^{\prime}\ge l$ and $\sum_{j}\phi_{n,1,j}=1$.

Now, take $U_{1}\sim\textrm{Uniform}[0,1]$ to be exogenous to $z_{1}(\cdot),z_{0}(\cdot)$,
and define 
\begin{align*}
q_{1,1}^{(m,L)} & :=\sum_{l=1}^{\bar{c}}\eta_{l}\mathbb{I}\left\{ \sum_{j\le l-1}\varphi_{1,j}<U_{1}\le\sum_{j\le l}\varphi_{1,j}\right\} ,\\
q_{0,1}^{(m,L)} & :=t_{1}-q_{1,1}^{(m,L)}.
\end{align*}
In the construction above, we truncate the sum at $\bar{c}$ since
$t_{1}=\eta_{\bar{c}}$ and $\phi_{n,k,l}=0$ for $l>\bar{c}$ by
the definition of $q_{n,1}(\cdot)$, so $\tilde{\phi}_{1,l},\psi_{1,l}$
and $\varphi_{1,l}$ must also be $0$ almost surely for $l>\bar{c}$.
To verify Condition C2, consider the event $\left\{ q_{1,1}^{(m,L)}=\eta_{l}\right\} $.
It is determined by $U_{1}$ and $\{\varphi_{1,j}\}_{j\le l}$, with
each $\varphi_{1,j}$ for $j\le l$ being a measurable function of
$\{s_{1,j^{\prime}}\}_{j^{\prime}\le j}\subseteq\{s_{1,j^{\prime}}\}_{j^{\prime}\le l}$
and $\{s_{0,j^{\prime}}\}_{j^{\prime}\le\bar{c}-j+1}\subseteq\{s_{0,j^{\prime}}\}_{j^{\prime}\le\bar{c}+1}$.
Since $\bar{c}+1\le(\bar{c}-l)+\bar{c}+1$, the event $\left\{ q_{1,1}^{(m,L)}=\eta_{l}\right\} $
is 
\[
\sigma\left\{ U_{1},\{s_{1,j}\}_{j\le l+\bar{c}+1},\{s_{0,j}\}_{j\le\bar{c}-l+\bar{c}+1}\right\} 
\]
measurable, and condition C2 therefore follows from (\ref{eq:sup_lemma_1}).
Finally, the conditional law of $q_{1,1}^{(m,L)}$ given $\{s_{a,l}\}_{a,l}$
is (almost surely) equivalent to the conditional law of $\tilde{q}_{1,1}^{(m,L)}$
given $\{s_{a,l}\}_{a,l}$. This is because the conditional laws of
$q_{1,1}^{(m,L)},\tilde{q}_{1,1}^{(m,L)}$ are uniquely determined
by the sets of conditional probabilities $\psi_{1,1},\dots,\psi_{1,\bar{c}}$
and $\varphi_{1,1},\dots,\varphi_{1,\bar{c}}$ which are almost surely
equivalent to each other. Hence, the joint law of $\left(\{s_{a,l}\}_{a,l},q_{1,1}^{(m,L)}\right)$
is equivalent to that of $\left(\{s_{a,l}\}_{a,l},\tilde{q}_{1,1}^{(m,L)}\right)$.

\begin{spacing}{2}
\noindent\textbf{\textit{Induction}}
\end{spacing}

Now, start with the induction hypothesis that $\left\{ q_{a,k^{\prime}}^{(m,L)}\right\} _{k^{\prime}}$
have been constructed in a manner that satisfies conditions C1-C3
for all $k^{\prime}<k$. We show how to extend the construction to
$k^{\prime}=k$ so that conditions C1-C3 continue to be satisfied.

It is useful to note that if $\tilde{q}_{1,k}^{(m,L)}=\eta_{l}$,
then $\tilde{q}_{0,k}^{(m,L)}=t_{k}-\eta_{l}=\eta_{k\bar{c}-l}$.
Denote 
\begin{align*}
\bm{\Phi}_{n}^{(k,l)} & \equiv\left\{ \phi_{n,\tilde{k},\tilde{l}}:\tilde{l}\le l,\tilde{k}\bar{c}-\tilde{l}\le k\bar{c}-l,\tilde{k}\le k-1\right\} ,\\
\tilde{\bm{\Phi}}^{(k,l)} & \equiv\left\{ \tilde{\phi}_{\tilde{k},\tilde{l}}:\tilde{l}\le l,\tilde{k}\bar{c}-\tilde{l}\le k\bar{c}-l,\tilde{k}\le k-1\right\} .
\end{align*}
Intuitively, $\bm{\Phi}_{n}^{(k,l)}$ represents the collection of
$\phi_{n,\tilde{k},\tilde{l}}$ variables that require less `information'
to determine than $\phi_{n,k,l}$. Define the $\sigma$-algebras 
\begin{align*}
\mathcal{\tilde{L}}_{k,l} & \equiv\sigma\left\{ \{s_{1,j}\}_{j},\{s_{0,j}\}_{j},\left\{ \tilde{q}_{a,\tilde{k}}^{(m,L)}:\tilde{k}\le k-1\right\} _{a}\right\} ,\\
\mathcal{\tilde{M}}_{k,l} & \equiv\sigma\left\{ \{s_{1,j}\}_{j},\{s_{0,j}\}_{j},\tilde{\bm{\Phi}}^{(k,l)}\right\} ,\\
\mathcal{\tilde{H}}_{k,l} & \equiv\sigma\left\{ \{s_{1,j}\}_{j\le l},\{s_{0,j}\}_{j\le k\bar{c}-l+1},\tilde{\bm{\Phi}}^{(k,l)}\right\} ,
\end{align*}
and let $B_{n,k,l},\tilde{B}_{k,l},B_{k,l}$ denote the events 
\begin{align*}
B_{n,k,l} & \equiv\left\{ q_{n,1,\tilde{k}}^{(m,L)}\le\eta_{l},q_{n,0,\tilde{k}}^{(m,L)}\le\eta_{k\bar{c}-l},\ \forall\ \tilde{k}\le k-1\right\} ,\\
\tilde{B}_{k,l} & \equiv\left\{ \tilde{q}_{1,\tilde{k}}^{(m,L)}\le\eta_{l},\tilde{q}_{0,\tilde{k}}^{(m,L)}\le\eta_{k\bar{c}-l},\ \forall\ \tilde{k}\le k-1\right\} ,\\
B_{k,l} & \equiv\left\{ q_{1,\tilde{k}}^{(m,L)}\le\eta_{l},q_{0,\tilde{k}}^{(m,L)}\le\eta_{k\bar{c}-l},\ \forall\ \tilde{k}\le k-1\right\} .
\end{align*}
Also, take $\mathcal{\tilde{L}}_{k,l}^{+}$ and $\mathcal{\tilde{L}}_{k,l}^{-}$
to be the $\sigma$-algebras corresponding to the restriction of $\tilde{\mathcal{L}}_{k,l}$
to $\tilde{B}_{k,l}$ and $\tilde{B}_{k,l}^{c}$, respectively. The
quantities $\mathcal{\tilde{M}}_{k,l}^{+},\mathcal{\tilde{M}}_{k,l}^{-}$
and $\mathcal{\tilde{H}}_{k,l}^{+},\mathcal{\tilde{H}}_{k,l}^{-}$
are defined analogously, as the restrictions of $\tilde{\mathcal{M}}_{k,l}$
and $\tilde{\mathcal{H}}_{k,l}$ to $\tilde{B}_{k,l}$, $\tilde{B}_{k,l}^{c}$.
Observe that $\mathcal{\tilde{L}}_{k,l}^{+}\equiv\tilde{\mathcal{M}}_{k,l}^{+}$.
This is because, when $\tilde{B}_{k,l}$ holds, all the $\left\{ \tilde{\phi}_{\tilde{k},\tilde{l}}\right\} _{\tilde{k}\le k-1}$
random variables outside the collection $\tilde{\bm{\Phi}}^{(k,l)}$
necessarily take on the value $0$, so they do not provide any additional
information. 

Define $\vartheta_{k,l}:=\mathbb{E}\left[\tilde{\phi}_{k,l}\vert\mathcal{\tilde{L}}_{k,l}\right]$,
$\psi_{k,l}:=\mathbb{E}\left[\tilde{\phi}_{k,l}\vert\tilde{\mathcal{M}}_{k,l}^{+}\right]$
and observe that
\begin{align*}
\vartheta_{k,l} & =\mathbb{I}\{\tilde{B}_{k,l}\}\mathbb{E}\left[\tilde{\phi}_{k,l}\vert\mathcal{\tilde{L}}_{k,l}^{+}\right]+\mathbb{I}\{\tilde{B}_{k,l}^{c}\}\mathbb{E}\left[\tilde{\phi}_{k,l}\vert\mathcal{\tilde{L}}_{k,l}^{-}\right]\ \textrm{a.s.}\\
 & =\mathbb{I}\{\tilde{B}_{k,l}\}\mathbb{E}\left[\tilde{\phi}_{k,l}\vert\mathcal{\tilde{M}}_{k,l}^{+}\right]+\mathbb{I}\{\tilde{B}_{k,l}^{c}\}\mathbb{E}\left[\tilde{\phi}_{k,l}\vert\mathcal{\tilde{L}}_{k,l}^{-}\right]\ \textrm{a.s.}\\
 & =\mathbb{I}\{\tilde{B}_{k,l}\}\psi_{k,l}\ \textrm{a.s.}
\end{align*}
The last step uses the fact $\tilde{q}_{a,k}^{(m,L)}$ is almost surely
non-decreasing in $k$ as it is the weak limit of $q_{n,a,k}^{(m,L)}$,
which is non-decreasing in $k$. Hence, conditional on $\tilde{B}_{k,l}^{c}$,
we have $\tilde{\phi}_{k,l}=0\ \textrm{a.s}$, implying $\mathbb{E}\left[\tilde{\phi}_{k,l}\vert\mathcal{\tilde{L}}_{k,l}^{-}\right]=0\ \textrm{a.s}$.

Set $\varphi_{k,l}:=\mathbb{I}\{\tilde{B}_{k,l}\}\mathbb{E}\left[\tilde{\phi}_{k,l}\left|\mathcal{\tilde{H}}_{k,l}^{+}\right.\right]$.
Now, in the actual experiment, the collection of random variables
\begin{align*}
 & \left\{ \phi_{n,k,l},\bm{\Phi}_{n}^{(k,l)},\{s_{n,1,j}\}_{j\le l},\{s_{n,0,j}\}_{j\le k\bar{c}-l+1},\mathbb{I}\{B_{n,k,l}\}\right\} \\
 & \textrm{and }\left\{ \{s_{n,1,j}\}_{j>l},\{s_{n,0,j}\}_{j>k\bar{c}-l+1}\right\} 
\end{align*}
are independent of each other. Combined with (\ref{eq:pf:Thm1:0.1})
and the properties of weak convergence, we conclude 
\begin{align*}
 & \left\{ \tilde{\phi}_{k,l},\tilde{\bm{\Phi}}^{(k,l)},\{s_{1,j}\}_{j\le l},\{s_{0,j}\}_{j\le k\bar{c}-l+1},\mathbb{I}\{\tilde{B}_{k,l}\}\right\} \\
 & \textrm{and }\left\{ \{s_{1,j}\}_{j>l},\{s_{0,j}\}_{j>k\bar{c}-l+1}\right\} 
\end{align*}
are also independent of each other for any $l$. Hence, by similar
arguments as in the initialization step, it follows that $\mathbb{E}\left[\tilde{\phi}_{k,l}\left|\mathcal{\tilde{H}}_{k,l}^{+}\right.\right]=\psi_{k,l}$
almost surely, and therefore: (i) $\varphi_{k,l}=\vartheta_{k,l}$,
and (ii) $\sum_{l}\varphi_{k,l}=\sum_{l}\vartheta_{k,l}=1$ since
$\sum_{l}\tilde{\phi}_{k,l}=1$, both almost surely. The Doob-Dynkin
theorem states that we may take $\varphi_{k,l}$ to be a measurable
function of 
\[
\{s_{1,j}\}_{j\le l},\{s_{0,j}\}_{j\le k\bar{c}-l+1},\tilde{\bm{\Phi}}^{(k,l)},\mathbb{I}\{\tilde{B}_{k,l}\}
\]
the random variables generating $\mathcal{\tilde{H}}_{k,l}^ {}$,
together with $\mathbb{I}\{\tilde{B}_{k,l}\}$. Denote this function
by $\varphi_{k,l}(\cdot)$, and note that by construction, it vanishes
when its last argument is $0$.

We define $q_{a,k}^{(m,L)}$ on a new probability space (i.e., separate
from the space in which $\tilde{q}_{a,l}^{(m,L)}$ reside) containing
$\left\{ z_{1}(\cdot),z_{0}(\cdot),U_{1},\dots,U_{2^{m}}\right\} $,
where $U_{1},\dots,U_{2^{m}}$ are iid $\textrm{Uniform}[0,1]$ and
independent of $z_{1}(\cdot),z_{0}(\cdot)$. Formally, given some
values for $\left\{ q_{1,\tilde{k}}^{(m,L)}\right\} _{\tilde{k}\le k-1}$---themselves
functions of $z_{1}(\cdot),z_{0}(\cdot),U_{1},\dots,U_{k-1}$ by the
induction hypothesis---we set:
\begin{align}
q_{1,k}^{(m,L)} & =\sum_{l=0}^{k\bar{c}}\eta_{l}\cdot\mathbb{I}\left\{ \sum_{j\le l-1}\varphi_{k,j}(\cdot)<U_{k}\le\sum_{j\le l}\varphi_{k,j}(\cdot)\right\} ,\nonumber \\
q_{0,k}^{(m,L)} & =t_{k}-q_{1,k}^{(m,L)}.\label{eq:sup_lem_0.4}
\end{align}
In the equation above, the functions $\left\{ \varphi_{k,l}(\cdot)\right\} _{l}$
now take as inputs the corresponding quantities on the new probability
space, i.e., $\varphi_{k,l}(\cdot)$ is applied on
\begin{equation}
\{s_{1,j}\}_{j\le l},\{s_{0,j}\}_{j\le k\bar{c}-l+1},\bm{\Phi}^{(k,l)},\mathbb{I}\{B_{k,l}\}\label{eq:sup_lem_0.5}
\end{equation}
where 
\begin{align}
\bm{\Phi}^{(k,l)} & :=\left\{ \phi_{\tilde{k},\tilde{l}}:\tilde{l}\le l,\tilde{k}\bar{c}-\tilde{l}\le k\bar{c}-l,\tilde{k}\le k-1\right\} ,\ \textrm{and}\nonumber \\
\phi_{\tilde{k},\tilde{l}} & :=\mathbb{I}\left\{ q_{1,\tilde{k}}^{(m,L)}=\eta_{\tilde{l}}\right\} .\label{eq:sup_lem_0.75}
\end{align}
Note that $\sum_{l\le k\bar{c}}\varphi_{k,l}(\cdot)$ is a fixed measurable
function of these inputs which equals $1$ almost surely under the
law of $\left\{ \{s_{a,l}\}_{a,l},\{\tilde{q}_{1,\tilde{k}}^{(m,L)}\}_{\tilde{k}\le k-1}\right\} $;
then, by the induction hypothesis (Condition C3) the inputs on the
new space have the same law, so $\sum_{l\le k\bar{c}}\varphi_{k,l}(\cdot)=1$
almost surely in the new space as well, and the construction (\ref{eq:sup_lem_0.4})
almost surely selects exactly one value of $\eta_{l}$. On the residual
null event, we set $q_{1,k}^{(m,L)}:=q_{1,k-1}^{(m,L)}$; but, as
all the $\sigma$-algebras involved are augmented, this convention
affects none of the claims below. 

We now verify that $q_{a,k}^{(m,L)}$, so defined, satisfies conditions
C1-C3.

Condition C1 is satisfied by construction since $q_{1,k}^{(m,L)}+q_{0,k}^{(m,L)}=t_{k}$.

Next, we verify Condition C2. To this end, we employ a decomposition
over past trajectories. For $\bm{r}_{k}=(r_{1},\dots,r_{k-1})$ with
$r_{0}=0$ and $r_{\tilde{k}}-r_{\tilde{k}-1}\in\{0,1,\dots,\bar{c}\}$
for each $\tilde{k}$, define 
\[
A_{\bm{r}_{k}}:=\bigcap_{\tilde{k}\le k-1}\left\{ q_{1,\tilde{k}}^{(m,L)}=\eta_{r_{\tilde{k}}}\right\} 
\]
as the representation of a potential past trajectory for $\{q_{1,\tilde{k}}^{(m,L)}\}_{\tilde{k}\le k-1}$
before $t_{k}$. Since both $q_{1,\tilde{k}}^{(m,L)}$ and $q_{0,\tilde{k}}^{(m,L)}$
are non-decreasing in $\tilde{k}$ (as verified from (\ref{eq:sup_lem_0.4})
and the vanishing of $\varphi_{\tilde{k},j}(\cdot)$ when $\mathbb{I}\{B_{\tilde{k},j}\}=0$
for any $\tilde{k}\le k$), the events $A_{\bm{r}_{k}}$ partition
the sample space up to a null set. On $A_{\bm{r}_{k}}$, each $\phi_{\tilde{k},\tilde{l}}=\mathbb{I}\{r_{\tilde{k}}=\tilde{l}\}\in\bm{\Phi}^{(k,l)}$
and $\mathbb{I}\{B_{k,l}\}=\mathbb{I}\{r_{k-1}\le l\le r_{k-1}+\bar{c}\}$
are constants; consequently, on this event, $\varphi_{k,l}(\cdot)$
reduces to a measurable function---say $\varphi_{k,l}^{\bm{r}_{k}}(\cdot)$---of
$\{s_{1,j}\}_{j\le l},\{s_{0,j}\}_{j\le k\bar{c}-l+1}$ alone, with
$\varphi_{k,l}^{\bm{r}_{k}}(\cdot)=0$ for $l\notin[r_{k-1},r_{k-1}+\bar{c}]$.
Say that a trajectory $\bm{r}_{k}$ is `plausible' for $l$ at $t_{k}$
if $l-\bar{c}\le r_{k-1}\le l$. Then, from (\ref{eq:sup_lem_0.4}),
\begin{equation}
\left\{ q_{1,k}^{(m,L)}=\eta_{l}\right\} =\bigcup_{\bm{r}_{k}\textrm{ plausible for }l}A_{\bm{r}_{k}}\cap\left\{ \sum_{j=r_{k-1}}^{l-1}\varphi_{k,j}^{\bm{r}_{k}}(\cdot)<U_{k}\le\sum_{j=r_{k-1}}^{l}\varphi_{k,j}^{\bm{r}_{k}}(\cdot)\right\} ,\label{eq:sup_lemma_1.5}
\end{equation}
with the sums starting at $r_{k-1}$ since all lower terms vanish
on $A_{\bm{r}_{k}}$. 

Fix $l$ and a plausible $\bm{r}_{k}$, and consider the measurability
of the right hand side of (\ref{eq:sup_lemma_1.5}). First, by condition
C1, each event $\left\{ q_{1,\tilde{k}}^{(m,L)}=\eta_{r_{\tilde{k}}}\right\} $
coincides with $\left\{ q_{1,\tilde{k}}^{(m,L)}\le\eta_{r_{\tilde{k}}},q_{0,\tilde{k}}^{(m,L)}\le\eta_{\tilde{k}\bar{c}-r_{\tilde{k}}}\right\} $,
so the induction hypothesis (Condition C2) implies it is 
\[
\sigma\left\{ U_{1},\dots,U_{\tilde{k}},\{s_{1,j^{\prime}}\}_{j^{\prime}\le r_{\tilde{k}}+\bar{c}+1},\{s_{0,j^{\prime}}\}_{j^{\prime}\le k\bar{c}-r_{\tilde{k}}+\bar{c}+1}\right\} 
\]
measurable. Now, $r_{\tilde{k}}\le r_{k-1}\le l$, and, as $\tilde{k}\bar{c}-r_{\tilde{k}}$
is non-decreasing in $\tilde{k}$ along any plausible trajectory,
$\tilde{k}\bar{c}-r_{\tilde{k}}\le(k-1)\bar{c}-r_{k-1}\le k\bar{c}-l$
using $r_{k-1}\ge l-\bar{c}$. Hence, $A_{\bm{r}_{k}}$ is measurable
with respect to 
\[
\sigma\left\{ U_{1},\dots,U_{k-1},\{s_{1,j^{\prime}}\}_{j\le l+\bar{c}+1},\{s_{0,j^{\prime}}\}_{j^{\prime}\le\tilde{k}\bar{c}-l+\bar{c}+1}\right\} .
\]
Second, for $j\in[r_{k-1},l]$, the function $\varphi_{k,j}^{\bm{r}_{k}}(\cdot)$
is $\sigma\left\{ \{s_{1,j^{\prime}}\}_{j^{\prime}\le j},\{s_{0,j^{\prime}}\}_{j^{\prime}\le k\bar{c}-l+1}\right\} $
measurable as noted earlier, where $j\le l$ and $k\bar{c}-j+1\le k\bar{c}-r_{k-1}+1\le k\bar{c}-l+\bar{c}+1$.
Combining, we conclude that every term of (\ref{eq:sup_lemma_1.5})
is 
\[
\sigma\left\{ U_{1},\dots,U_{k},\{s_{1,j^{\prime}}\}_{j^{\prime}\le l+\bar{c}+1},\{s_{0,j^{\prime}}\}_{j^{\prime}\le k\bar{c}-l+\bar{c}+1}\right\} 
\]
measurable. In view of (\ref{eq:sup_lemma_1}), this completes the
verification of Condition C2. 

It remains to verify Condition C3. Note that $U_{k}$ is independent
of $\{s_{a,l}\}_{a,l}$ and $\{q_{a,\tilde{k}}^{(m,L)}\}_{\tilde{k}\le k-1}$.
Hence, for each $l$ and plausible $\bm{r}_{k}$, the construction
of $q_{1,k}^{(m,L)}$ implies 
\[
\mathbb{P}\left(\left.q_{1,k}^{(m,L)}=\eta_{l}\right|\{s_{a,j}\}_{a,j},A_{\bm{r}_{k}}\right)=\varphi_{k,l}^{\bm{r}_{k}}(\cdot),
\]
which, as a function of $\{s_{a,j}\}_{a,j}$, is the same (a.s.) as
the conditional probability of $\left\{ \tilde{q}_{1,k}^{(m,L)}=\eta_{l}\right\} $
given $\{s_{a,j}\}_{a,j}$ and $\tilde{A}_{\bm{r}_{k}}:=\bigcap_{\tilde{k}\le k-1}\left\{ \tilde{q}_{1,\tilde{k}}^{(m,L)}=\eta_{r_{\tilde{k}}}\right\} $;
indeed, the latter is $\vartheta_{k,l}$ evaluated on $\tilde{A}_{\bm{r}_{k}}$,
and $\varphi_{k,l}=\vartheta_{k,l}$ almost surely.\footnote{In essence, this is a statement about almost sure equivalence of Markov
kernels: we are asserting that if the conditioning trajectories $\left\{ q_{a,k^{\prime}}^{(m,L)}:k^{\prime}\le k-1\right\} _{a}$
and $\left\{ \tilde{q}_{a,k^{\prime}}^{(m,L)}:k^{\prime}\le k-1\right\} _{a}$
have the same value, then the conditional distributions of $q_{1,k}^{(m,L)}=\eta_{l}$
and $\tilde{q}_{1,k}^{(m,L)}=\eta_{l}$ would also be the same. } At the same time, $\left\{ q_{1,k}^{(m,L)}=\eta_{l}\right\} $ never
occurs on $A_{\bm{r}_{k}}$ for non-plausible $\bm{r}_{k}$, matching
the fact that the conditional probability of $\left\{ \tilde{q}_{1,k}^{(m,L)}=\eta_{l}\right\} $
on $\tilde{A}_{\bm{r}_{k}}$ for non-plausible $\bm{r}_{k}$ is 0.
So, overall, we conclude that the conditional law of $q_{1,k}^{(m,L)}$
given $\{s_{a,l}\}_{a,l}$ and $\left\{ q_{a,k^{\prime}}^{(m,L)}:k^{\prime}\le k-1\right\} _{a}$
is the same (a.s.) as the conditional law of $\tilde{q}_{1,k}^{(m,L)}$
given $\{s_{a,l}\}_{a,l}$ and $\left\{ \tilde{q}_{a,k^{\prime}}^{(m,L)}:k^{\prime}\le k-1\right\} _{a}$.
Combined with the induction hypothesis, it then follows that the joint
law of $\left(\{s_{a,l}\}_{a,l},q_{1,1}^{(m,L)},\dots,q_{1,k}^{(m,L)}\right)$
is equivalent to that of $\left(\{s_{a,l}\}_{a,l},\tilde{q}_{1,1}^{(m,L)},\dots,\tilde{q}_{1,k}^{(m,L)}\right)$.
\end{proof}

\section{Equivalence of out-of-sample regret\label{sec:Representations-for-out-of-sample}}

In contrast to in-sample regret, out-of-sample (or simple) regret
measures the expected difference between the welfare from a chosen
action and that of the optimal action, evaluated on new, unseen data.
Specifically, suppose that at the end of the adaptive experiment,
the DM is tasked with specifying a treatment decision $\bm{\delta}_{n}\equiv(\delta_{n,1},\delta_{n,0})\in\mathcal{S}^{2}$
to be applied on the entire population. Here $\mathcal{S}^{2}$ denotes
the $2$-dimensional simplex. The out-of-sample frequentist regret
of this decision is then defined as 
\[
W_{n}^{o}(\bm{h})=\sqrt{n}\left\{ \max_{a}\mu_{n,a}(\bm{h})-\sum_{a}\mu_{n,a}(\bm{h})\mathbb{E}_{n,\bm{h}}[\delta_{n,a}]\right\} .
\]
Note that $\bm{\delta}_{n}$ must be $\mathcal{I}_{n,1}:=\mathcal{G}_{n,q_{n,1}(1),q_{n,0}(1)}$
measurable. 

Analogously, in the limit experiment, the out-of-sample frequentist
regret is defined as 
\[
W^{o}(\bm{h})=\max_{a}\dot{\mu}_{a}^{\intercal}\bm{h}-\sum_{a}\dot{\mu}_{a}^{\intercal}\bm{h}\mathbb{E}_{\bm{h}}[\delta_{a}],
\]
where $\bm{\delta}\equiv(\delta_{1},\delta_{0})\in\mathcal{S}^{2}$
is $\mathcal{I}_{1}$-measurable, and $\dot{\mu}_{a}(\cdot)$ is defined
in Section \ref{sec:Application-2:-Equivalence}. We then have the
following analogue to Theorem \ref{Thm: Welfare}. 

\begin{thm} \label{Thm:Simple-regret}Suppose Assumptions 1, 2 and
4 hold. Let $\{x_{a}(\cdot),q_{a}(\cdot)\}_{a}$ denote the weak limit
of $\{x_{n,a}(\cdot),q_{n,a}(\cdot)\}_{a}$ under a sequence of policy
rules $\{\pi_{n,j}\}_{j}$ in the actual experiment, and let $\{\bm{\delta}_{n}\}_{n}$
be a sequence of treatment decisions with associated out-of-sample
regret $W_{n}^{o}(\bm{h})$. Then, there exists a subsequence $\{n_{k}\}_{k}$
and a limiting treatment decision $\bm{\delta}^ {}$ such that:\\
(i) $\bm{\delta}$ depends only on the terminal states $\{x_{a}(1),q_{a}(1)\}_{a}$;
and\\
(ii) For each $\bm{h}$, $\lim_{k\to\infty}W_{n_{k}}^{o}(\bm{h})=W^{o}(\bm{h})$,
where $W^{o}(\bm{h})$ is the out-of-sample regret associated with
$\bm{\delta}$ in the limit experiment.\end{thm} 
\begin{proof}
As $\{\bm{\delta}_{n}\}_{n}$ is uniformly bounded, it is tight. Since
$\left\{ x_{n,a}(1),q_{n,a}(1)\right\} _{a}$ converge in distribution
due to Theorem \ref{Thm: ART} and Assumption 2, it follows that the
joint $\left(\bm{\delta}_{n},\left\{ x_{n,a}(1),q_{n,a}(1)\right\} _{a}\right)$
is also tight. Hence, by Prohorov's theorem, given any sequence $\{n\}$,
there exists a further sub-sequence $\{n_{k}\}$---represented as
$\{n\}$ for ease of notation---such that 
\[
\left(\bm{\delta}_{n},\left\{ x_{n,a}(1),q_{n,a}(1)\right\} _{a}\right)\xrightarrow[\mathbb{P}_{n,0}]{d}\left(\bar{\bm{\delta}},\left\{ x_{a}(1),q_{a}(1)\right\} _{a}\right),
\]
where $\bar{\bm{\delta}}\in[0,1]$ is some tight limit of $\bm{\delta}_{n}$.
Combined with (\ref{eq:SLAN property}), this implies
\begin{align}
\left(\begin{array}{c}
\bm{\delta}_{n}\\
\hat{\varphi}(\bm{h};q_{n,1}(1),q_{n,0}(1))
\end{array}\right) & \xrightarrow[\mathbb{P}_{n,0}]{d}\left(\begin{array}{c}
\bar{\bm{\delta}}\\
\ln V
\end{array}\right);\ \textrm{where}\label{eq:pf:Thm-simple-regert:weak convergence-2}\\
V\sim & \exp\sum_{a}\left\{ h^{(a)\intercal}I_{a}^{1/2}x_{a}(1)-\frac{q_{a}(1)}{2}h^{(a)\intercal}I_{a}h^{(a)}\right\} .\nonumber 
\end{align}
Therefore, by similar arguments as in the proof of Theorem \ref{Thm: Point estimation},
\begin{equation}
\bm{\delta}_{n}\xrightarrow[\mathbb{P}_{n,\bm{h}}]{d}\mathcal{L};\ \textrm{where }\mathcal{L}(B):=E\left[\mathbb{I}\{\bar{\bm{\delta}}\in B\}V\right]\ \forall\ B\in\mathcal{B}(\mathbb{R}^{2}).\label{eq:pf:Thm-simple-regret-weak convergence 2}
\end{equation}
Define $\bm{\delta}=E\left[\bar{\bm{\delta}}|\{x_{a}(1),q_{a}(1)\}_{a}\right]$.
By construction, $\bm{\delta}$ is a valid treatment policy in the
limit experiment, and it is also $\mathcal{I}_{1}$-measurable. Furthermore,
by (\ref{eq:pf:Thm-simple-regret-weak convergence 2}), 
\begin{align*}
\lim_{n\to\infty}\mathbb{E}_{n,\bm{h}}[\delta_{n,a}] & =E\left[\bar{\delta}_{a}e^{\sum_{a}\left\{ h^{(a)\intercal}I_{a}^{1/2}x_{a}(1)-\frac{q_{a}(1)}{2}h^{(a)\intercal}I_{a}h^{(a)}\right\} }\right]\\
 & =E\left[\delta_{a}e^{\sum_{a}\left\{ h^{(a)\intercal}I_{a}^{1/2}x_{a}(1)-\frac{q_{a}(1)}{2}h^{(a)\intercal}I_{a}h^{(a)}\right\} }\right]=\mathbb{E}_{\bm{h}}[\delta_{a}]\ \forall\ a,
\end{align*}
where the second equality follows by the law of iterated expectations,
and the last equality follows by the Girsanov theorem. 

We have thereby shown that $\mathbb{E}_{n,\bm{h}}[\delta_{n,a}]\to\mathbb{E}_{\bm{h}}[\delta_{a}]$
for each $\bm{h},a$. Combined with Assumption 4, this implies $\lim_{n\to\infty}W_{n}^{o}(\bm{h})=W^{o}(\bm{h})$
for each $\bm{h}$, which proves the desired claim.
\end{proof}
Theorem \ref{Thm:Simple-regret} is stronger than Theorem \ref{Thm: Welfare}
in that it reduces the set of sufficient statistics for treatment
assignment to $\{x_{a}(1),q_{a}(1)\}_{a}$, i.e., only the terminal
values of the score and allocation processes are relevant. 

\section{Additional results on anytime-valid inference\label{sec:Additional-results-on-anytime-inference}}

\subsection{Verifying the conditions of Corollary \ref{Cor:mGRO} for $\varepsilon_{n}^{*}(\cdot)$
from (\ref{eq:asymptotic_mGRO})\label{subsec:Verification-of-the-conditions}}

We show that the asymptotically mGO optimal e-process $\varepsilon_{n}^{*}(\cdot)$
satisfies the requirements for Corollary \ref{Cor:mGRO} under the
following primitive conditions: 

\begin{asm6} (i) There exists $p>0$ independent of $n,\bm{h}$ such
that $\mathbb{E}_{n,\bm{h}}\left[\left|\psi(Y_{i}^{(a)})\right|^{2+p}\right]<\infty$
for each $a,\bm{h}$.\\
(ii) For each $a$, $\sqrt{n}\mathbb{E}_{n,\bm{h}}\left[\psi(Y_{i}^{(a)})\right]=I_{a}h^{(a)}+\delta_{n}\vert h^{(a)}\vert$,
where $\delta_{n}\to0$ is independent of $\bm{h}$.\\
(iii) The weighting function $w(\cdot)$ satisfies $\int e^{c\bm{h}^{\intercal}\bm{h}}dw(\bm{h})\le M<\infty$
for some $c>0$.\end{asm6}

Assumption 6(i) is a mild regularity condition. Assumption 6(ii) follows
from quadratic mean differentiability (Assumption 1). Assumption 6(iii)
requires the weighting function to have sub-Gaussian tails. This is
natural since $\varepsilon_{n}^{*}(\cdot)$ is based on integrating
an exponential term with respect to $w(\bm{h})$. 

We verify the various requirements for Corollary \ref{Cor:mGRO} below:

\subsubsection*{Weak convergence}

Based on the form of $\varepsilon_{n}^{*}(\cdot)$, Theorem \ref{Thm: ART}
and standard weak convergence arguments imply $\{z_{n,a}(\cdot),\varepsilon_{n}(\cdot,\cdot)\}_{a}$
converges weakly under $\mathbb{P}_{n,\bm{0}}$. This verifies the
first part of Assumption 4. 

\subsubsection*{Uniform Integrability}

We now show that $\sup_{q_{1},q_{0}}\ln\varepsilon_{n}^{*}(q_{1},q_{0})$
is uniformly integrable under $\mathbb{P}_{n,\bm{h}}$, which verifies
the second part of Assumption 4. 

Denote $\tilde{\bm{z}}_{n}(\bm{q})=[z_{n,1}(q_{1})^{\intercal}I_{1}^{1/2},z_{n,0}(q_{0})^{\intercal}I_{0}^{1/2}]^{\intercal}$,
$\bm{I}=\textrm{diag}(I_{1},I_{0})$ and $\bm{I}_{\bm{q}}=\textrm{diag}(q_{1}I_{1},q_{0}I_{0})$.\footnote{Here, $\textrm{diag}(A,B)$ denotes the block diagonal matrix with
diagonal (matrix) elements $A,B$.} Then, 
\begin{align*}
\varepsilon_{n}^{*}(q_{1},q_{0}) & =\int e^{\bm{h}^{\intercal}\tilde{\bm{z}}_{n}(\bm{q})-\frac{1}{2}\bm{h}^{\intercal}\bm{I}_{\bm{q}}\bm{h}}dw(\bm{h})\\
 & =\int e^{\bm{h}^{\intercal}\tilde{\bm{z}}_{n}(\bm{q})-\frac{1}{2}\bm{h}^{\intercal}(\bm{I}_{\bm{q}}+2c\cdot\textrm{Id})\bm{h}}\cdot e^{c\bm{h}^{\intercal}\bm{h}}dw(\bm{h}),
\end{align*}
where $\textrm{Id}$ denotes the identity matrix. Define $\bm{\Lambda}=\bm{I}_{\bm{q}}+2c\cdot\textrm{Id}$,
and note that by completing the square,
\begin{align*}
\varepsilon_{n}^{*}(q_{1},q_{0}) & =e^{\frac{1}{2}\tilde{\bm{z}}_{n}(\bm{q})^{\intercal}\bm{\Lambda}^{-1}\tilde{\bm{z}}_{n}(\bm{q})}\int e^{-\frac{1}{2}\left|\bm{\Lambda}^{1/2}\bm{h}-\bm{\Lambda}^{-1/2}\tilde{\bm{z}}_{n}(\bm{q})\right|^{2}}e^{c\bm{h}^{\intercal}\bm{h}}dw(\bm{h})\\
 & \le e^{\frac{1}{2}\tilde{\bm{z}}_{n}(\bm{q})^{\intercal}\bm{\Lambda}^{-1}\tilde{\bm{z}}_{n}(\bm{q})}\int e^{c\bm{h}^{\intercal}\bm{h}}dw(\bm{h})\\
 & \le Me^{\frac{1}{2}\tilde{\bm{z}}_{n}(\bm{q})^{\intercal}\bm{\Lambda}^{-1}\tilde{\bm{z}}_{n}(\bm{q})}\le Me^{\frac{1}{4c}\tilde{\bm{z}}_{n}(\bm{q})^{\intercal}\tilde{\bm{z}}_{n}(\bm{q})}
\end{align*}
where the third step follows by Assumption 6(iii), and the last step
makes use of the fact that $\bm{I}_{\bm{q}}$ is positive semi-definite
for any $q_{1},q_{0}$. 

Hence, $\sup_{q_{1},q_{0}}\ln\varepsilon_{n}^{*}(q_{1},q_{0})$ is
uniformly integrable under $\mathbb{P}_{n,\bm{h}}$ as long as 
\[
\sup_{q_{1},q_{0}}\tilde{\bm{z}}_{n}(\bm{q})^{\intercal}\tilde{\bm{z}}_{n}(\bm{q})=\sup_{q_{1}}\left|\frac{1}{\sqrt{n}}\sum_{i=1}^{\left\lfloor nq_{1}\right\rfloor }\psi(Y_{i}^{(1)})\right|^{2}+\sup_{q_{0}}\left|\frac{1}{\sqrt{n}}\sum_{i=1}^{\left\lfloor nq_{0}\right\rfloor }\psi(Y_{i}^{(0)})\right|^{2}
\]
is uniformly integrable under $\mathbb{P}_{n,\bm{h}}$. It therefore
suffices to show that each term, $\sup_{q_{a}}\left|n^{-1/2}\sum_{i=1}^{\left\lfloor nq_{a}\right\rfloor }\psi(Y_{i}^{(a)})\right|^{2}$,
is uniformly integrable. Take $a=1$ without loss of generality and
note that by Assumption 6(ii), 
\begin{align}
 & \sup_{q_{1}\in[0,1]}\left|\frac{1}{\sqrt{n}}\sum_{i=1}^{\left\lfloor nq_{1}\right\rfloor }\psi(Y_{i}^{(1)})\right|^{2}\nonumber \\
 & \le2\sup_{q_{1}\in[0,1]}\left|\frac{1}{\sqrt{n}}\sum_{i=1}^{\left\lfloor nq_{1}\right\rfloor }\left\{ \psi(Y_{i}^{(1)})-\mathbb{E}_{n,\bm{h}}\left[\psi(Y_{i}^{(1)})\right]\right\} \right|^{2}+4\left|I_{a}h^{(a)}\right|^{2}+4\delta_{n}\vert h^{(a)}\vert^{2}.\label{eq:UI:proof}
\end{align}
Define $A_{n,i}:=\psi(Y_{i}^{(1)})-\mathbb{E}_{n,\bm{h}}\left[\psi(Y_{i}^{(1)})\right]$
and observe that $M_{k}:=n^{-1/2}\sum_{i=1}^{k}A_{n,i}$ is a martingale
under $\mathbb{P}_{n,\bm{h}}.$ Then, for any $p>0$, 
\begin{align*}
 & \mathbb{E}_{n,\bm{h}}\left[\sup_{q_{1}\in[0,1]}\left|\frac{1}{\sqrt{n}}\sum_{i=1}^{\left\lfloor nq_{1}\right\rfloor }A_{n,i}\right|^{2+p}\right]=\mathbb{E}_{n,\bm{h}}\left[\sup_{k\le n}\left|M_{k}\right|^{2+p}\right]\\
 & \le\left(\frac{2+p}{1+p}\right)^{2+p}\mathbb{E}_{n,\bm{h}}\left[\left|M_{n}\right|^{2+p}\right]\apprle\mathbb{E}_{n,\bm{h}}\left[\left|A_{n,i}\right|^{2+p}\right]<\infty,
\end{align*}
where the second step follows by Doob's maximal inequality, the penultimate
step uses the Marcinkiewicz--Zygmund inequality, and the last step
follows from Assumption 6(i). This proves that $\sup_{q_{1}\in[0,1]}\left|n^{-1/2}\sum_{i=1}^{\left\lfloor nq_{1}\right\rfloor }A_{n,i}\right|^{2}$
is uniformly integrable, and therefore, in view of (\ref{eq:UI:proof}),
that $\sup_{q_{1}}\left|n^{-1/2}\sum_{i=1}^{\left\lfloor nq_{1}\right\rfloor }\psi(Y_{i}^{(1)})\right|^{2}$
is uniformly integrable as well. 

\subsubsection*{Upper bound on GRO value}

We first show that the GRO value, $R_{n}(\varepsilon_{n}^{*};\bm{h},\{q_{n,a}(t)\}_{a})$,
of $\varepsilon_{n}^{*}(\cdot)$ is always non-negative for $n$ large
enough. To this end, observe that by Jensen's inequality, 
\[
\ln\varepsilon_{n}^{*}(q_{n,1}(t),q_{n,0}(t))\ge\int\sum_{a}\left\{ h^{(a)\intercal}I_{a}^{1/2}z_{n,a}(q_{n,a})-\frac{q_{n,a}}{2}h^{(a)\intercal}I_{a}h^{(a)}\right\} dw(\bm{h}).
\]
Consequently, Fubini's theorem implies 
\[
R_{n}(\varepsilon_{n}^{*};\bm{h},\{q_{n,a}(t)\}_{a})\ge\int\sum_{a}\left\{ h^{(a)\intercal}I_{a}^{1/2}\mathbb{E}_{n,\bm{h}}\left[z_{n,a}(q_{n,a})\right]-\frac{\mathbb{E}_{n,\bm{h}}[q_{n,a}]}{2}h^{(a)\intercal}I_{a}h^{(a)}\right\} dw(\bm{h}).
\]
Now, Wald's identity and Assumption 6(ii) imply 
\[
\mathbb{E}_{n,\bm{h}}\left[z_{n,a}(q_{a})\right]=I_{a}^{-1/2}\frac{n\mathbb{E}_{n,\bm{h}}\left[q_{n,a}\right]\mathbb{E}_{n,\bm{h}}\left[\psi(Y_{i}^{(a)})\right]}{\sqrt{n}}=\mathbb{E}_{n,\bm{h}}\left[q_{n,a}\right]\left(I_{a}^{1/2}h^{(a)}+\epsilon_{n}\vert h^{(a)}\vert\right),
\]
where $\epsilon_{n}\to0$. Since $h^{(a)\intercal}I_{a}h^{(a)}\ge2\epsilon_{n}\vert h^{(a)}\vert^{2}\ \forall\ a$
for $n$ large enough, we thus have
\[
R_{n}(\varepsilon_{n}^{*};\bm{h},\{q_{n,a}(t)\}_{a})\ge\sum_{a}\int\frac{\mathbb{E}_{n,\bm{h}}[q_{n,a}]}{2}\left(h^{(a)\intercal}I_{a}h^{(a)}-2\epsilon_{n}\vert h^{(a)}\vert^{2}\right)dw(\bm{h})\ge0.
\]

Finally, to obtain an upper bound, note that (\ref{eq:UI:proof})
and the steps below it imply 
\[
R_{n}(\varepsilon_{n}^{*};\bm{h},\{q_{n,a}(t)\}_{a})\le\mathbb{E}_{n,\bm{h}}\left[\sup_{q_{1},q_{0}}\ln\varepsilon_{n}^{*}(q_{1},q_{0})\right]\lesssim\left(1+\left|I_{a}h^{(a)}\right|^{2}\right).
\]
Combined with the lower bound, this implies that the statement of
Corollary \ref{Cor:mGRO} can be verified by setting $g(\bm{h})=C(1+|\bm{h}|^{2})$
for some $C<\infty$.

\subsection{Locally REGROW optimal e-processes}

In this section, we construct e-processes that are locally optimal
against fixed values of $\{q_{a}\}_{a}$. These e-processes therefore
achieve REGROW optimality with respect to non-adaptive sampling designs
where the allocations are fixed in advance. 

For a given pair $(q_{1},q_{0})$, equation (\ref{eq:l.f distribution - REGROW - dual})
implies that the optimal weighting function $w_{(q_{1},q_{0})}^{*}$
is obtained by solving 
\begin{equation}
w_{(q_{1},q_{0})}^{*}=\argmax_{w\in\Delta(\mathcal{H}_{1})}\textrm{KL}_{(q_{1},q_{0})}\left(p_{\bm{h}}\cdot w\mid\mid p_{w}\cdot w\right)=\argmax_{w\in\Delta(\mathcal{H}_{1})}I_{(q_{1},q_{0})}\left(w;p_{w}\right),\label{eq:l.f distribution - REGROW - fixed_value}
\end{equation}
where $\mathbb{P}_{\bm{h}}^{(q_{1},q_{0})}$ corresponds to an independent
bivariate-normal distribution 
\[
\prod_{a}\mathcal{N}(I_{a}^{1/2}q_{a}h^{(a)},q_{a}).
\]
The channel capacity and $w^{*}$ depend on the structure of the compact
set $\mathcal{H}_{1}$. 

As a leading example---following \citet{grunwald2024safe}---consider
the case where each $h_{a}$ is scalar and
\[
\mathcal{H}_{1}\equiv\left\{ \bm{h}:\vert h^{(a)}\vert\le K\ \forall a\right\} ,
\]
for some constant $K<\infty$. In this setting, the optimization problem
(\ref{eq:l.f distribution - REGROW - fixed_value}) factorizes across
arms, yielding a product form solution $w^{*}=\prod_{a}w_{a}^{*}(h^{(a)})$,
and the overall channel capacity becomes the sum of individual channel
capacities for each arm. Classical results from \citet{smith1971}
establish that the least-favorable distribution $w_{a}^{*}(\cdot)$
is always discrete. For low signal to noise ratios, specifically,
when $K/I_{a}^{1/2}q_{a}\lessapprox1.4$, $w_{a}^{*}(\cdot)$ reduces
to a symmetric two-point prior supported on $\{-K,K\}$. However,
when $q_{a}$ falls below the critical threshold $K/(1.4I_{a}^{1/2})$,
the support of $w_{a}^{*}$ expands to include more than two points
and must be computed via a finite-dimensional convex program, as described
in \citet{smith1971}. 

Importantly, because $w^{*}$ depends on $(q_{1},q_{0})$, there is
no e-process that is simultaneously REGROW optimal for all possible
allocation pairs. Nevertheless, when $(q_{1},q_{0})$ are sufficiently
large, $w^{*}$ stabilizes at the symmetric two-point prior on $\{-K,K\}$.
The corresponding e-process
\[
\varepsilon^{*}(q_{1},q_{0})=\prod_{a}\frac{1}{2}\sum_{h=\pm K}\exp\sum_{a}\left\{ hI_{a}^{1/2}z_{a}(q_{a})-\frac{q_{a}}{2I_{a}}h^{2}\right\} 
\]
is therefore REGROW optimal at these values. As before, a corresponding
finite-sample approximation can be constructed by replacing $z_{a}(\cdot)$
with its empirical counterpart $z_{n,a}(\cdot)$. 

\subsection{Testing parameters corresponding to individual arms}

Mirroring the setup of Section \ref{subsec:Illustrative-example-(contd.)-anytime},
suppose we are interested in conducting an anytime-valid test of the
null hypothesis $H_{0}:\theta^{(1)}=\bar{\theta}$ against the two-sided
alternative $H_{1}:\theta^{(1)}\neq\bar{\theta}$. Since $\theta^{(0)}$
is unrestricted, this corresponds to a testing problem with a composite
null $\Theta_{0}\equiv\{(\bar{\theta},\theta^{(0)}):\theta^{(0)}\in\mathbb{R}\}$
and a composite alternative $\Theta_{1}\equiv\{(\theta^{(1)},\theta^{(0)}):\theta^{(1)}\neq\bar{\theta},\theta^{(0)}\in\mathbb{R}\}$. 

We take the reference parameter vector to be $\bm{\theta}_{0}=(\bar{\theta},\bar{\theta})$,
and consider local alternatives of the form $\bm{\theta}=\bm{\theta}_{0}+\bm{h}/\sqrt{n}$.
The partitions $\Theta_{0},\Theta_{1}$ of the space of $\bm{\theta}$
induce a corresponding partition of the local parameter space which
we denote by
\[
\mathcal{H}_{0}\equiv\{(0,h^{(0)}):h^{(0)}\in\mathbb{R}\}\quad\text{and}\quad\mathcal{H}_{1}\equiv\{(h^{(1)},h^{(0)}):h^{(1)}\neq0,h^{(0)}\in\mathbb{R}\}.
\]

Assume a prior (or weight function) $w_{1}(h^{(1)})$ is placed over
the alternative values of $h^{(1)}$. Fix any $h^{(0)}\in\mathbb{R}$.
Then, by (\ref{eq:asymptotic_mGRO}), the unique mGRO-optimal e-process
for testing the simple null $\bar{H}_{0}:\bm{h}=(0,h^{(0)})$ against
the composite alternative $\bar{H}_{1}:\bm{h}\in\{(h^{(1)},h^{(0)}):h^{(1)}\neq0\}$,
given $w_{1}(\cdot)$, is independent of observations from arm 0 and
takes the form:
\[
\varepsilon^{*}(q_{1}(t))=\int\exp\left\{ h^{(1)\intercal}I_{1}^{1/2}z_{1}(q_{1}(t))-\frac{q_{1}(t)}{2}h^{(1)\intercal}I_{1}h^{(1)}\right\} dw_{1}(h^{(1)}).
\]

This process is clearly a valid e-process for testing the composite
null $\bm{h}\in\mathcal{H}_{0}$ against the composite alternative
$\bm{h}\in\mathcal{H}_{1}$. Moreover, since $\varepsilon^{*}(q_{1}(t))$
is mGRO-optimal for each fixed value of $h^{(0)}$, it follows that
it is also mGRO-optimal for testing $\bm{h}\in\mathcal{H}_{0}$ against
$\bm{h}\in\mathcal{H}_{1}$, when there is a weight function $w_{1}(\cdot)$
over $h^{(1)}$.

\subsection{Additional simulation results}

\begin{figure}
\includegraphics[height=5cm]{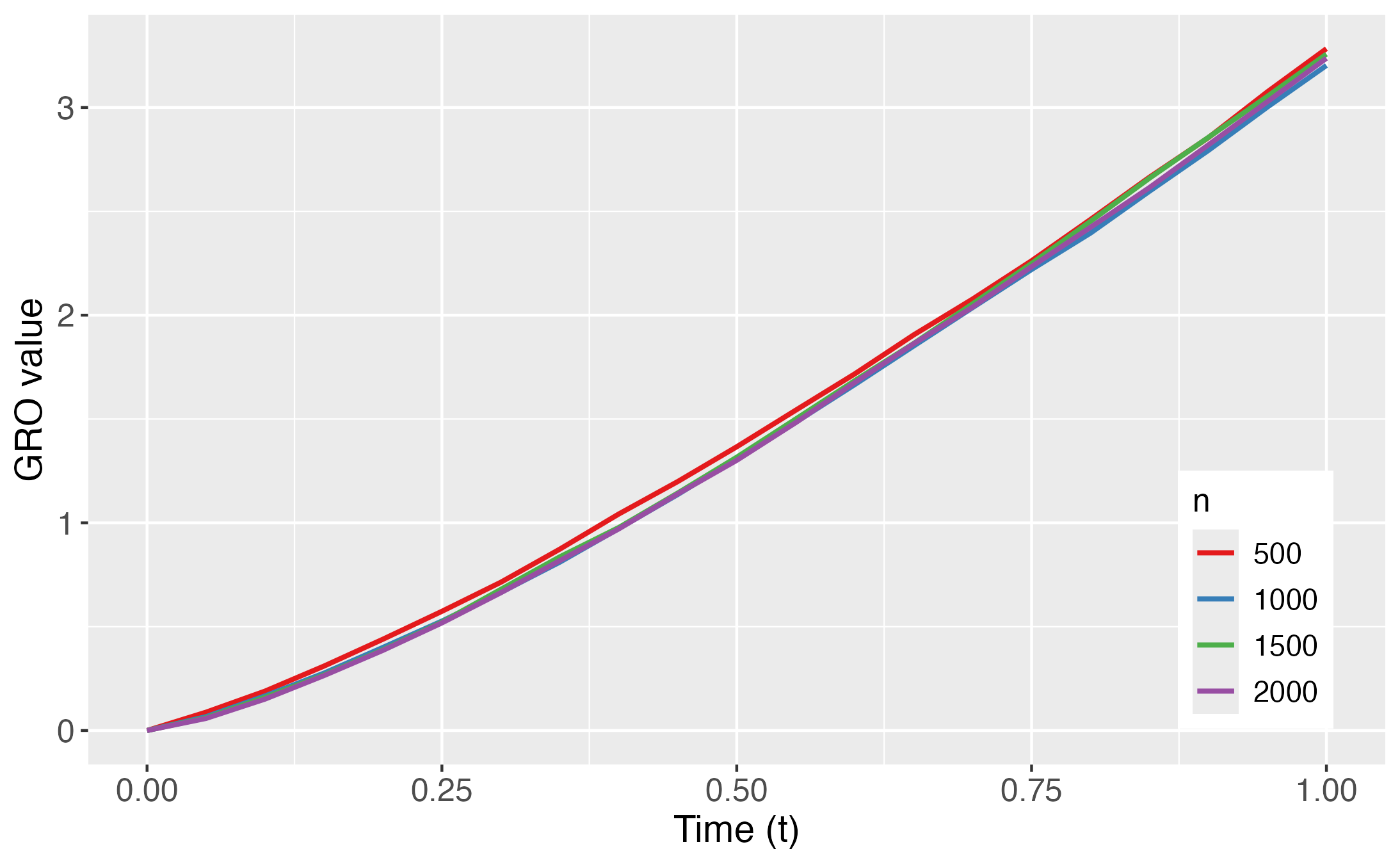}{\scriptsize}\\
{\scriptsize Note: The figure displays the GRO value as a function
of time for the e-process in (\ref{eq:e-process_application}) under
Thompson Sampling, at the local alternative $(h^{(1)},h^{(0)})=(1/\sqrt{n},0)$.}{\scriptsize\par}

\caption{Evolution of GRO under Thompson Sampling\label{fig:Any_time_valid_inference-TS}}
\end{figure}

Figure \ref{fig:Any_time_valid_inference-TS} replicates the analysis
in Panel B of Figure \ref{fig:Any_time_valid_inference}, but replaces
the UCB algorithm with Thompson Sampling (TS). The resulting curves
again exhibit striking stability across values of $n$. Notably, the
e-process defined in equation (\ref{eq:e-process_application}) achieves
uniformly higher GRO values under TS than under UCB. This difference
arises from the fact that we only consider alternatives where $\theta^{(1)}=0.1+1/\sqrt{n}$
and $\theta^{(0)}=0.1$. Under this class of alternatives, UCB allocates
fewer observations, on average, to arm 1---the arm being tested---than
TS does. As a result, the TS-based allocation yields a more powerful
e-process in this setting.

\end{document}